\documentclass[12pt]{article}
\usepackage{amssymb,graphicx}

\newcommand{\bo}{{\bar o}}

\def\h{\eta}

\def\o{\omega}

\def\q{\theta}

\def\bo{{\raise.15ex\hbox{\large$\Box$}}}               % D'Alembertian

\def\face{{\raise.2ex\hbox{$\displaystyle \bigodot$}\mskip-2.2mu \llap {$\ddot
        \smile$}}}                                      % happy face
\def\leftrightarrowfill{$\mathsurround=0pt \mathord\leftarrow \mkern-6mu
        \cleaders\hbox{$\mkern-2mu \mathord- \mkern-2mu$}\hfill
        \mkern-6mu \mathord\rightarrow$}       % <--> double differential
\def\dvec#1{\vbox{\ialign{##\crcr
        \leftrightarrowfill\crcr\noalign{\kern-1pt\nointerlineskip}
        $\hfil\displaystyle{#1}\hfil$\crcr}}}           % <--> accent
\def\beq{\begin{equation}}
\def\eeq{\end{equation}}

\def\beqx{\begin{displaymath}}
\def\eeqx{\end{displaymath}}

\def\beql{\begin{eqnarray}}
\def\eeql{\end{eqnarray}}

\newcommand{\bea}{\begin{eqnarray}}
\newcommand{\eea}{\end{eqnarray}}

\newcommand{\half}{\frac 12 }

\def\Im{\rm Im \;}

\def\[{\left [}
\def\]{\right ]}
\def\({\left (}
\def\){\right )}
\def\C{\Gamma}

\def\omit#1{ }

\newif\ifPreprint
\Preprinttrue
\def\+{\oplus}

\begin{document}

%\vspace*{0.15in}
\begin{titlepage}

\ifPreprint
\hbox{\hskip 12cm NIKHEF/2008-011\hfil}
\hbox{\hskip 12cm July 2008  \hfil}
\hbox{\hskip 12cm hep-th/yymmnnn \hfil}
\fi

\vskip .5in

\begin{center}
{{\Large \bf  The Emperor's Last Clothes?}\\ \vskip .1in
\large \it  Overlooking the String Theory Landscape\footnote{Based on colloquia in DESY, Hamburg, and at the University of Utrecht. This is the
extended version of an invited contribution to  ``Reports on Progress in Physics" \cite{RPP}.}}

\vspace*{.5in}
{ A.N. Schellekens}% \footnote{t58@nikhef.nl} 
\\
\vskip .2in

{\em NIKHEF Theory Group, Kruislaan 409, \\
1098 SJ Amsterdam, The Netherlands} \\

\vskip .2in

{\em IMAPP, Radboud Universiteit,  Nijmegen}
\vskip .2in

{\em Instituto de F\'\i sica Fundamental, CSIC, Madrid}

\end{center}

%\begin{center}
\vspace*{0.5in}
\centerline{{\bf Abstract}}
\vspace*{0.1in}
\noindent We are in the middle of a remarkable
paradigm shift in particle physics, a shift of opinion that
occurred so slowly that some even try to deny that they changed their minds at all.
It concerns a very basic
question: can we expect to derive the laws of particle
physics from a fundamental theory? The Standard Model of particle physics
as well as the 1984 string theory revolution provided ample food for thought
about this.
The reason this was ignored for so long can be traced back to an old fallacy: 
a misguided idea about our own importance.

%\end{center}
\end{titlepage}

\vskip 1in

\noindent

\newpage

\section{Introduction}

In the last five years, a big debate has erupted concerning 
 a very basic question: are the laws of physics we observe unique, or
are there various distinct possibilities? Understanding this requires a definition of what
is meant  by ``the laws of physics" and what could be non-unique about them, and
we will get to that later. But even before going into detail it should be obvious
that the answer to this question is of fundamental importance. It would have interested
Einstein very much, judging from
his famous phrase ``{\it I wonder if God had any choice when he made our
universe.}" 
The answer to this question touches on deep existential questions that
have troubled mankind for centuries. If different laws of physics are
possible, we -- or anything else of comparable intelligence -- might not exist in some of
those  different circumstances. 

Physicist tend to get nervous about this sort of statement, and not without reason. 
Ideas of this  kind go by the generic name ``anthropic principle", but there are many
different formulations, ranging from tautological to just plain ridiculous. 
But 
advances in string theory in the past three decades have moved these questions
from the area of philosophy into physics and mathematics. 
String theory suggests a very concrete answer to this question: it suggests that the laws  of physics
(and in particular the properties of fundamental particles and interactions)
are  highly non-unique, but in a precise and quantifiable way. 
Whether this is indeed a correct statement about string theory 
can be decided without invoking philosophy.
It is a matter of using the traditional methods of
physics: finding controllable 
approximations, performing calculations, and making use of the powerful machinery
of mathematics.

While it is more and more widely accepted that string theory does indeed have 
this property, it remains to be demonstrated
that string theory is indeed the fundamental
theory of all interactions that it is hoped to be, and of course the answer a theory gives is
irrelevant if that theory is wrong. But it already deserves credit for forcing us
to ask the question about uniqueness of the laws of physics in a much more precise way.
This was never done before, because we never had a candidate theory that was so
ambitious.

Initially, when string theory was touted as the ``theory of everything" around 1984,
there were hopes it would lead to exactly the opposite: a unique derivation of all
the laws of physics. 
Evidence that quite the opposite was true
  started emerging almost immediately after 1984, but
most people chose to ignore it.
In 2003, after important
additional evidence had been found,
Leonard Susskind published a 
paper \cite{ALS} entitled ``The Anthropic Landscape of String Theory", which finally started a debate
that should have started fifteen years earlier. What is at stake in this debate is not 
only the uniqueness of our universe, but also the fate of string theory as a fundamental
theory of all interactions.

In my opinion string theory gives the right answer, and the fact that it does adds
to the evidence in its favour. 
I can say this
without being accused of trying to put a positive spin on the recent developments, because
I actually wrote in 1998 \cite{DutchText} that I hoped string theory would ultimately lead to a huge number
of  possible choices for the laws of physics, a point of view I have been advocating
since the late eighties. \ifPreprint I reached that conclusion 
after having been involved in one of the first papers \cite{Lerche} pointing out that
the number of possibilities was humongous.  
Observing that the
number of possibilities was huge, in the approximation used at the time, was a relatively
simple matter. But it was not obvious what to conclude from that. 
Most people blamed on our lack of understanding of the theory.
Indeed,
there were, and still are, many serious problems to be addressed.  

My own conclusion was that 
we were
in the process of uncovering a fundamental property of nature. String theory \else
Almost from the start (which for me was around  1985) string theory \fi
suggested a fantastic solution to a deep mystery:  why
the laws of physics we observe seem engineered especially for us to exist, even though
theoretically there seem to be many other possibilities.   We all hope to live during a time
when big things are happening in our field, and I have never doubted that this is
one of those things.
I have spent the last
twenty years trying to convey my sense of excitement to my colleagues, 
but with little success. 
But in the last few years I have been delighted to see more evidence coming in supporting
this point of view, so that 
the mood has started to change.
I hope this is the right time to make one more attempt.

Since I have been 
defending
this point of view for a long time already, I knew it was controversial.
Therefore I was not surprised that many people rejected Susskind's conclusion or called
it ``premature". This is a reaction I can understand.
Scientific debates about the validity of approximations or of the validity of a  theory
are normal.
However, there was another kind of reaction to Susskind's paper that inspired
me to choose the title of this paper.
Some string theorists claimed that they had reached the same 
conclusion already a long time ago, but did
not find it worth mentioning. 
Apparently they did not find it worthwhile to point out that many 
great physicists had the wrong expectations about fundamental physics.
They do no think it is of any interest to know that our universe is just one of a huge
number  of theoretically possible ones, and that
it is fundamentally impossible to derive it  in full detail
from a fundamental theory. \ifPreprint They do not seem to  care about the existential consequences
of this large number of possibilities.
They do not find it important that a theory they have
invested years of their life in has lead us to such a startling conclusion, and they
do not seem to  care about the implications for the correctness of that theory. \fi
%\ifPreprint
%It has even been suggested to me that the reason Susskind and I find the result so
%important is that we 
%are old people
%who grew up in a period where the Standard Model of elementary particles
%was believed to be unique. 
%It is certainly true that the older generation of physicist have more difficulties accepting
%that this notion is wrong than the younger ones, but I think this is mostly because a large group
%of string theorist has gotten used to the idea that string theory allows many 
%options for matter and non-gravitational interactions. This
%does not mean that they have accepted the ultimate consequences of that observation. Is this feature 
%a defect of
%(our understanding of) string theory or a fundamental property of nature? 
%\vskip 100cm
%\fi 
Or do they secretly hope that the conclusion will turn out to be wrong after all?
Sadly, since nobody  found it worthwhile
to speak or write about this, we will never know what they really thought.

% If I talk to younger people, from before the we-knew-it-already-but-did-not-tell-anyone
%generation, and who have not yet been 
%exposed to string theory, or experimentalists who follow string theory
%only from a distance,
%I find that they are equally shocked as I was when they realize what the consequences are, and
%often reject the whole idea for that reason.

\ifPreprint
String theorists who say that the conclusion is premature could simply be right.
String theorists who claim that they think it is correct, but that
they ``knew this already" and did not find it worth mentioning 
have missed one of the most fundamental scientific discoveries of the last decades. 
Unless 
it is all wrong, but that seems hardly an attractive alternative for a 
string theorist. 
\fi

The inspiration for the title of this paper is the well-known fairy tale about the emperor
and his new clothes, advertised to be invisible only to stupid and incompetent people.
Since nobody wanted to belong to that category, all claimed to be able to see the clothes,
until a small child said ``but he is not wearing anything". Then suddenly everyone claimed
to have noticed this already. 

%I am sure the reader understands who plays the role of the child in
%the fairy tale analogy, but there is more to be said.

\subsection{Unification versus uniqueness}\label{SecUU}

But there is another moral  that fits the recent events.
Two concepts that played an important role in the
history of physics are {\it unification} and {\it uniqueness}. Sometimes these concepts
seem to be inadvertently confused.

Unification means reducing the laws of physics to
fewer and fewer fundamental principles. This has been
a fantastic success. Many breakthroughs in physics can be viewed in terms
of unification. The gravitational force acting on object on the surface of the earth
turned out to be the same as the ones governing
the motion of planets; inertial mass the same as gravitational mass;
electricity was found to be related to magnetism, and the unified
theory of both forces is responsible for such diverse phenomena as
light and atomic physics, chemistry and the structure of matter.
Furthermore, in the second half of last century 
we learned that two other forces of nature, the weak force
(responsible for radioactive decays) and the strong force 
(binding quarks into nucleons), are based on the same
principles as electromagnetism, the principle of local gauge invariance.
If string theory is correct the  success of unification continues in a spectacular way.
It unifies all interactions with gravity, and all matter with interactions.

Uniqueness concerns the question if the laws of physics we
observe in our universe are the only ones possible. 
Historically, whenever alternatives were imaginable, the hypothesis
of uniqueness has almost systematically been a failure. 
The origin of this fallacy can often 
be traced back to an -- explicit or implicit --  belief that there is something special
about us, or about our planet, or our solar system.   
For centuries it was believed that the entire universe was a unique
structure with the earth in its center. This was corrected by
Copernicus, who put the sun in the center of the solar system. But he also put
the sun in the center of the universe, a similar fallacy.
The idea that the solar system was something special is evident in the work of
Kepler, who tried to derive the sizes of the planetary
orbits from a mathematical principle. He was even moderately successful in
identifying the ratios of these  sizes for the first five planets using nested platonic
solids.  
\ifPreprint 
The fact that he tried this implies that he was thinking of the solar system in terms 
of a
mathematically unique system.  Giordano Bruno speculated about other solar systems with
other kinds of life in them,
and paid dearly for that (although not only for that).
This  particular issue is of course still undecided, but it illustrates the strong sentiments that arguments
against our uniqueness evoke. Fortunately they have become a  bit less emotional more recently.
In 1920 there was the  so-called  ``Great Debate" 
(it got the name because it concerned the scale of the universe)
between the
astronomers Shapley  and Curtis. The former defended the point of view that our galaxy was
the entire universe, but that our solar system was not in its center, while the latter claimed
that the then observed  ``spiral nebulae" were actually other galaxies. However, Curtis
also defended the point of view that the Milky Way was centered  around our solar
system. Each defended an anthropocentric statement, and was fully aware of that. 
Each of them later turned out to be wrong on that statement. 
The mistaken concept of human uniqueness not only plagues physics and
astronomy, but also biology, and is at least to some
extent the origin of the strong resistance against evolution, which took the human
being from the pedestal where it once placed itself. \fi

The debate that is the subject of this article fits nicely in
the foregoing list.  Although it was triggered by string theory,
it could and should
have started independently. It is the structure of the Standard Model itself that raises the
question. String theory suggest an answer
that would fit perfectly in the brief history presented above.

Of course a historical
analogy can never be used as a proof.  However, it can be used
to understand the importance of a conclusion, and the resistance against it. 
In the course of history, 
ideas of this kind have never
been received with  much enthusiasm, as 
Galileo and Giordano Bruno found out. Unification sells much better than non-uniqueness.

This is the second moral: the emperor is us, and the beautiful clothes are the strange
ideas we have about our place in the universe.  Repeatedly we have been shown to be
naked, and I am convinced that in the last decades this has been happening again.

\ifPreprint Let me make it clear that I am not trying 
to portray great scientists like Ptolemy, Copernicus, Kepler and others
as fools. My point is quite the opposite: to show that even the greatest minds of their time
can make the obvious anthropocentric mistake of thinking that what we see around us
is all there is. I might be making that mistake myself somewhere in this article, and I will
try to point out where that could be the case. \fi

\subsection{Contents}

This article has two parts, devoted to two essential ingredients of the debate. The
first part follows a ``bottom-up" way of thinking. It contains a brief introduction to the
Standard Model of particle physics, which leads us naturally and inevitably to the question
how it might come out of a fundamental theory, and whether we can expect this to happen in a 
unique way. I will explain in particular why the Standard Model forces us to think about that
question much more seriously than any of its predecessors. The second part takes a ``top-down"
point of view. I will consider string theory as an example of what a fundamental theory
of quantum gravity might have to offer as an answer to this question. In between
these two topics I will discuss a concept that inevitably comes up in connection with
these discussions, and that tends to confuse the issue and stir up strong emotions: The ``anthropic principle".

The main focus of this article is on the Standard Model of particle physics and 
the anthropic implications of its embedding in
string theory. Other essential ingredients
in the history of anthropic arguments, in particular from cosmology, will only be mentioned briefly. 

The assumed background knowledge is quantum mechanics, some basic group theory and 
some facts about elementary particle physics. A few parts on the latter subject are perhaps a bit
more advanced, but they can be skipped without missing the main points.

This article summarizes twenty years of passionate and vigorous discussions.
It contains answers to  most of the questions that came up during these debates or
afterwards. I wish to thank everyone who contributed their point of view and
thereby influenced mine.

\section{The Standard Model}

\subsection{Towards the Standard Model}

In the beginning of last century, after Einstein formulated his theory
of gravity, General Relativity, it may have been natural to assume that 
a unique and unified theory might be emerging. Both Maxwell's theory 
of electromagnetism and General Relativity have  an air of uniqueness 
about them. Could the laws of physics be anything else than this? \ifPreprint The
quest for a unified theory dominated the rest of Einstein's scientific career,
and it is plausible that he also hoped or expected the result to be unique.\fi

But then the experimental facts started pointing in a different direction. New interactions
were discovered that did not seem to have an equally simple and elegant
description: the strong and the weak interaction.
New particles were discovered that nobody needed, such as the muon. And
new parameters emerged that did not seem to have a simple explanation, such
as the masses of all these particles. 
\ifPreprint The resistance against new particles was so strong
that when Dirac discovered the positron he was so reluctant to introduce
a new particle that he tried to identify it with the proton, even though he knew that the
positron and the proton had totally different masses. 
\fi

An important theoretical development also put the idea of uniqueness into doubt. Yang-Mills theory,
a generalization of electrodynamics, was discovered. Electrodynamics
is a ``gauge theory" based on the group $U(1)$. For readers not familiar with the concept
of a gauge theory the only important point to note is that $U(1)$ is merely an example
of a group that can be used in their construction. But there are many other possibilities that
work equally well from a theoretical point of view: for example $SU(N)$ and $SO(N)$ for any integer $N$.
It just so happens that the first example of a gauge theory we encountered
in nature was based on $U(1)$, but that does not make it unique \ifPreprint (actually, the physics of 
the other possibilities is more complicated, making them much harder to observe directly)\fi.

Indeed, nature does make use of these other possibilities: $SU(3)$ for the strong
interactions and $SU(2)$ for the weak interactions. This result emerged during the sixties
and seventies, and put all the interactions once again on a common conceptual footing:
all turned out to be based on the principle of ``local gauge invariance", which is
a close relative of the concept of general coordinate invariance that underlies
General Relativity. This unification of concepts generated some new dreams of
unification and uniqueness, but on the other hand it was clear that the Standard
Model did not really look like something mathematically unique. 

\subsection{The Standard Model}

Let us have a closer look at the Standard Model. The only feature that
I want to emphasis is the choices involved in its definition. We have already seen
that the first choice is a gauge group, which is $SU(3) \times SU(2) \times U(1)$.
There is no theoretical justification for this choice. It is just what we observe\ifPreprint, albeit
rather indirectly and through tremendous experimental efforts. Each factor implies the existence of an interaction, which may be thought of as
a kind of generalization of electrodynamics. Indeed, the first factor is responsible for
the theory of the strong interactions, quantum chromodynamics (QCD), whereas the other two
combine to yield the weak and electromagnetic interactions. \else .\fi

%It is of course the
%theory of the strong, weak and electromagnetic interactions between quarks and leptons. 
%Specifying it requires several choices. First we choose a ``gauge group", which
%in this case is $SU(3) \times SU(2) \times U(1)$. The precise meaning
%of this concept will not be explained here, but it is undoubtedly clear to the reader that
%other choices are mathematically allowed. We could have chosen $SU(m)$ or $U(n)$ for any
%other value of $m$ and $n$, we could have started with an arbitrary number of factors, and
%there are other groups than the unitary ones that happen to occur in the description of the
%Standard Model. Each factor implies the existence of an interaction, which may be thought of as
%a kind of generalization of electrodynamics. Indeed, the first factor is responsible for
%the theory of the strong interactions, quantum chromodynamics (QCD), whereas the other two
%combine to yield the weak and electromagnetic interactions. 

Next we must specify matter for these interactions to couple to. This is done simply by
specifying a spin and a set of charges. The charges, more generally ``representations"
of the Standard Model gauge group $SU(3) \times SU(2) \times U(1)$, specify how the
corresponding particle couples to the three fundamental interactions. 
\ifPreprint
All fundamental
matter particles observed so far have spin $\frac12$. They fall into two classes: those
that have QCD interactions (the quarks) and those that do not (the leptons). 
Mathematically speaking 
there are infinitely many possible choices for the charges/representations, and just one of them
agrees with experimental observation.  It is usually written as follows\footnote{Right-handed neutrinos
have been ignored here.}
\beq
\label{SMrep}
({\bf 3},{\bf 2},\frac16)_L+({\bf 3},{\bf 1},\frac23)_R+({\bf 3},{\bf 1},-\frac13)_R+({\bf 1},{\bf 2},-\frac12)_L+({\bf 1},{\bf 1},-1)_R \ ,
\eeq
where the three entries indicate the coupling to $SU(3)$, $SU(2)$ and $U(1)$ respectively,
and $L$ and $R$ denote left- and right-handed. Although I am showing these representations
in detail, I do not expect all readers to understand what  the notation means. I only
want to emphasize that this is just a choice. Indeed many other choices are possible;
one may change all the numbers appearing in (\ref{SMrep}), subject to a few consistency conditions.
These conditions allow an infinity of possibilities, and there is nothing in the formulation of the
\else
The choice we observe, quarks and leptons, is one of 
the many theoretical possibilities, and there is nothing in the formulation of the \fi
Standard Model that specifies this particular choice. Furthermore,
for unknown reasons, this entire set is repeated three times: the matter particles belong
to three ``families". Each of these families has identical couplings to the Standard Model
interactions; each consists of two quarks and two leptons. 

We may add particles of spin 0 (scalar particles), and they may also have all
kinds of charges. So far no such particle has been observed, but one is expected: the
famous Higgs particle. \ifPreprint To play the r\^ole of  the 
Higgs particle it must have a very definite set
of charges, but mathematically there is once again an infinity of possibilities. 
The gauge group itself implies the existence of particles of spin 1; these are
completely fixed and correspond to the photon (responsible for all
electromagnetic phenomena), the gluons (particles that bind quarks together
to form protons and neutrons) and the $W$ and $Z$ particles (vector bosons with a 
mass of about eighty times the proton mass that are responsible for weak decays).\else 
Its properties are dictated by phenomenology, but not by fundamental theory.
\fi

There is one final set of ingredients to be specified. The spins and charges of the particles
determine precisely how they can couple to each other, up to some numerical constants,
the so-called coupling constants. These are real numbers, and any value is in principle
allowed. The Standard Model has about\footnote{The precise number depends
on assumptions regarding the number of right-handed neutrinos and  the existence of Majorana masses.\ifPreprint
The most common set of assumptions (three right-handed neutrinos and the existence of both Dirac and Majorana
mass components)  gives three masses and six mixing angles for  neutrinos, and  a total of 28 parameters. \fi} 28 of them. \ifPreprint Only a few of them really play a r\^ole
in the rest of the story, but for completeness I will list all  of them:
\begin{itemize}
\item{Three gauge couplings $g_1$, $g_2$ and $g_3$.}
\item{The nine masses of the quarks $u,d,c,s,t,b$ and the charged leptons $e$,$\mu$,$\tau$.}
\item{Four weak interaction mixing angles of the quarks.}
\item{The Higgs boson mass parameter $\mu^2$ and the Higgs self-coupling $\lambda$.}
\item{The strong interaction CP-violation parameter $\theta$.}
\item{Neutrino mass and leptonic weak interaction mixing parameters.}
\end{itemize}
Apart from the neutrino sector, where
a lot is still unknown, the only parameter that has not been measured yet is a combination  of $\mu^2$ and $\lambda$.
We do know the value of the combination $v=2\sqrt{-\mu^2/\lambda}$, known as the ``vacuum expectation value of the
Higgs field".  It has the dimension of a mass, and in the conventional units of particle physics  its value is  
246 GeV\rlap.\footnote{I will
follow the standard conventions of particle physics and express all energies in electron-Volts (eV).
Furthermore all masses and lengths are also converted to eV by inserting appropriate
powers of $\hbar$ and $c$. For comparison: the proton mass is about one GeV (938 MeV) in these units.}
If and when the Higgs boson is  discovered we will know its mass, which is
given by $\sqrt{-2\mu^2}$ (note that $\mu^2 < 0$, although the notation might suggest otherwise), and then  of course
we also  know $\lambda$. 

The quark and charged leptons owe their masses  to their coupling to the Higgs bosons. 
These are called ``Yukawa couplings" and are in fact
dimensionless 3 x 3 matrices mixing the particles of equal charge in the
three families.
Not all parameters in these matrices are observable, but their
eigenvalues are.
Denoting these  as $\lambda_x$
we may write the mass $m_x$ of particle $x$ as $\lambda_x v$.

A combination of $g_1$ and $g_2$, $e=g_1 g_2 / \sqrt{g_1^2+g_2^2}$, yields the famous fine-structure constant
$\alpha=\frac{e^2}{4\pi}$, with a value of about $\frac{1}{137.04}$.
Most of the remaining parameters
are responsible (via the Higgs mechanism) for the masses of the quarks and leptons. 

It should be emphasized that the number of parameters of the Standard Model is
amazingly small in comparison to its achievements. Most physical systems have an
infinite number of parameters (even though in practice only a finite number is relevant).
For example any  function specifying the shape of objects has an infinite number of parameters.
Nevertheless most people think 28 is too many, although 
few people have 
expressed themselves very explicitly on their ultimate expectations. 
One often encounters 
phrases like the following one from the Wikipedia item on the Standard Model:

{\it However, the Standard Model falls short of being a complete theory of fundamental interactions, primarily because of its lack of inclusion of gravity, the fourth known fundamental interaction, but also because of the large number of numerical parameters (such as masses and coupling constants) that must be put ``by hand" into the theory (rather than being derived from first principles)}.

No-one would disagree with the first part of that statement\rlap.\footnote{Except that there are
other ways in which the Standard Model falls short of being complete, most notably
the absence of particles that can account for the large amount of dark, non-standard-model
matter that  appears to be present in our universe.} Most people would also agree that it is
unacceptable (although not for
very concrete reasons) 
that parameter values must be put in by hand. But it is the last, parenthetical remark that  will be especially important:
should we expect that their values can be ``derived from first principles", and what exactly
would that mean?

\fi 

Feynman said the following about 
one of the 28 parameters, the fine-structure constant $\alpha \sim 1/137.04$:

{\it There is a most profound and beautiful question associated with the observed coupling constant, e, the amplitude for a real electron to emit or absorb a real photon. It is a simple number that has been experimentally 
determined [....]. \ifPreprint It has been a mystery ever since it was discovered more than fifty years ago, and all good theoretical physicists put this number up on their wall and worry about it. \fi
Immediately you would like to know where this number for a coupling comes from: is it related to $\pi$ or perhaps to the base of natural logarithms? Nobody knows. It's one of the greatest damn mysteries of physics: a magic number that comes to us with no understanding by man. You might say the ``hand of God" wrote that number, and ``we don't know how He pushed his pencil."}

\ifPreprint

Indeed, some theorists did put the numerical value of $\frac{1}{\alpha}$ on their office door.
Others produced surprisingly accurate formulas for this value. A quick search on the internet
easily produces a handful (the fact that there is more than one undermines the significance
of each of them). There are similar attempts for other quantities, such as the the proton/electron
mass ratio, or quark or lepton mass ratio's. 

In fact, our current understanding of the Standard Model makes it  very unlikely that a parameter like $\alpha$
(as well as the other Standard Model parameters)
has a mathematically defined numerical value. We know that all these parameters are
in fact functions of energy (see section \ref{QFT}). In particle collisions where the incoming particle beams
have energies of order 100 GeV, the value of $\alpha$ is about $\frac{1}{128}$. 
Towards lower energies, the function $\alpha$ would tend to zero, except that its decrease is stopped
by the lowest value that its argument can have, the energy corresponding to the electron mass
(the lightest charged particle).
It would be very surprising if this value could be written as a simple formula
in terms of $e$ or $\pi$. However, it is still possible that the {\it function} is itself unique, and
that it has a nice mathematical value at some special, presumably high, energy.

\fi
\ifPreprint
\else
Many people have tried to find formulas for $\alpha$, as if it were a mathematically unique quantity. 
A quick search on the internet
easily produces a handful (the fact that there is more than one undermines the significance
of each of them). There are similar attempts for other quantities, such as the the proton/electron
mass ratio, or quark or lepton mass ratio's. \fi

\subsection{Beyond the Standard Model: Two paradigms}

\ifPreprint After a breathtaking century of discoveries, to which I did insufficient justice above,
 our understanding of the
fundamental laws of nature has advanced to a remarkable level.
 We now
have a theory of all known non-gravitational interactions that seems almost perfect. 
This theory is called, not very glamorously, the ``Standard Model". One might also
call it ``The theory of almost everything" \cite{Oerter:2006iy}. \fi The story of last century concerned the
exploration of matter at ever shorter distances, from atoms to nuclei, protons and neutrons 
and finally quarks.
How will this story continue to unfold in the present
century? 

%The first few years of this century may have given us an interesting clue that to
%many people was surprising, even shocking or unacceptable.

There are basically two points of view regarding physics at shorter distances. One is that it is like
peeling away the layers of an onion. In this picture new physics should be expected
whenever we go up several orders of magnitude
in energy, or down several orders of magnitude in distance.
This is  certainly what happened during last century, and it is {\it a priori} not 
unreasonable to expect the story to continue that way. \ifPreprint Most ideas in this direction
focus on the Higgs boson, because it is the least understood part of the Standard Model,
and it is responsible for most of its parameters.\fi

However, during the last decades
of last century a different point of view has emerged: perhaps we can peel off the
last layer of the onion in one bold step. If that is the case, one would expect this
bold step to solve also the longstanding problem of consistently combining General Relativity
and quantum mechanics.
Assuming that our universe has three spatial dimension at
all lengths scales,
this puts the energy scale of the new
physics at an inconceivably large value of $1.2 \times 10^{19}$ 
GeV,
the Planck energy\rlap,\footnote{The Planck energy is given by 
$E_{\rm Planck}=\sqrt{\frac{\hbar c^5}{G_N}}$, where $G_N$ is Newton's constant. For comparison: the most powerful
particle accelerator currently under construction, the LHC (CERN, Geneva), will collide protons with each other with
energies of a mere 7 TeV.
\ifPreprint
I will not consider
the possibility of large but unobservable extra dimensions, which effectively lower the
scale of quantum gravity. This does not affect the main points.\fi} corresponding to a length scale of
$1.6 \times 10^{-33}$ cm.

There are several arguments in favor of this
point of view. \ifPreprint The first one is purely practical (and not very convincing).  Looking deeper
and deeper  into the structure of matter becomes more 
and more difficult and expensive, and if
there are too many onion shells still ahead of us, the ``final theory" may elude us forever.
There is, of course, no scientific reason why we should be able to find such a theory, assuming
it exists. It seems almost ridiculous to believe that we,  primitive beings on a speck of dust in 
the universe, should be able to figure out all the laws of physics with our
severely limited equipment and mental abilities. There may be crucial information that we are simply unable to get.
However, \else First of all, \fi
the structure of the Standard Model itself {\it does} give us some reason to hope that such a final bold step may be possible. 
\ifPreprint
It looks like it is more than simply
the next step in a sequence: periodic system, nuclear physics and hadronic physics. \fi
I will discuss this in the next section. 
\ifPreprint 
There is a third
reason.
More or less 
 \else
 The second reason is that
 \fi
 by coincidence we have stumbled on a theory with magical properties
called ``string theory", which looks like a perfect candidate for the center of the onion, the final theory.
This will be the subject of the second half of this paper.

\ifPreprint When one phrases the problem in this way it seems rather natural to expect the
center of the onion to be something unique. One starts dreaming of a ``theory of
everything" from which all known physics can be derived, at least in principle.  
Indeed, we have always been able to move backward in the hierarchy of onion shells.
For example we can derive the spectrum and interactions of protons and neutrons
 from quarks and quantum-chromodynamics,
the fundamental theory of the strong interactions.
We can derive nuclear physics from the strong force mediated by pions 
(themselves quark-anti-quark bound states), atomic physics
from the electromagnetic force between nuclei and electrons. In practice we can do these 
computations 
only in a few simple cases and with sometimes brutal approximations, 
but the principle is not in doubt. 
\fi

The question regarding uniqueness of the Standard Model becomes unavoidable
if one adopts the second paradigm. 
This is not really an issue if one just thinks in terms of moving to the next shell
of the onion. 
%
%In the past this has always led to a reduction of the number of parameters.
%The parameters of Nuclear Physics consist of the masses of a few hundred nuclei, but
%also their magnetic moments, form-factors, etc. If we go down to the next onion shell, hadronic
%physics, many of the parameters of nuclei, such as their masses, can be derived from properties of the constituents,
%protons and neutrons. Clearly the masses of the nuclei are not independent parameters anymore.
%Nevertheless, hadronic physics itself still has an infinite number of parameters. For example the
%proton has a form-factor which all by itself amounts to an infinity of parameters. If we go to the
%next shell, quantum chromodynamics and the rest of the Standard Model, we are down to
%the aforementioned 28.
%
%According to the onion shell paradigm, it would be natural to expect that at the next level
%some of these 28 parameters will be related. Experience also suggests that new parameters
%might appear at that next level, for example as form-factors of composite quarks or leptons. 
%However, one may hope that a new theory emerges that describes all this in terms of
%fewer fundamental parameters. It might even happen that the number of parameters goes up.
%There is always a next onion shell  that might reduce it again.
%
But if one is aiming for the ``fundamental theory of all interactions", this would be
the last chance. There is, by definition, nothing beyond such a theory. 
\ifPreprint
Should we really expect that all the 
parameter values
of the Standard Model can be ``derived from from first principles", as the Wikipedia text suggests?\fi

%No-one would disagree with the first part of that statement\rlap.\footnote{Except that there are
%other ways in which the Standard Model falls short of being complete, most notably
%the absence of particles that can account for the large amount of dark, non-standard-model
%matter that  appears to be present in our universe.} The first test that any candidate
%for a ``complete theory of fundamental interactions" must pass is that it should contain gravity,
%and
%indeed string theory passes that test.

\subsection{Parameters in Quantum Field Theory}\label{QFT}

A fundamental feature of Quantum Mechanics is that one must sum over all
possibilities, weighted with an exponential factor. Here ``all possibilities" includes 
literally everything, including physics that we know nothing about. It includes 
the possibility that particles are created and annihilate again, at arbitrarily short
time and length scales. These particles may be quarks and leptons, but also  any particle
we have not yet observed. 
\ifPreprint
Even for a finite set of particles there is an infinite set
of possibilities: many particles may be created and they may interact and recombine
 in any possible way. 
 \fi
 How can we possibly get precise answers out of such a theory?

The reason is that in certain ``good" theories, including the Standard Model, all these
unknowns can be lumped together in a finite number of parameters.  \ifPreprint These parameters
are not specified by the theory, indeed, they are ``infinitely unknown", but that does not
matter. We only have to measure each of them once and we can compute any quantity
of interest.

One can distinguish three kinds of parameters: dimensionless ones, parameters with a 
positive mass dimension and parameters with a negative mass dimension. The latter 
become more and more relevant at high energies, since they contribute to physical
quantities with powers
of the dimensionless quantity $\xi E^n$, where  $\xi$ is a  parameter 
of mass dimension $-n$
and $E$
the energy of the process of interest. In every quantum field theory one can write down
an infinity of such parameters, and as soon  as one is known to be non-zero, one should
expect an infinite number of others to be non-zero as well, because of incalculable quantum
corrections that contribute to them. Consequently, for energies larger than the  lowest  mass scale
set by the negative dimension parameters the predictive power of a theory breaks down
completely. Positive dimension parameters are irrelevant at sufficiently high energies, but important
at low energies.  
Dimensionless parameters are relevant at all energy scales. A given quantum field theory
has a  finite number of parameters of  positive or zero dimension. The Standard Model has
one parameter of positive mass dimension (the mass of the still to be discovered Higgs particle)
and the remaining 27 are dimensionless. No parameters of negative mass dimension are known.

The feature that unknown quantum  corrections can be lumped together in a  finite
number of parameters goes by the name of ``renormalizability".
It is  a  theoretical fact (for which 't Hooft and Veltman were awarded the Nobel prize in 1999) that
the Standard Model has  this feature, but it will always remain an experimental  question
whether this really corresponds to nature.  Indeed, it is possible that  the next generation
of experiments, and in particular the LHC (Large Hadron Collider, currently in the
last phase of construction at CERN, Geneva), finds evidence of new parameters of negative
mass dimension. In this case the Standard Model looses its validity at the corresponding
energy scale, and the extrapolation towards the Planck scale that is required to make
contact with a fundamental theory of quantum gravity is unjustifiable. 
If such parameter is ever discovered, it would essentially invalidate the argument presented in the next chapter, or at best
postpone the discussion by many decades\rlap.\footnote{This does not mean that any kind of
``new physics" at shorter distance scales necessarily undermines the argument. New phenomena that
can be consistently extrapolated to the Planck scale themselves, or that only affect the Standard Model in a marginal
way, or involve entirely new particles are not relevant. Anything that ruins the extrapolation of gauge couplings
 or quark and lepton masses would however seriously weaken the argument.}

\else
The property I am referring to here is known as ``renormalizability".
\fi
There are two important features of renormalizable gauge theories that are important in the 
present context. The first is that they retain their consistency for a large range of parameter choices 
other than those that we observe in our universe. The second important feature is the fact that the properties
of the theory are independent of the details of short distance physics.

The meaning of  ``consistency" in the foregoing paragraph is that we can carry out computations of quantum
corrections with unlimited precision without running 
into infinities and without needing new information
that was not present in the original definition of the theory. In particular, no new parameters will be needed.
This statement is most clearly correct for strong interaction ($SU(3)$) part of the Standard Model,
 Quantum Chromo Dynamics (QCD). Paradoxically, this is the part of the Standard Model where computations are {\it least}
precise, bur this is due to practical limitations and not matters of principle. 

\ifPreprint
The other two interactions are only slightly more problematic. They are also renormalizable, but
even for renormalizable theories
there is something else that could go wrong. It turns out that the dimensionless parameters
also give rise to energy dependence, although not by powers of energy but only
logarithmically. This is known as ``running" of coupling constants, although
``crawling" would perhaps be a better word. 
The energy dependence is typically as follows (to lowest order in the expansion in quantum corrections)
\beq
\label{RunC}
g^n(E)=\frac{g^n(E_0)}{1-b_0 \hbox{~ln}(E/E0)} \ .
\eeq
Here $g$ is some coupling, $E$ the energy, $E_0$ some reference energy scale, an $n$ can be 1 or 2 depending
on the kind of coupling considered. The coefficient $b_0$ is a numerical constant that depends on the particle content
of the theory, and that can be straightforwardly  computed.  The sign of $b_0$ is important. It turns out to  be negative in QCD, and
this implies that the coupling becomes smaller at higher energies, eventually approaching zero.  This phenomenon is
called ``asymptotic freedom", and Gross, Wilczek and Politzer received the 2004 Nobel prize for its discovery.
If  $b_0$ is  positive the corresponding
coupling becomes infinite at the energy $E_0$ if we 
extrapolate it to higher energy (this is known as a ``Landau pole"). In the Standard Model
this happens for the coupling of the $U(1)$ factor, and for the self-coupling of the Higgs bosons.
The latter can only be determined if and when the Higgs boson is found experimentally. In both cases,
the Landau pole limits the precision of the theory to  effects of order $E/E_0$.
For the $U(1)$ coupling
the pole is far beyond the Planck scale, and for the Higgs self-coupling it depends on the still
unknown mass of the Higgs particle: if that mass is less than about 180 GeV the Higgs Landau pole
is also beyond the Planck scale. 

Obviously this is therefore another issue where LHC results could change the entire picture.
The entire justification for assuming that  we can go directly to
a fundamental theory of all interactions without passing through intermediate ``onion shells"
would break down if the Higgs boson were too heavy. This would imply that we will encounter
a pole in the Higgs self-coupling before reaching the Planck scale, and the existence of that
pole would provide strong evidence of an intermediate onion shell.
%It would not falsify string theory, but would certainly make it a whole
%lot less interesting for the immediate future.

Because of their Landau poles and the still unknown Higgs sector
 the other two Standard Model interactions are on a less strong theoretical footing
than QCD, but in reality neither of the three parts of the Standard Model can be expected to be
infinitely precise. The reason is that they all couple to quantum gravity, which will lead to
presently incalculable corrections of order $E/E_{\rm Planck}$.
Quantum gravity is precisely the problem that the
has to be solved anyway by the fundamental theory we are looking for, and therefore one might say
that from this point of view all three Standard Model interactions are on comparable solid footing. \else
The other two Standard Model interactions are only slightly more problematic, since their couplings
become infinite when we extrapolate them to very short distances. However, everything we know at this moment
suggests that this will not happen until we reach distances much shorter than the length scale of quantum
gravity, the Planck length. Quantum gravity is precisely the problem that the
has to be solved anyway by the fundamental theory we are looking for, and therefore one might say
that from this point of view all three Standard Model interactions are on comparable solid footing.  \fi

More importantly, we can write down an infinity of other theories that are on equally
solid footing theoretically.
This puts us in a unique situation in the history of quantum physics. Nobody would make such a 
statement about the theory of nuclear physics or hadronic physics. Both of them have plenty of parameters
(particle masses, hadronic for factors, nucleon-nucleon potentials etc.)
but it would be foolish to suggest that one still obtains a consistent theory for arbitrary changes of these
parameters. \ifPreprint Even for the parameters we observe in our own universe the concept of high precision 
quantum computations is senseless in these theories. Hence they contain no clue about
how they could possibly be modified. We {\it can} give a meaning to alternatives to nuclear and
hadronic physics now that we know how to obtain these theories from gauge theories. This tells us precisely
what we can vary (in particular the quark masses and the strength of the couplings) without affecting
their implicit consistency. \fi

This notion of what can be modified in the laws of physics we observe plays an essential   
r\^ole in the following. Gauge theory defines a huge set of possibilities \ifPreprint
which I am tempted
to call the ``Gauge Theory Landscape", but this would be an inappropriate name for two reasons. First of all 
it may be confused with the ``String Theory Landscape" to be discussed later, 
and secondly it lacks an essential feature for
a true landscape: a height function. So instead I will refer to it as the \else
that I will refer to as \fi
``Gauge Theory Plane". 
\ifPreprint
String theory will provide a height function on that plane to turn it into a landscape.\fi

The fact that thirty-five years ago we seem to  have landed in a point of the Gauge Theory Plane 
that extrapolates consistently to the Planck scale is remarkable, but
not a guarantee that this is the last onion-shell before the Planck scale. However, the principles on which
gauge theories are based are so powerful that it seems reasonable to assume that eventually we
will have to land somewhere in that plane. 
\ifPreprint Indeed, practically all new ideas of physics
``beyond the Standard Model" are based on other points in the Gauge Theory Plane. This also addresses another
obvious objection one might have, namely the fact that it only seems to describe less than a quarter
of all the matter in the universe. The vast majority of matter in the universe does not interact with photons, and
a large fraction of this ``Dark Matter" is believed not to consist of the Standard Model particles we know.
We are in obvious danger of making
another anthropocentric mistake if we assume that the matter we have not observed
yet has properties similar to the small percentage of matter that we are made of. 
Nevertheless, I have enough confidence in the theoretical underpinnings of gauge theory
to assume that
those interactions are probably eventually
described by gauge theories as well.
\fi

So if we have to land eventually somewhere in the Gauge Theory Plane, we can only hope that nature is kind to us
and puts us directly in the right point, rather than confusing us with a theory that just mimics
a well-behaved gauge theory for no apparent reason. This is an important point, because it can be tested experimentally, and we do not have to wait very long for this to happen. Experiments at
the LHC (Large Hadron Collider), which is scheduled to start running this year at CERN, Geneva, may force
us to conclude
that indeed nature has been confusing us, and that there is no well-behaved extrapolation of currently known
physics
to the Planck scale. 
Then everything written here is, at best, premature.

\section{Two ``gedanken" calculations}\label{TGC}

So is our beloved Standard Model derivable from  first principles, {\it i.e.} from pure mathematics? I was confronted
with that question  around 1987, when string theory started hinting at an answer. I arrived at a conclusion
that I started explaining to my colleagues, scribbling pictures like the one below on napkins in the CERN cafeteria.
The gist of the argument was that for most alternatives we
would not exist, and neither would anything else of interest, but it is worthwhile to make that a bit
more precise. \ifPreprint Since string theory will be discussed later, this presentation is not
in chronological order. The reason for doing it like this is that
I  am convinced that someone should have arrived at this conclusion
even before string theory forced us to think about it.\fi 

\subsection{Physics in the Gauge Theory Plane}

Consider all theories 
\ifPreprint
we can think about theoretically, with
different kinds of interactions, quarks and leptons, and arbitrary masses for all these
particles, in other words the 
\else
in
\fi
``Gauge Theory Plane" mentioned above.
This is a huge space, with discrete and continuous parameters. The important point is
that all of these make as much sense theoretically as the Standard Model. 
We are going to consider two ``gedanken calculations". 
These are calculations we cannot really do, but that we can at least think about without
being accused  of transcending the boundaries of science.

 The first computation is the possible existence of observers, or intelligent life,
 in a universe governed by a given gauge theory. 
This is a ``gedanken" calculation because we are technically unable to do it. 
We certainly would not be able at present to derive the existence of observers from the fundamental 
properties and interactions of particles in our own universe.
Even the definition of life or observers is a problem.  However, we can certainly  imagine quantities
that  are computable in principle and that are almost certainly necessary for the existence of observers.

\ifPreprint
The definition of such quantities is tricky because of the risk of making precisely the kind
of anthropocentric mistakes I warned against in section (\ref{SecUU}). It is the danger of assuming
too quickly that what we see in our own environment must be necessary for life in general.
Indeed,  arguments  of this kind tend to be based on assumptions such as the necessity  of Carbon,
galaxies and stars, and some go as far as assuming the special physical properties of water
as essential. 
\fi

\ifPreprint
However, it seems reasonable to assume, for example,
\else
For example it seems reasonable to assume
\fi
that some level of complexity of the spectrum is necessary for life, 
and on the other hand complexity is something that can be defined and computed, at least in principle.
Furthermore it seems plausible that the building blocks of this complexity should be sufficiently
abundant, at least locally, and that there should be sources of energy available.
 
   So one can at least imagine drawing contours of some appropriately defined
 ``observer function" on the space of all gauge theories.
 \ifPreprint
  Ideally such a  function might
 specify the average total number of observers that exist in a universe during its entire
 lifetime, and the contour may be drawn at the value 1. 
  I will insist on
working exclusively in the Gauge Theory Plane, because that is where parameter changes are certain to
make sense. Any other  variables, in particular cosmological ones, may simply be put at their optimal
values. If it turns out that those values can never by attained, the contours in the Gauge Theory Plane
can only be smaller than we first thought they were. The same is true if we use too liberal a definition
of what constitutes an ``observer". 

\fi
Each such contour divides the space into two regions, one where observers  might exist, and
 one where they most definitely do not exist. 
 Both regions are non-empty: it is not hard to construct gauge theories that have completely 
uninteresting and hence ``lifeless" spectra,
 and obviously the Standard Model provides an example within the contour.
 \ifPreprint
 The computation would involve nuclear physics and chemistry for gauge theories that are different from ours, with different
 interactions, coupling to different kinds of particles. 
It is not doable in practice, but well-defined in principle.  
\fi
The figure below illustrates what such contours might look like
in some slice  of the Gauge Theory Plane. On the axes one may put some of the  Standard Model parameters, such as quark or
lepton masses. The small circle indicate the current experimental data. It lies entirely within an observer contour,
because our own existence is a piece of experimental data. 
\ifPreprint
These contours are not based on real computations (one could in
fact do a lot better than this), but are drawn like this to illustrate some important points and address a few common
misconceptions that I will discuss later.
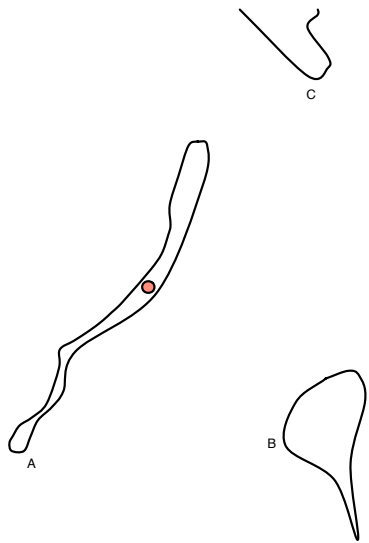
\fi

\begin{center}
\includegraphics[width=5cm]{IOPpic1.pdf}
\end{center}

\subsection{A unique gauge theory?}

    Now consider the second gedanken computation. It would be to compute
 the unique gauge theory resulting from some fundamental theory.
  This is a gedanken computation because no such theory is known.
    String theory was once hoped to be such a theory, but  does not seem to have this
    ``uniqueness" property. But let us imagine that one day we find some theory that
    does give a unique answer. Deriving that answer would involve some computation
    involving properties of new physical phenomena relevant to distances much shorter than
    what can be observed today, for example the Planck distance.
     
   The problem is now that the unique result of the second computation has to 
fall within the narrow contour of the first, even though these are
 completely unrelated computations. 
    
   The two gedanken computations are unrelated precisely because gauge
   theories, as explained above, are insensitive to the physics of very short distances. So
   even if the second gedanken computation picks only one point as the only theory that
   is consistent for arbitrarily short distances, this does not invalidate the first
   gedanken computation. Nobody would doubt that we could repeat all the
   computations in nuclear and atomic physics and chemistry with different values of
   the electron and quark masses or the strength of the electromagnetic coupling, and
   that all the results are correct, even if such a theory would turn out to be inconsistent
   at distances the size of the Planck length. 
   
   There is an obvious way out of  this puzzle.
   There would be no mystery about the special features 
   of our own universe if the fundamental theory  had a huge number of ``solutions". I am using
   the world ``solution" because I am thinking of some set of equations derived from the hypothetical
   fundamental theory. Equations can of course have more than on solution, even if the equations
   themselves are unique. In order to demystify the apparent fine-tunings we see
   in our environment, the fundamental theory should have, in addition to our own universe, a huge
   number of other solutions, densely covering the Gauge Theory Plane (or at least a large neighbourhood
   around us) with points.

\subsection{Our own neighbourhood}\label{Oon}

Although the complete computation of the contours in fig 1. is far beyond our technical capabilities,
part of our own neighbourhood has been mapped out already. 
There exist several arguments  showing that small modifications in some of the parameters 
of the Standard Model would be fatal for our kind of life and most likely any kind of life. 
There are many arguments of this kind, 
see for example \cite{BT},   \cite{Cahn:1996ag}, \cite{Agrawal:1997gf},  \cite{Hogan:1999wh}, \cite{Donoghue:2007zz} and \cite{Hall:2007ja}. 
A rather obvious one
is the fact that the proton is slightly lighter than the neutron. If you turn that around, the proton would be
unstable against decay to a neutron, a positron and a neutrino. In the Standard Model, we can tune that
mass difference by increasing the mass of the up quark (a proton consist  of
two up quarks and a down quark, a neutron of two down quarks and an up quark;  up and down quarks are also
denoted as u and d, and have charges $\frac23$ and $-\frac13$ respectively), keeping everything else fixed. 
\ifPreprint
There is a small
range in parameter space  where a free proton  would be unstable, but where it is stable in certain nuclei,
just as the neutron in our universe. But if we continue to increase the up quark mass a bit more, we loose
all charged stable nuclei. \else 
If we increase the up-quark mass sufficiently, we loose all charged stable nuclei. \fi

If we increase the electron mass from its observed value of .511 MeV, 
keeping everything else fixed, pretty soon
it becomes larger than the proton-neutron mass difference of  1.293 MeV. This would make the neutron
stable. Then the hydrogen atom becomes unstable against electron capture and we get to  a universe dominated
by neutrons, with few protons and electrons. Another remarkable fact is the lightness of the up and the down
quark.   If we increase their masses, the pion become heavier. This in its turn reduces the range of the
strong interactions,  which fall off exponentially with the a length scale set by the inverse pion 
mass.                   
Soon we start loosing the stable heavier nuclei\rlap,\footnote{This assertion is too simplistic, 
as it relies on the popular, but 
incorrect
idea that the interactions between protons and neutrons in a nucleus can be largely attributed to single pion exchange. The correct statement is that our understanding of nuclear
physics from the point of view of QCD  is surprisingly limited, and that even for a nucleus as simple as the deuteron it
is not known how the binding energy depends on the quark masses.  I am grateful to E. Witten for pointing this out to me, as well as
for other, appreciative remarks about this paper. For a recent discussion on the dependence of nuclear physics  on quark masses see 
R. Jaffe, A. Jenkins, and I. Kimchi, Phys. Rev. D 79, 065014 (2009). This footnote was added in July 2010. The original text has not been
modified.}
until eventually Carbon is not
stable anymore. Similarly, if we increase the electromagnetic fine structure constant $\alpha$, we increase
the repulsion of protons in nuclei, and this would also destabilize heavier nuclei. 
\ifPreprint

Opponents  of this kind of argument will say that the burden of proof is on me
to show that beyond these thresholds  nothing resembling
intelligent life could possibly exist. But instead
 \else
There are many arguments of this kind, 
see for example \cite{BT},   \cite{Cahn:1996ag}, \cite{Agrawal:1997gf} and \cite{Hogan:1999wh}. 
Some of them are based on specific human-friendly details of our own environment in the Gauge Theory Plane,
such as 
the chemical properties of DNA or water. This seems much too anthropocentric to me.
There may well be life elsewhere in the Gauge Theory
plane, equally intrigued about the friendly features of their own environment as we are about ours.
But on the other hand, one cannot simply dismiss the entire bulk of evidence 
in this way. Instead
\fi of arguing about the precise position of the contours in the gauge
theory plane, one should consider the opposite question: is it plausible that  intelligent life is a dominant feature
over most of the Gauge Theory Plane?  If  one looks at the entire picture, rather than zooming in on a particular
part of the contour, it seems difficult to argue that life is generic. 
\ifPreprint
There may well be life elsewhere in the Gauge Theory
plane, equally intrigued about the friendly features of their own environment as we are about ours.
But on the other hand, one cannot simply dismiss the entire bulk of evidence 
in this way. \fi

\ifPreprint

\subsubsection*{Scales}

The most obvious peculiarity of the Standard Model that seems relevant for the existence of life
is its mass scale. In fact, 
the Standard Model has {\it two} mass scales that are -- to the best of our knowledge -- independent.
The first one is the QCD scale. It is the mass scale $E_0$ where the running QCD coupling constant, given by (\ref{RunC})
with $b_0 < 0$, has a pole.
This scale is a few hundred MeV, and it determines the masses
of most  hadrons, in particular the proton. An important exception is the pion, whose mass
-- unlike that of the proton an neutron -- goes to zero with the up and down quark masses: it is
proportional to $\sqrt{m_{\rm u} + m_{\rm d}}$.
The second mass scale we know of is the weak interaction scale $v=246\ {\rm~GeV}$. It determines the
masses of the W and Z bosons as well as those of all the quarks and leptons, with the
exception of the neutrino masses, which may well require a third mass-scale (this is usually
regarded as ``beyond the Standard Model physics" and hence not counted among the
Standard Model mass scales). Both the QCD and the weak scale are much smaller than the Planck scale, and in
both cases this is clearly essential for the 
existence of observers: If the proton or the electron had Planckian masses, gravity would
be much too strong in comparison with the electromagnetic or hadronic and nuclear forces. Most people
consider the problem of the smallness of the QCD scale ``solved". I will comment on this in
appendix C.  

\subsubsection*{Discrete choices}

The discrete choices in defining the Standard Model, the gauge
groups and representations, are undoubtedly important for the existence of life.
This quickly gets us into uncharted territory, because most cases 
are to complicated to work out. But if a few things are fairly obvious. Removing either the $SU(3)$ factor or the
$U(1)$ factor removes the strong or electromagnetic interactions, which both seem essential, not just
for our life but for any kind of complexity. 
The $SU(2)$ factor
responsible for the weak force is less essential.  Indeed, it has been argued that life could still
exist if we remove it altogether \cite{Harnik:2006vj}, provided we make some other changes. 
However, in such a universe our world
would look even more fine-tuned.
The quark and lepton masses are dimensionless numbers time a natural scale of
about 246 GeV, the weak interaction scale.
Removing $SU(2)$ amounts to send the natural mass scale towards the Planck mass
which replaces the miracle of a small weak scale by the bigger miracle of small up-quark, down-quark
and electron masses. 
Nevertheless, the computations in \cite{Harnik:2006vj} are a very nice example of a computation of
part of the contours in the Gauge Theory Plane, slightly away from our own place. 
Although it is argued in \cite{Clavelli:2006di}
that in these weak-less parts of the Gauge Theory Plane we would be starved of oxygen, that argument strikes me
as perhaps a little too anthropomorphic. Perhaps other kinds of life could exist in such a universe. If true,
it would simply yield another contour in the Gauge Theory Plane.

I have drawn some examples of other contours
in the figure. Their existence does not undermine the argument in any way. It does not even matter if
there is another contour that looks ``larger" than ours, because it does not make it less
surprising that a unique solution managed to hit precisely our own, locally narrow, contour.

%The purpose of \cite{Harnik:2006vj} was to argue against the the overly simplistic statement: ``the
%weak scale is so small because otherwise we would not exist". I will discuss statements of this kind 
%below under ``Determining parameter values".

There are other changes in the  discrete choices that have  drastic consequences. Changing the number of colors from
three to any even number turns the analog of protons and neutrons into bosons, eliminating the r\^ole of the Pauli exclusion
principle in nuclear physics. 
Changing the choice of representations
by choosing instead of Eqn. (\ref{SMrep}) something else from the myriad of possibilities has consequences that are hard to analyse, but most of them would undoubtedly yield
something pretty boring.

\subsubsection*{Quark and lepton masses}

Let us take a closer look at the quark and charged lepton masses.
Measuring quark masses is a tricky business, because they cannot be liberated from their bound states. To first
approximation
the mass
of the proton has nothing to do with the quark masses. The proton mass is determined mainly by the intrinsic energy scale  
of QCD. To observe the actual quark masses we have  to probe the proton at high energies, where the quarks start approaching a
free particle behaviour. I will use here the official estimate of the ``Particle Data Group" \cite{Yao:2006px}.
 For the lightest quark, the
u-quark, they give 1.5 to 3 MeV. 
The next quark
with the same electric charge $\frac23$, the c quark, weighs 1.25 GeV, or at least 400 times more. The mass of the next quark
of charge $\frac23$, the top (t) quark is about 173 GeV, larger  by a factor of 138. For the three quarks of charge $-\frac13$ these masses are
respectively  5 MeV, 95 MeV and 4.5 GeV. For the
charged leptons we find .511, 105, and 1800 MeV, with ratios of 200 and 17 respectively. 

In the Standard Model all quark and lepton masses are given by
the Higgs vacuum expectation value of 246 GeV, multiplied by a dimensionless number, the Yukawa coupling. 
In the following I will keep the Higgs vacuum expectation value fixed, as well as the strength of the
electromagnetic and strong  interactions, and vary the Yukawa couplings; see
the subsection ``determining parameter values" below for a justification.

In terms of their natural units of 246 GeV, the up  and down masses as well as the electron mass look ridiculously small; the electron mass is
smaller by a factor half a million.  Note
that the mass-ratios of the charge 2/3 and charge -1/3 quark are inverted in the two heavier families, which
makes especially the lightness of the up quark rather peculiar. The lightness of the up-quark is
important for at least two reasons: to keep the pion light, and to make the proton lighter than the neutron. The
mass difference between the latter is not only determined by the differences of the quark masses that they
are made of, but also by electromagnetic effects. Although this is hard to compute, looking at the charges of the quarks
it is easy  to  see that the electromagnetic contribution tends to make the neutron lighter than the proton.
The pion mass is proportional to $\sqrt{m_u+m_d}$, whereas
the neutron-proton mass difference contains a term $m_d-m_u$ and a negative electromagnetic contribution.

The optimal way to deal with these constraints would be to make the up-quark nearly massless, to
make the down quark mass large enough to overcome the electromagnetic contribution, to hope that then
the pion is still light enough, and then to choose the electron mass smaller than the neutron proton mass
difference. To make this quantitative would
 require
 a lot more work\rlap.\footnote{There are other bounds one may take into account. For example Hogan \cite{Hogan:1999wh}
 takes into account that the conversion of two protons to deuterium, $p+p \to d + e^{+}  + \nu_e$ should be exothermic; Weinberg \cite{Weinberg:2005fh} points
 out that this may not be necessary, because the reaction $p+p+e^- \to d+ \nu_e$ can take its place. It may not
 be easy to pin down the exact boundary of the inhabitable island in parameter space, but
 in any case
 it is true that the parameter values we observe are remarkably close to thresholds beyond which the
 necessary conditions for life change drastically, and need to be re-examined carefully.}
 because we need to know precisely how heavy we can make the pion without destabilizing too many
 heavy nuclei.  But qualitatively, 
 it is hard to avoid the conclusion that the light quark and lepton masses have been pushed precisely
 to this corner of parameter space, away from any imaginable natural distribution. 

What would such a ``natural distribution" look like? If we were to assume that the Yukawa couplings are number of
order 1, the amount of tuning required to get the right up-, down-quark and electron masses looks enormous.  But
the distribution of the other masses does not suggest that they are
``randomly" chosen numbers of order 246 GeV. 
Indeed, there is a clear hierarchy, with masses of quarks and leptons going up by one or two orders of magnitude
each time we move to the next family. This hierarchy (the ratios of subsequent quark masses of the same charge) is
not very sharply defined, and does not look universal: it is considerably larger for the up quarks than for the down quarks. If  we assume that
such a hierarchical  distribution of Yukawa couplings is fundamental, the amount of tuning is of course reduced.
But even taking that into account, It is hard to feel
comfortable with the idea that this entire structure, including the hierarchy itself, its non-universality and the large deviations from it,
could be the unique solution to some mathematical equation, if our existence depends so  crucially on it.

Not all continuous parameters of the Standard Model have serious implications for our existence. The most obvious
ones are the up and down quark mass, the electron mass, and the strong and electromagnetic coupling constant.
Other parameters may not be as irrelevant as they seem. Consider the top quark mass. The top quark is 
about 180 times heavier than
the proton and has an extremely short life-time. It may nevertheless be important since
the large mass implies a large Yukawa coupling, which in its turn
implies that it makes important contributions to the running of all other couplings. 
In particular the Higgs self-coupling
is strongly affected, and since all quarks and leptons owe their mass to the Higgs boson, one may expect
drastic consequences if we modify the top-quark mass while keeping everything else fixed. Indeed, in the 
supersymmetric\footnote{Supersymmetry
is a hypothetical and at best approximate symmetry between fermions and bosons. Because fermions and bosons 
contribute with opposite signs to quantum corrections, it greatly helps in controlling those corrections. The upcoming LHC experiment
may tell us whether it is merely  a powerful theoretical tool, or if nature makes use of its power as well.}
extension of the Standard Model the large top quark coupling drives 
the symmetry breaking mechanism that
produces the quark masses. We do not know enough yet about the Higgs boson to make general statements about
the r\^ole of the top quark, but is remarkable that
its mass of about 173 GeV is close to
the theoretical upper limit for quark masses, about 230 GeV: if we push the mass beyond that point its Yukawa coupling
will have a Landau pole below the Planck mass. Neutrino masses are also relevant for various features of our universe
that are important for the existence of life, but that is a much more complicated story.

%

%These are just some examples showing that as we  move some of the parameters of the Standard Model away from their
%observed value we cross certain thresholds. Beyond these thresholds, some of the particles that are important to us
%become unstable, whereas other particles may become stable, disturbing the abundances of particles we need. 

\subsubsection*{Secondary effects}

The first gedanken computation would include not only the effect of direct variations of the gauge theory parameters, but
also the secondary effect these variations have on the parameters of nuclear physics and chemistry. 
A good example is an excited state of the Carbon nucleus, correctly predicted by Fred Hoyle.  He was worried how
stars could have produced as much Carbon as we observe. 
The obvious way to produce Carbon is that two $\alpha$-particles fuse to form Beryllium, which
in its turn fuses with another $\alpha$ particle to form Carbon. The trouble is that the first step fails, because
Beryllium-8 is not stable. Hoyle then conjectured that there should exist an excited state of Carbon
appearing as a resonance in the triple-$\alpha$
channel, so that the normally suppressed process fusing three $\alpha$-particles to Carbon is strongly enhanced.
He urged experimentalists to look for that excited state, and they found it.
The energy of the excited Carbon state he conjectured is of course not a variable. But nuclear physics
itself varies over the Gauge Theory Plane, mainly as a function of the up and down quark masses and the strength of the 
QCD coupling. These variations would affect the abundance contour, and end up reducing the size of the region where life
can exist. According to Weinberg \cite{Weinberg:2005fh} the Hoyle resonance is not as remarkable as it may seem. He argues that it has the same
origin as the instability of Beryllium, and can be understood entirely 
in terms of unstable bound states of $\alpha$-particles. If Weinberg
is right this would imply that the Hoyle resonance would persist over a large part of the Gauge Theory Plane, and hence does not
contribute much to the reduction of the contours. But either way, it can be taken into account.

There are many other secondary effects. In particular \cite{BT} is a treasure trove of remarkable facts about nuclear physics and
chemistry, relevant for our existence, such as the properties of water (the fact that ice is lighter than water) or the 
properties of the DNA molecule
(supposedly very sensitive to the electron/proton mass ratio, 
a claim attributed in \cite{BT} to T. Regge). Many of these arguments are not terribly convincing
since they focus on {\it our} kind of life instead of observers in general. For example, even if {\it our} DNA replication fails for
a different proton/electron mass ratio, it would still be extremely hard to argue that no life could possibly exist for another value.
The essential complexity of chemistry is unchanged if we change this ratio, and there may well exist entirely different molecules
playing a similar r\^ole if the parameter values are different.
We should not be too surprised to find that nuclear
physics and chemistry are remarkably suitable to {\it our} kind of life, but that does not exclude the possibility of entirely
different kinds of life, presumably equally puzzled about their own environment. This kind of remark is
often used to dismiss {\it any} argument concerning apparent fine-tuning of the laws of physics. But that is
not reasonable either. If we look at all the evidence without prejudice, it does look
plausible that we live in a privileged place in the  Gauge Theory Plane, but there is no reason
to assume it is the only one. Indeed, if it were, that would raise another disturbing puzzle.

These secondary effects are a favorite target for critics. After all, we can compute them! We can compute the resonances of
carbon, the properties of water, and the properties of DNA, at least in principle. Does this not show that it is senseless
to worry about such apparent coincidences in our environment?

Although we can indeed compute them, the result depends on the parameters of the Standard Model, and as such contributes
to the shape of the contours.  
If these secondary arguments turn out to be correct ({\it i.e.} if they
hold for generic observers, rather than just us) the contours might
become very small indeed. In some cases there is even a risk of over-determination, in the sense that several unrelated
arguments constrain the same set of Standard Model parameters. If we could really do these computations, it seems
more likely to me that the contours would decrease is size in comparison to current guesses, than that they would increase.  

\subsubsection*{Determining parameter values}

To which extent do these arguments {\it determine} or {\it explain} the values of some of
the Standard Model parameters? One has to be very careful in drawing such a conclusion. A common,
and valid, criticism of arguments of this kind is that one cannot simply vary a parameter
while keeping all others fixed. Consider the example of the electron mass. Can we conclude that
 it is so small because if it where three times heavier the neutron would be stable?  
Even in our own neighbourhood, one would first have to examine if there are ways to increase the
up and down quark mass difference. The range is limited by the pion mass, which in its turn is limited by
the required range of the strong interactions, which compete in nuclei with the electromagnetic interaction. 
So one would really have to consider the full variation of all these parameters to see if the electron mass
can be pushed up significantly. But if we go further away from our own neighbourhood, there are other possibilities.
For example in the ``weakless universe" proposed in \cite{Harnik:2006vj} the neutron is stable and the problem of a neutron dominated
universe is avoided by having stable charged pions, which produce protons when colliding with neutrons. So then this
particular limit on the electron mass disappears altogether.

Obviously the correct way of obtaining a true observer window for a single variable is to project
the contours on a single axis (including contours for different discrete choices of groups and 
representations, as long as one can identify quarks and leptons). In practice, this will be very difficult.

But this is not what I am trying to do. I merely wish to argue that our own contour is remarkably narrow in
certain directions. To show that, it suffices to scan it along a few axes, where an axis is defined by
keeping all but one variable fixed. As fig 1. suggests, our own region can be narrow, but obviously its projection on
the two axes can be large. If our contour is narrow in certain directions
(essentially a lower-dimensional sub-manifold), it would indeed by
a puzzle if a unique answer from a fundamental theory ended up precisely in that narrow strip,
regardless of the fact that the strip might be very elongated in other directions.

\fi
\ifPreprint
\subsection{Two questions}

The advantage of presenting the arguments in terms of contours in the Gauge Theory Plane is
that it cleanly separates the issue into two fundamental questions, one that we may be able
to answer, and one for which more information is required.
 
The first is why non-empty contours exist at all: why is
there life {\it anywhere} in the Gauge Theory Plane? The second question is how a first principle computation 
could ever end up within the contour, instead of in one of the far more abundant boring places in the Gauge Theory Plane. 
This is the question addressed here. 
 
I do not have much to say about the first question. The problem becomes more
worrisome if our island turned out to be the only one. If all the arguments based solely
on our {\it own} existence are taken seriously, one might be tempted to conclude that.
But such an conclusion
would fit perfectly in the series of
anthropocentric mistakes listed in the first chapter. 
In any case, the answer to this question is clearly out of reach. 
Even in our own universe we have no idea what other forms of life might exist.

The most attractive way out would be if the contours themselves
were abundant, so that also in this respect we are not unique.  The most attractive way
out of {\it both} questions would be that some form of life is generic in all gauge theories. But that
seems very implausible, and would still leave us with the question why it was realized in
such a complicated way in our own environment. 
\fi
\ifPreprint

\subsection{Boundaries and measures}
   
In order to make all the foregoing mathematically precise, one would like to compute
the relative size of the region in the Gauge Theory Plane that can support the existence
of observers. If we could prove that this is true only in a portion of relative size $\epsilon$,
it would require the existence of at least $\frac{1}{\epsilon}$ ``solutions" in a fundamental theory in order
to demystify our existence. 
Unfortunately such a computation requires  the definition of a boundary  and of a measure, and
the Gauge Theory Plane is not naturally equipped with such quantities.
Indeed, it is of infinite size in many directions, although the requirement that a theory
should remain consistent until the Planck scale cuts it down. There is also no natural measure, and
this concept becomes even harder to define if we consider the discrete choices. For example,
what is the relative weight of $SU(3)\times SU(2)\times U(1)$ with respect to other
possible gauge groups?

On the other hand, if we choose any boundary and
measure without prior knowledge of the Standard Model,
we would undoubtedly find that the window for the existence of observers is small. 
Mathematically speaking,
one can always deform any measure in such a way that the relative size gets arbitrarily
close to 1, but only if the location of the observer window is already known. 

%But this
%is precisely something a Planck scale fundamental theory is not likely to know about.

A fundamental theory with a large number of discrete ``solutions" would define a measure for us,
namely the continuum limit of the distribution of those solutions. But once the correctness
of such a theory has been established, there is no need anymore to argue against uniqueness.
On the other hand, if a fundamental theory with a unique solution is found, we will undoubtedly
be immensely puzzled how it could ever land precisely within the observer contour, but there
is no way to make that puzzlement quantitative. This is not significantly different if we consider
our planet or our solar system. Undoubtedly in that case nearly everyone would agree that our
understanding of their very fortunate features is greatly enhanced by the fact that we know that there
are huge numbers of other possibilities. But there is no way to quantify how fortunate we are without
a theory of the formation of solar systems and planets (the analog of the ``fundamental theory" above),
and even then it is very hard. In the time of Kepler such a theory was not available, so if he had chosen
to think of the analog of observer contours on the space of all possible solar systems or planets, he would
have faced a similar measure problem. 

To illustrate this, assume that 
some fundamental theory has solutions with a flat distribution 
between 0 and 1, for all dimensionless parameters and ratios of mass scales with the Planck scale.
Under this assumption we can compute how many
solutions it must have in order that, statistically, there is one within our contour. From   
the ratio of the QCD scale over the Planck scale we get
$10^{-19}$, the weak scale over the Planck scale gives an additional $10^{-17}$. 
The $u$, $d$ quark masses and the electron mass give a factors of about $10^{-4}$ to $10^{-5}$ each, when expressed
in terms of their natural scale, the weak scale: $v=246 \rm{~GeV}$. This does not yet include
the value of the fine structure constant or the discrete choices of gauge groups and representations.
It seems that we would need at least $10^{50}$ solutions. However, in at least one case it is not likely that
the distribution will be as assumed here: instead of the ratio of the QCD scale over the Planck scale
it would perhaps be more natural to use the strength of the QCD coupling constant $g_3$. The relation between
these two is logarithmic, as in (\ref{RunC}), with $g=g_3$ and $n=2$.  Clearly, if we assume a flat distribution
for $g_3$ instead of the value of the pole, $E_0$, the relative size of the observer contour is not reduced
by a factor $10^{-19}$ but by a factor of order $10^{-1}$. Note that this argument does not by itself explain the
small ratio, nor determine the measure, but it does illustrate the importance of
understanding the correct parametrization of the variables (see also Appendix C).
 There may be other cases where Standard Model
variables can be reparametrized in order to blow up the observer contour. The weak scale
is a candidate, although no convincing mechanism exists in this case. The quark and lepton masses also
do not look as if they originated from a flat distribution. Indeed, as discussed above, they display a clear, though not very precise
hierarchy. All of these effects may conspire to make the Standard Model look less rare than
1 in $10^{50}$. 
But on the other hand, the number ${1\over \epsilon}$ is just a lower limit that would
statistically give rise to just one point within our own contour. It would be yet another example
of a potential anthropocentric mistake to assume that our universe is the only fundamentally allowed one within
our contour. Furthermore, if the parameters that are relevant to us can take many values, it
is reasonable that all the others can do so as well. It is  not easy to scatter just a few hundreds of points over a 
a multi-dimensional space so that they are nicely spread around our neigbourhood without prior knowledge of its existence.
We cannot expect the fundamental
theory to produce a distribution that is just barely dense enough and just scattered 
in the minimal way to meet
 the requirement that we exist. That would
be another serious puzzle, although a slightly less severe one.

%If in the entire parameter landscape only one point exists with
%intelligent life, namely our Standard Model, then I was just lucky to guess the existence of
%a large ensemble,  and mankind is lucky to exist. 

%One could add other contours 
%to indicate abundances of the essential building blocks for life in a universe based on a given gauge theory,
%and yet other contours indicating the possibility for energy sources, playing the r\^ole of stars in our
%own universe. Of course adding requirements borrowed from our own environment enhances the risk of walking
%into the anthropomorphic trap, and one would have to think very deeply about the need for such requirements. 
%In any case, it is not necessary to know the exact contours very far from our own spot in the gauge theory
%plane. It will be sufficient to convince ourselves that we live on a relatively small island in that plane, or
%actually a narrow strip, because parameter variations in several directions (for example the masses of most of the
%heavy quarks and leptons)
%are of little relevance. 

\fi

\subsection{Against uniqueness}

   One may look at the problem in the following way. We have at our disposal a
   discretely and continuously infinite set of possible theories, all making as much
   sense theoretically as any other: the Gauge Theory Plane. In our own universe we observe only one of those.
   Let us assume the existence of some fundamental theory that selects a number $N$
   of points in this infinite set. We would like to know the value of $N$.
   
   I have presented above arguments indicating that we should expect $N$ to be large.
   These arguments cannot, and will never be, mathematically rigorous. Even if we were
   technically able to do all relevant gauge theory computations, we are still
   faced with the problem of the precise measure on this space, something
   that can only be determined once we already know the fundamental theory. 
   So this is the best that can be done: a plausibility argument, not a mathematical proof.
   
   But now let us look at the other side of the issue. What are the arguments for believing
   that $N=1$? Well, there aren't any. There are a few
   empirical arguments in favor of relations among some Standard Model features,
   but on closer examination 
   \ifPreprint
   (see appendix B)
   \else
   (see \cite{LongVersion}) 
   \fi
   even these do not really point towards
   uniqueness. 
   In fact, the argument for uniqueness  of the Standard Model is not better
   than Kepler's belief that the ratios of planetary orbits should be computable. 
   
   At this point some people will say that they never really believed in uniqueness anyway,
   but just hope that there will be ``very few". But this reveals the true
   nature of their argument: it is nothing else than wishful thinking. 
   The concept of ``very-few-ness", unlike uniqueness,  has no mathematical definition. 
   All this says is that it would be nice if we could enumerate all the solutions. Indeed, that
   would be nice. But there is no reason to believe that it will be true.
   
   \ifPreprint
   Believers in uniqueness might appeal to a hope for some underlying, yet to be
   discovered principle of nature. Indeed, I admit that I am hoping for a principle of nature
    that fixes the underlying theory (as opposed to the ``solutions" of that theory) uniquely. But
    as soon as one gives up on absolute uniqueness, there is no way of telling what the
    total number will be. In our universe we observe 1, and that is the maximum we are
    able to observe with foreseeable technology. All this says about the total is that it
    is larger than or equal to 1.
   \fi
   
\ifPreprint

\subsection{Entropic selection?}\label{Statistics}

Let us assume that we find a fundamental theory that has a large, discrete set of solutions.
Then we end up with
some kind of statistical distribution, perhaps peaked in somewhere in the Gauge Theory Plane. There are
two obvious ingredients to such  distributions: the density of points in a certain region of interest,
and the probability of ending up  at a particular point. The former can at least in principle be obtained from
the underlying fundamental theory.  The second contribution depends on details of the model for populating
the distribution, and may depend on initial conditions beyond our control. It is not even clear
if the notion of probability makes any sense in the context, or how to define it. 

But let us assume for the sake of the argument that this can all be defined and computed. 
Some people may hope that this will restore what seemed to be lost, namely the
possibility to derive the Standard Model. Could it be the maximum of the probability
distribution? 

To me, that would seem to be even less attractive than a unique solution.
The computation of the maximum of a statistical distribution is analogous to the
second gedanken computation of section \ref{TGC}. It would be an equally incomprehensible mystery
why such a computation would end up in the observer contour.

Furthermore, unlike a unique vacuum, this cannot even be called an explanation for 
the parameter values we observe. 
If only one value is allowed by fundamental physics, there is nothing else to
discuss. But if an ensemble of values is allowed, some of which are outside the observer 
window, statistics does not help. 
If the majority of values is within the observer window,
do we find ourselves there because it is more likely, or because no observers can exist outside
that window? On the other hand, if the distribution has a huge peak outside the observer window,
would that be a reason for concern?

This should not be confused with a different application of statistical arguments, which try
to determine our place {\it within} an observer window. If there is a well-defined
notion of probability, we may hope that our universe is the most likely one, 
within the set of inhabitable ones. In particular, if there is a multitude of observer
contours, one may hope that there is no other that is far more densely covered
than our own. This would  cast doubt on the correctness
of the fundamental theory producing such a distribution.

\fi

\subsection{Discrete or continuous?}\label{DiscCont}

In the previous discussion I was assuming that the second gedanken calculation will
either yield a unique solution, or else
produce a discrete set of solutions. Obviously the demystification works equally well if
there are solutions with continuous parameters that cover part of the observer contour.
One could even ask if we need
a fundamental theory at all? If indeed we need a large ensemble of theories
to explain why there exist theory within the contour, perhaps the
Standard Model itself, with its 28 parameters, is just what we need. Perhaps we were
wrong in regarding the existence of this relative small set of parameters as a problem.  
Perhaps the ``theory of everything" we are looking for is simply the Standard Model,
combined with a (hypothetical) theory of Quantum Gravity that is completely insensitive to the
Standard Model and its parameters.

The trouble is not only that this is esthetically unappealing, but more importantly it
does not allow for the possibility of moving to other points in the parameter space.
The Standard Model with its observed value of the fine-structure constant, $\alpha=1/137.04$, has
nothing to do with a similar theory with a different value of $\alpha$. 
It is of course imaginable that the parameter values are simply fixed to random values
at the initial moment of our universe, but this would be highly unsatisfactory, to say the least.

Therefore the parameters
must become, in some way ``dynamical". 
Once the parameters are dynamical, it makes sense to compare theories with different values for
parameters like $\alpha$. 
In particular one would expect their ground state energies to depend on
the parameter values, if we can make sense of them. Ground state energy in quantum field theory is a 
highly divergent quantity, which furthermore is not physically relevant for Standard Model physics.
It starts being important if one includes gravity, because only gravity couples to it. 
In theories of gravity, the Standard Model contribution to the
vacuum energy is part of a long-standing problem, the cosmological constant problem,
which looks even more serious if one attempts to quantize gravity.  I will say more about this below.
If $\alpha$ and
other parameters are just constants, one may postpone discussing this issue until it is understood
in a consistent theory of quantum gravity, and simply set the ground state energy to zero in the meantime.
But if we allow $\alpha$ to vary, we cannot expect to be able to set it to zero for all values of $\alpha$. 
If we assume that ultimately a theory of quantum gravity turns vacuum energies into meaningful quantities,
they would seem to produce a kind of potential, and what are called ``solutions" above would actually
correspond to minima of this potential. These may just be local minima. They need not be absolutely
stable with respect to quantum tunneling, as long as they live long enough. This obvious fact has been
understood for many years already, but the full implication has only been appreciated in 
recent years. 

\ifPreprint
Such a potential picture would almost inevitably imply the possibility of fluctuations around the minima. 
The description of these fluctuations in quantum field theory necessitates the introduction of new fields, which
can take constant values in one of the minima without breaking Lorentz invariance. Such fields are called
scalar fields.
Nothing I have said so far requires the
minima to be fully discrete. 
There might exist elongated valleys along which the parameters could vary. 
But it is not likely that we find
ourselves in one of those elongated valleys, for at least three reasons.  
First of all for observational reasons: such a valley would imply
the existence of massless scalar fields (the mass corresponds to the
second derivative of the bottom of the valley), which would mediate a ``fifth force" with has not been observed.
Secondly, a scalar field is not expected to be massless unless there is a symmetry that requires it; in other
words, generically one would not expect a valley connecting 
physically distinct values of $\alpha$ to be absolutely flat. This would only be possible
if there were miraculous cancellations between various quantum contributions to the vacuum energy\rlap.\footnote{Such
cancellation do occur in supersymmetric theories, but supersymmetry is at best an approximate symmetry in nature.}
Finally,
our kind of life would not develop if crucial parameters could vary significantly over space or time. \fi

\ifPreprint
The previous paragraphs were written with string theory hindsight, and perhaps there are different
ways in which an ensemble of 
special parameter values in the Gauge Theory Plane could be selected by a fundamental theory, but
it seems unlikely that any theory would produce a continuum. In chapter (\ref{StringTheory}) I will discuss how precisely such a picture
emerges from string theory.  It is called the ``String Theory Landscape". 
\fi

\ifPreprint
\subsection{The danger of circularity}\label{circ}

In this section I want to compare this kind of arguments for the standard mode parameters with
 those for the cosmological constant. First I will briefly summarize some basic facts about the latter.
For more details see {\it e.g.} \cite{Bousso:2007gp}.

 \subsubsection*{The Cosmological Constant} \label{CC}
 \else
 \subsection{The Cosmological Constant}
 \fi
 The vacuum energy discussed above may be thought of as an additional axis orthogonal to the 
 Gauge Theory Plane. One can extend the contours into this additional parameter dimension, and
 work out where life is possible. It turns out that, at least in our own neighbourhood, the contour is
 an incredibly tiny strip around  zero. The reason vacuum energy affects the existence of life at all is
 that gravity couples to it.  It enters into the  Einstein equations for gravity via the cosmological
 constant  $\Lambda$, that appears in the following way
 \beq
 R_{\mu\nu}-\half g_{\mu\nu} R +\Lambda g_{\mu\nu} = 8\pi G_N T_{\mu\nu}\ ,
 %S= -\int {\rm d}^4 x \sqrt{-g}\left( \frac{1}{2\kappa^2} R - \Lambda\right)
 \eeq
 where $G_N$ is Newton's constant and $T_{\mu\nu}$ the energy momentum tensor
 of matter. The left-hand side contains the Ricci tensor $R_{\mu\nu}$ and the curvature scalar $R$,
 both related to the curvature tensor $R^{\mu\nu\rho\sigma}$. The physical principle behind
 the precise form of the left-hand side is general coordinate invariance. If one arbitrarily imposes
 an additional requirement, namely that the left-hand side should be linear in $R$, one gets the
 standard form of the Einstein equations, with $\Lambda=0$. But there is no physical reason for
 such a constraint, and on general grounds one should therefore expect the left-hand side
 to contain terms of higher order in $R$ (which are irrelevant in normal circumstances)  as
 well as a term of zeroth order, $\Lambda$. 
 %Einstein understood that, but it should not be
 %necessary to repeat the ``biggest blunder"  story here. 
 
 Precisely such a term is generated by the 
 the energy density $\rho_{\Lambda}$ of the empty universe, namely $\Lambda=8\pi G_N \rho_{\Lambda}$. 
 This is a quantity of no relevance in the absence of gravity,
 which is why it is legitimate to adjust the zero-point of the energy scale in non-gravitational physics.
 But gravity sees everything.
 
The coefficient $\Lambda$, if present, has important implications for the fate of the universe. If it
is negative, the universe collapses, whereas if it is positive the universe undergoes accelerated
expansion. The energy density $\rho_{\Lambda}$ has dimension
[length]$^{-4}$, which in the standard units of particle physics (with $\hbar=c=1$) is equivalent
to [mass]$^4$.  In a theory of quantum gravity, where the only fundamental mass scale is
the Planck mass, its natural value would 
seem to be of order $M_{\rm Planck}^4$. More precisely, $\rho_{\Lambda}$ receives quantum corrections due
to virtual particle creation and annihilation which diverge with the energy of the virtual
particles. If we assume that quantum gravity provides a natural cut-off for these divergences,
one would indeed expect a result of order $M_{\rm Planck}^4$. But even if we plead ignorance 
about the cutoff mechanism and the scale at which they occur, we cannot put that scale lower than
the energy scale we have explored experimentally, because then we would have discovered
the mechanism already. Known physics, cut off  just above the Standard Model scale still
contributes about $10^{-56} M_{\rm Planck}^4$ (see  \cite{Bousso:2007gp}, \cite{Weinberg:2005fh} and  \cite{Polchinski:2006gy}, 
and references therein for recent discussions of this issue).

\ifPreprint Furthermore, as noted above, $\rho_{\Lambda}$
 receives
contributions from any re-adjustment of the zero-point of the energy scale. 
The most notorious example
is the Higgs mechanism. In the standard classical picture, this involves a shift in the value
of a scalar field from a local maximum to a local minimum of a potential density. 
The difference of the values of the potential density at the maximum and the minimum is
a contribution to $\rho_{\Lambda}$. This contribution is also of order $10^{-60} M_{\rm Planck}^4$.\fi
The problem is that the observed value is about $+10^{-123} M_{\rm Planck}^4$.
The first way in which this was observed, around 1998, was from the fact that we appear to be accelerating
away from distant supernovae. 

\subsubsection*{CC versus SM}

It goes without saying that  if $\rho_{\Lambda}$ were 60 or even
120 order of magnitude larger we would have noticed this long before 1998. In fact, if it were
just a few orders of magnitude larger there would have been nobody to notice it. The universe
would have collapsed before we even existed (if $\Lambda$ were negative), or would have
been ripped apart in a  catastrophic acceleration, if $\Lambda$ were positive.

These bounds where made more precise by Barrow and Tipler \cite{BT} and by Weinberg \cite{Weinberg}.
The former authors assumed that main sequence stars are essential for our existence, and
required that the universe would not collapse before they were formed. This gives a 
negative lower
bound. Weinberg assumed that
galaxies (as star factories)
are crucial, and from the fact that they cannot form if the universe expands too
rapidly obtained a positive upper bound. Both of these bounds have an order of magnitude 
of about $10^{-120}$  (the precise factor depends on various assumptions), and leave an allowed
region that is considerably smaller on the negative axis then on the positive one.

Although both bounds can be criticized for possibly be too anthropocentric(are stars and
galaxies just essential for life in our universe, or in any imaginable universe?), it is quite
clear that an observer window exists which is many order of magnitude smaller than the natural
value of the cosmological constant. This window is the analog of the first gedanken 
calculation in section \ref{TGC}. In this case the word ``gedanken" can safely be omitted. It is also
clear that the  amount of tuning required is far larger than what is needed for
the Standard Model. For this reason this argument convinces some people more of the need for a
large ensemble
than the analogous argument for the Standard Model parameters.

But there are two other issues to worry about that look more convincing for the
Standard Model case. The first is the risk of circularity. It is important to know that physics
continues to make sense  if we
consider different values for a parameter than those observed in our Universe. In the case
of the cosmological constant the problem is that it becomes relevant only if we couple
it to gravity, which is precisely the theory we are still struggling to understand. Suppose some
alternative theory of gravity does not couple to $\Lambda$, or that  $\Lambda$=0. If that theory were correct, we
had no right to consider $\Lambda \not=0$ in the first place. The observer window becomes
meaningless, and only the value $\Lambda=0$ would be acceptable.  This is not
the case for observer windows in the Gauge Theory Plane. This is because we can discuss 
gauge theories at the quantum level without understanding gravity. Let us imagine that
some fundamental theory of gravity singled out one particular value of the electron mass
or the fine-structure constant, by being consistent or finite only for those particular values.
We could still do meaningful chemistry computations with
different values different values  of those parameters.  All computations that produce
the observer contours in the Gauge Theory Plane remain valid. Hence the puzzle produced if
a unique fundamental set of values landed precisely within the observer contour
remains genuine.

Furthermore it would not be equally
surprising if a fundamental theory managed to land exactly within the ridiculously small observer
contour along the cosmological constant axis, since this contour contains the mathematically special number zero. Of course, this would
require an alternative explanation of the apparently observed non-zero value for $\Lambda$,  but this possibility
will continue to attract those that have strong objections to an explanation in terms of  a large ensemble.
For these reasons I am more strongly convinced of the  need for a large ensemble covering the 
Gauge Theory Plane than
of the need for such an ensemble for the cosmological constant, even if the degree of 
apparent fine-tuning in the latter case is much larger.

\ifPreprint

\section{The Anthropic Principle}

\else

\subsection{The Anthropic Principle}

\fi
\ifPreprint

%The time has come to mention the ``A-word", which I have carefully tried to avoid until now.

The term ``Anthropic Principle" was coined by Brandon Carter in 1973, and 
for a long time it received limited attention, and mainly from cosmologists and relativists whose
aim was to understand certain ``large number problems", such as the age of the universe in Planck units.

\fi

\ifPreprint

%The name ``anthropic principle" is an unfortunately accident of history, just as
%``Standard Model" and ``string theory" (as we will see later). My problem with
%the name is not primarily the word ``principle", but the word ``anthropic", which 
%suggests a crucial r\^ole 
%for human beings.  
%I am reluctant to give a generic definition, because
%I know from experience that
%whatever I write will be misinterpreted by someone to mean the exact opposite of what I intended.
The basic idea is that we cannot possibly observe any other universe than a universe which allows
us to exist. Therefore our observations are necessarily biased by our own existence. Depending on the
interpretation one gives this to these phrases this can be turned either into a meaningless tautology
or a disturbing form of anthropocentrism. Indeed, initially the anthropic principle was presented as a reaction
to the Copernican Principle. 

\subsubsection*{Anthropic principles}

There seem to be as many anthropic principles (called WAP, SAP, FAP, etc) as there are letters in the
alphabet, but two main classes can be distinguished: 
one where the existence of humans, 
intelligent life or observers is considered essential for the existence of physics and/or the cosmos,
and one where human beings, from a cosmic point of view, are a minor chemical perturbation,
of no special significance, except to ourselves. The first is profoundly anthropocentric, the other just the opposite.
Remarkably, many of the early discussions of the ``anthropic principle" discuss both of these  extreme possibilities as viable
alternatives. 
 I will not attempt here to describe the various forms of the anthropic principle
in detail because it is often hard to understand what people really mean.
Language is an extremely limited
tool for expressing ideas. Consider for example the following definition of the ``Strong Anthropic Principle"
by Barrow and Tipler \cite{BT}: 
``The Universe must have those properties which allow life to develop within it at some stage in its history."
What is the meaning of ``must"? And what is ``{\it The} Universe"? Our Universe, or any possible Universe? 
Why would anyone even want to formulate such a ``principle"?
 Reading the context of such statements usually does not help much either. Occasionally, however, one
finds a formulation that leaves no room for misinterpretation, such as Barrow and Tipler's ``Final Anthropic Principle":
``Intelligent information-processing must come
 into existence in the Universe, and once it comes into existence, it will never die out". Perhaps this should be understood
as the extreme limit of anthropocentric thinking: if the universe exists just for us, why would we die out?

When I first started discussing the point of view explained in
section (\ref{TGC}), around 1987 or 1988, I had a  vague
awareness about something called the ``Anthropic Principle", but because
I did not know exactly what people meant by that, I usually avoided
that term. 
I am still not sure that what I am talking about here is what other people call
``the anthropic principle". But in any case I hope I made it clear that for me it  does
{\it not} mean that human beings have some sort of essential r\^ole in the cosmos.

\subsubsection*{Disgust, Denial, Derision}

Some of the versions of the anthropic principle may sound preposterous, but 
so are some of the reactions it has received.

During the  many discussions I have had on this subject, 
I learned that it is against some unwritten rule of particle physics
to think like this.
Someone even called it ``religion", a word some physicists use as the 
superlative of ``disgusting".  This strong resistance
exists until today. Just last year (2006) a respected colleague vowed to
quit his job if  the anthropic principle would ever play a r\^ole, and he was  not the first
to make such a statement.
In Physics Today Nobel prize winner B. Richter refers to
people thinking in this way as ``creationists" \cite{Richter}. I could fill an entire section with similar
remarks. 

What is even more remarkable
than the emotional language
used in these criticisms is the fact that -- in the area of Particle Physics -- 
nearly all of them date from the last few years.
Before that time, particle physicist and string theorists
 tried to ignore the issue altogether. An interesting example
is \cite{Cahn:1996ag} by R. Cahn from 1996, containing a very nice discussion
about the contours of our region
in the Gauge Theory Plane, pointing out that many aspects are crucial for our existence, all
without any mention of the anthropic principle (by then a well-known subject, at least
 in cosmology), not even to reject it.  
  
 It is still not uncommon, even now, to hear an entire
 talk about the cosmological constant (usually described as one of the deepest mysteries of
 physics),  in which several solutions are suggested that even by the speaker's own admission
 don't make any sense, without even any mention of the possibility of an anthropic explanation,
 and when it is mentioned at all it is usually with some sarcasm, as in
{\it  ``The situation is so desperate that anthropic arguments have been advanced"}
\cite{Sarkar:2007cx}.

As far as I know, the first particle physics paper
openly
discussing  anthropic arguments for Standard Model parameters
is  Agrawal et. al. \cite{Agrawal:1997gf} from 1997, almost twenty-five years
after Carter's talk, and more than ten years after
Linde and Vilenkin advocated it in the context of inflation. 
In string theory, it took even longer. Although string theory
clearly pointed towards an ``anthropic ensemble" of solutions since 1985, mentioning
this in public was ``not done". 
Even nowadays, one can attend an entire conference on ``string phenomenology"  without 
any mention of anthropic arguments, except occasionally in a derogatory way.
Remarkably, the aforementioned paper by Agrawal et. al. does not
even mention string theory!
I know of just three papers before the year 2000 that
contain both the terms ``anthropic" and ``string theory"\rlap.\footnote{LINDE}
The first, by Gibbons and Townsend in 1993 \cite{Gibbons:1993sv},
mentions anthropic selection or chance  as  options in the introduction; the second, a review
of string phenomenology by Quevedo from 1996 \cite{Quevedo} discusses it in a footnote as a possibility we
may have to face if the quest for uniqueness fails. My own text from 1998 \cite{DutchText}
advocates a huge ensemble as the most desirable outcome, but was written in Dutch and
not translated until 2006 \cite{Schellekens:2006xz} . 
 The attitude of denial is nicely illustrated by the title of \cite{Kane:2000ya} from January 2000, the
 first paper on the preprint archives to confront the issue directly: 
  ``{\it The beginning of the end of the anthropic principle}". But contrary to what the
 title may suggest, the authors talk about the possibility of many solutions, and conclude the following:
{\it  Then life would actually arise in those minima that were approximately ``just
so". Thus the ``just so" issue is resolved by having a large number of possible
vacua in which universes can end up.}  That statement is precisely what some people  
would call the anthropic principle. Another ironic twist of history is that three months later
a seminal paper by Bousso and Polchinski \cite{Bousso:2000xa} appeared, that marked the ``beginning
of the beginning" of the anthropic  principle in string theory. 

% Looking at this episode with
%the benefit of hindsight, one must conclude that the authors of \cite{Kane:2000ya} 
%came amazingly close to writing
%a visionary paper, but they chose an unfortunate point of view.

\subsubsection*{So what is new?}

Much of what there is to say about the ``anthropic principle" can already be found in the
classic book on the subject by Barrow and Tipler \cite{BT}. If nothing had changed since then,
my reaction might be similar to those described above. But a few
things have changed in an essential way. They include:
\begin{itemize}
\item{Focus on non-uniqueness as the main principle, instead of considering various 
philosophical options more or less on equal footing.
The fact that our observations are necessarily biased by our own existence is of interest 
only
if there could be something else: the essential issue 
is non-uniqueness of the laws of physics. If the laws of
physics are really unique, there is not much left to discuss, as far as I am concerned.}
\item{The Standard Model. Many of the ``anthropic coincidences" described in \cite{BT} are
based on nuclear physics or chemistry, without a clear notion of what can be varied. The
Standard Model, viewed as a point in the Gauge Theory Plane changes all that.}
\item{The Cosmological Constant, which by Weinberg's argument is bounded from above,
and from observations appears to have a non-zero value.}
\item{String Theory. This offers a concrete example of a fundamental theory that fixes
the possible parameter values, and seems to do so in exactly the right way.
One could try to discuss such a theory purely hypothetically,
as I did in the last section, but I doubt that it would be as convincing if we did not have this explicit example.}
\end{itemize}

%In comparison with this semantic confusion
%about the the meaning of the words ``Anthropic Principle" 
%the ``string theory Landscape" is a breath of fresh air. 
%Any discussion of the AP requires the consideration of other universes than our own. At least now we 
%know which ones are possible. If the fundamental theory only allows just one value for a certain parameter,
%it makes no sense to ask if it has that value for ``anthropic" reasons. If on the other hand the parameter
%has a range of possible values, some of which clearly do not allow anything of interest to exist, then 
%it is obviously true that we will measure a value within that ``anthropic window", since outside
%the window there cannot be anyone to measure it.
%Here ``anything of interest"
%means anything that might plausibly be called an observer. The precise definition of this concept may be troublesome,
%but it is clearly not {\it a priori} restricted to human beings, carbon-based life or even quark/lepton based life. This
%is why I said that the name ``anthropic" principle is unfortunate.
%Precisely
%to avoid the troublesome definition of an observer I decided to enlarge the anthropic window to something that
%most people would agree on as a necessary condition: complexity. 

\subsubsection*{An (anti)-anthropic check list}

Perhaps it is useful to divide the argument in steps, so that everyone can identify their
exact point of disagreement. This is a list of statements in decreasing order of acceptability in the particle
physic community:
\begin{itemize}
\item{The Standard Model may not be the {\it unique} mathematical solution of any fundamental theory.}
\item{Not all alternative solutions allow observers.}
\item{The total number of solutions should be sufficiently large to make the existence
       of a solution with observers plausible.}
\item{We live in the most probable universe which allows observers.}
\end{itemize}

Some people already see the first statement as something close to an act
of treason against the goals of physics.  We are supposed to assume that every number
we encounter can be computed from first principles. One does not have to go back very far
in history to encounter  statements indicating this. For example in Randall's book ``Warped passages" from
2005 \cite{Randall:2005xy} the author states with obvious disapproval:
{\it Some String Theorists no longer try to find a unique theory}. The phrase suggests that most
string theorists {\it are} still trying, but I am not sure if this is true.

Accepting point one but rejecting point two is an interesting way out of anthropic
reasoning, but not a popular one. If all alternatives allow observers anyway, there is
not much reason to reject uniqueness.
People who accept the first two points have already
accepted a form of the anthropic principle. If there exist (at least mathematically) 
universes different from ours in which no life is possible, it is obvious why we don't live there.

Point four is the most controversial one. It is not clear how probabilities are defined
and if we can compute them, and even if we can, our Standard Model is just one sample out
of an ensemble, a nightmare scenario for statistics.
This is another favorite line of attack against anthropic arguments. However,
it is useful to distinguish two frontiers in the current debate: one that concerns point four, and
one that concerns point three (or even just one and two). Some people are still hoping for
a unique outcome, while others take non-uniqueness for granted, and are marching ahead
trying to master probabilities. Even if they fail, it does not bring us back to the unique outcome
many people hoped for two decades ago.
 In this article I am just advocating point three. I do not really care whether that
 is properly called ``the anthropic principle". That is what others called it in discussions 
 (usually as an accusation). 
 
Another way of stating this is to distinguish two kinds of anthropic arguments: anthropic
window arguments and anthropic probability arguments. The first just states that parameter
values have to lie within certain ranges to allow observers, and the second kind of argument
tries to derive our precise position within such a window. The first is non-tautological only
in the context of a fundamental theory with
an ensemble of parameter values in- and outside the window. 
In that context  it really explains why we find ourselves within that window, whereas otherwise it
explains nothing. Once a fundamental theory  with such an ensemble is established beyond
any doubt, the anthropic principle as defined in point three is only of secondary relevance. It is obviously true that the anthropic principle
explains why the size of the orbit of the earth is neither extremely large or extremely small, but this is usually not worth
more than a footnote in a treatise on planetary orbits. 
We have not reached the same level of understanding of a fundamental theory of all interactions
 as of the theory of planetary orbits. Anthropic arguments tell us something about the properties such a theory should have.

The cosmological constant provides a nice illustration of this point. If the fundamental theory
just has one solution, or only solutions within the anthropic window, the anthropic argument explains
nothing. If it has many solutions spread in and outside the window, one can indeed claim that the
anthropic argument explains why we measure a value inside the window. This fact becomes
less miraculous as the density of the solutions increases. If the solutions
have a known distribution one can move to point 4, and try to derive {\it where} we should expect to be
within that window.

Note that I am implicitly assuming discrete solutions here, for reasons explained earlier. 
Apparently some people (in particular the authors of \cite{Kane:2000ya}) have reserved the term ``anthropic
principle" for the continuous case only. Their logic is that if the distribution is discrete, we 
will eventually be able to determine experimentally which one describes our universe. Indeed, the tiny
circle in fig. 1 is monotonically decreasing with time, as long as we continue to do
experiments, and eventually will contain only a single point. But does not take
anything away from the anthropic argument I am advocating here. 

\subsubsection*{Throwing in the towel?}

An often-heard objection to anthropic thinking is that it is like ``giving up" or 
``throwing in the towel". But what are we giving up? These statements are based on the
assumption that the ideal outcome would be that everything is derived from an underlying
theory. As I tried to argue above, that would be a disastrous result. This is not a battle I would like
to pursue.

The objection would be valid if someone just declared that we live in a point in the Gauge Theory
plane where life is apparently possible, and that there is nothing left to understand. 
Perhaps this is what some supporters of
anthropic principles have been advocating in the past. But this
is not what I a advocating here. The crucial difference is the requirement of
finding a fundamental theory that produces the required distribution of solutions in the Gauge Theory Plane.
This goal is actually {\it more} ambitious than many ideas just aiming at partial understanding, in the
form of relations between gauge theory parameters.

\subsubsection*{Superintellects?}

When  Fred Hoyle saw his prediction of a resonance in Carbon confirmed he was so
impressed that he
stated ``A common sense interpretation of the facts suggests that a superintellect 
has monkeyed with physics, as well as with chemistry and biology".
After surrendering the solar system and loosing the battle concerning the origin of species, 
people seeking scientific evidence for their favorite superintellect hold on to
anthropic
coincidences as their next straw.
While it is logically possible that in the course
of science we discover that the only possible explanation of some phenomenon
is the involvement of some sort of superintellect,
it is not what most scientists would aim for, because almost by definition that would be the end
of the road towards a deeper understanding. 

But the question I am addressing here  does not hold much promise for evidence for
superintellects. A 
fundamental theory with a huge number of solutions solves the anthropic coincidence
problem without any need to monkey with the laws of physics.  After many frustrating
debates with physicists, I was delighted to see that biologist Richard Dawkins understood
that point immediately \cite{Dawkins}  (among physicist there are notable exceptions, including Weinberg, 
Linde and Susskind).

But even if the other side in the debate is right, there is not much left for a  superintellect to do.
Even though a unique solution creates a deep mystery, and even though the hope
for such a solution smells like medieval anthropocentric thinking, a unique solution 
would be mathematically fully determined, just as the value of $\pi$. Even a superintellect
cannot change mathematics. 

\subsubsection*{The Multiverse}

If the fundamental theory has a huge number of solutions we avoid the 
anthropic fine-tuning problem in the mathematical sense.  It would not be surprising
anymore that the two gedanken computations of the previous section overlap.
Our universe would be based on one of those solutions. The others would just be
possibilities that are not realized in our universe. 

But could the others be realized in other universes? Whatever that statement might mean,
something like it is an almost inevitable  consequence of the foregoing. If our universe 
were unique (as opposed to its laws of physics being unique), we would be faced with
another puzzle: how was our particular set of laws of physics selected from the huge
set of possibilities?  Here there would still  appear to be a potential for involvement of superintellects.

But there is an obvious way out: that our
universe is {\it not} unique, but it is created multiple times by some process that
samples the set of possibilities. I have always taken something like that for granted, without
focusing on a specific mechanism. 
The whole idea of non-uniqueness of the laws of physics just does not make any
sense without it.

Fortunately other people {\it did} focus on specific mechanism,  which led them
to the same conclusion from a different direction. 
The concept of a ``multiverse" goes back a long time, and originated from a variety of ideas in cosmology.
In the
early eighties, the idea of cosmic inflation was proposed, and naturally
led to the possibility of separate patches of space-time blowing up into distinct universes, a notion
that was especially emphasized by Andrei Linde and Alexander Vilenkin. This leads to a concept known
as ``eternal inflation".
Linde and Vilenkin 
concluded that such a multiverse would offer a scientifically acceptable realization of the ``anthropic principle".

The multiverse  idea is close to, but not exactly the same as the main topic of this paper. One could
have a multiverse consisting of universes with identical fundamental laws of physics (gauge theory
choices), and one could even imagine a single universe where the laws of physics were
chosen randomly from a set of mathematically allowed possibilities. But clearly  the last option would not
suit my goal, as it replaces a mathematical coincidence (the matching of the two gedanken computations)
by a statistical miracle. Obviously, both ingredients are essential: the multiverse and eternal inflation on the one hand
and  a plethora of solutions of a fundamental theory on the other hand. 
Cosmology and inflation will not be discussed here in detail since I wish to focus on 
two other ingredients whose importance has been somewhat underrated: the Standard Model and string theory. For
more information and references I refer the reader to \cite{Linde:2007fr}.

The multiverse raises some obvious questions. 
Do these other universes ``exist"? Can we observe them, even in principle? If not, have we
left the boundaries of science? For me, a sufficient answer would be that they exist as solutions
to some fundamental theory that also contains our own universe as a solution. Our task  is
to gather evidence for the correctness of that fundamental theory from the only
universe we are likely to have access to: our own. This evidence may be experimental, but it
can also be based on consistency of the theory itself.
The best one can hope for is that this
fundamental theory makes a prediction about  our own universe that is so convincing that
we also have confidence in the extrapolation to other universes. 
 But it is logically  possible
that our own universe does not contain enough information to crack this problem. We may be
left with  a huge number of unrelated candidates for a fundamental theory, each (of course!)  with
a huge number of solutions, without being able to distinguish between these fundamental theories. 
Fortunately, at this
moment there is only one candidate.

\fi

\section{String Theory}\label{StringTheory}

Even a basic introduction into string theory would be too long for this article, and there
are many available elsewhere. I will limit myself to a few points that are relevant in this
context. Let me begin by stating the most important unsolved problem.

We are used to specifying a physical theory by writing down, for example
a Hamiltonian. In most problems of interest in physics  this Hamiltonian is too complicated to deal with.
Then we may resort to perturbation theory. We write the Hamiltonian $H$ as 
$H = H_0 + \epsilon H_1 $, where $H_0$ is  a manageable
part that gives the dominant contribution, and $\epsilon$ is small. Then we
expand in $\epsilon$. 

In string theory we are in the bizarre situation that we know the perturbative
expansion, but we do not know what $H$ is. 
\ifPreprint 
That this can happen is easy to understand, since
there are functions that do not have a Taylor expansion, for example ${\rm~exp~} (\frac{1}{\epsilon})$.
Such a contribution can never be detected perturbatively. 

\fi
String theory got its name because $H_0$ describes the propagation through space-time
of string-like objects, and $H_1$ describes splitting of such objects into other strings, or
the joining of two strings into one string. \ifPreprint \else Remarkably there exist six distinct 
choices for $H_0$ and $H_1$ that seem all to correspond to the same $H$. One of these
perturbative expansions cannot even be described in terms of interacting strings. \fi

\ifPreprint
Quantum mechanics 
imposes severe restrictions on the space-times in which strings can propagate
consistently. The simplest theories can only exist in ten dimensions.
There are in fact five such theories, with different $H_0$ and $H_1$,  apparently defining five
distinct fundamental Hamiltonians $H$. However, there are reasons to believe that in fact all five
are different perturbative expansions of the {\it same} theory. This is expected because we know
that all five ten-dimensional are closely related. These relations typically involve replacing
one or more expansion parameters $\epsilon$ by its inverse $\frac{1}{\epsilon}$. Such relations
are known as {\it dualities} and caused a great deal of excitement around 1995, because
they pointed towards a possible uniqueness: instead of five theories there seemed to be just one,
with different perturbative expansions.  
Furthermore, to most people's surprise,  there turned out to be a sixth
perturbative expansion of $H$ that is not described  by string propagation at all, and 
that corresponds to a theory in 11 dimensions. 
\fi

The proper definition of string theory would be to specify $H$. At present, nobody knows how
to do that, or even if it is possible. This is not a new problem and also not one that string
theorists have overlooked or tried to hide. Indeed, the problem is mentioned prominently
on page 27 of the classic string theory book from 1987 by Green, Schwarz and Witten  \cite{Green:1987sp}, where it
appears as a big question mark in a figure. The fact that twenty years later this question
mark is still there is worrisome. At this point one may see this either as a challenge or a fatal
flaw. It may mean that some of the ingredients we are using are just wrong: this is undoubtedly
true for the concept of a Hamiltonian which I used for simplicity above, but may also be true
for the idea of an action, quantum
mechanics, the continuity of space-time or the very concept of space-time or. Indeed, 
the fact that in different limits it gives expansions
with distinct space-time dimensions may give a hint: it is not
even clear in how many space-time dimensions $H$ should be formulated. In fact, many
people think that the definition of $H$ should not involve space-time at all, and the 
space and time are ``emergent" concepts, in a sense that remains to be made precise.
Hence the name ``string theory" is misleading in two respects: it is not yet a theory, and
it is only described by propagating strings in some special limits.  Sometimes one uses the
name ``M-theory" for this entire connected ensemble, but I will simple continue to call
it string theory despite these caveats.

Every statement we make about this would-be theory is based on results in one of the
perturbative regions, with additional insights from some non-perturbative effects that can
be taken into account, and symmetries such as duality and supersymmetry. Attempts to make these insights mathematically
rigorous tend to be hampered by the fact that at some point full control of all non-perturbative
effects is required. However, what we have learned by studying string theory
using the limited tools at our disposal is amazing. The spin-offs alone
(various results in mathematics and quantum field theory, and especially the
so-called AdS/CFT correspondence) have 
made the effort worthwhile so far. Remarkable results in quantum gravity, 
in particular an understanding -- in certain cases -- of the 
microscopic origin of black hole entropy suggest that something is right
about string theory.

Even without explicitly knowing $H$,  we can
deduce which answers the theory will give to fundamental questions, such as the one that is the main
topic of this article, the uniqueness of the laws of physics formulated in terms of gauge theories. 
The answers it gives to questions like this one
 will tell us whether it is worthwhile to try and unveil the mysteries hidden in $H$.
Two decades ago that was clearly true, and the pursuit of this goal has
led to a bounty of beautiful results that all by themselves made the effort worthwhile. Let us see
 what it says about the central theme of this article, and decide if the answer is sufficiently
 encouraging to continue.

%I do not intend to present a  string theory advertisement, or
%pretend that we already know that it is the ``theory of everything".
%The point of view I will try to explain here is that to some of the big questions of physics,
%and even science in general, string theory suggests spectacular answers. To me these answers
%seem precisely right. This may either mean that string theory is right, or that we should start
%looking for an alternative that gives similar answers.
% For the moment, however, no such alternative
%is in sight.

\subsection{Quantum Gravity and Particle Physics}\label{QGPP}

String theory may  -- and must -- provide the solution to many 
outstanding problems, but
there is one problem that it addresses directly: Quantum Gravity. Basically, the goal is to put Quantum Gravity
on equal footing with the other three interactions we know: to turn it into an
unlimited precision theory. When one tries to quantize gravity naively, one is
confronted with exactly the same problem we encountered in the quantization
of the Standard Model: quantum contributions that depend on unknown short
distance physics. The difference is that in the case of quantum gravity these unknown
contributions do not ``decouple": they are  proportional to powers of $E/M_{\rm Planck}$,
and since $M_{\rm Planck}$ itself is expressed in terms of Newton's  constant, one
cannot send it to infinity if one discusses gravity. \ifPreprint In the case of the Standard Model,
 such
contributions, should they exist, are proportional to  $E/M_{X}$, where $M_X$ is an unknown scale which
is not linked to the parameters of the Standard Model itself\rlap.\footnote{The only
exception are the Landau poles described in section (\ref{QFT}), which however are
beyond the Planck scale, if the Higgs boson  is not too heavy.}
Hence we can, at least
formally, consider the limit $M_X \rightarrow \infty$. At present, we do not know
of any such corrections to the Standard Model, but if we couple it to gravity, 
corrections  proportional to powers of $E/M_{\rm Planck}$ may be expected to the 
Standard Model parameters as well. To make matters worse, there  are infinitely many
independent corrections of this kind.\fi

\ifPreprint
One could consider various options. Perhaps some principle 
of nature prohibits unlimited precision, in the spirit of the uncertainty principle. 
Or perhaps there simple exists an infinity of parameters, and the goal of unlimited
precision can only be reached if we measure all of them (for practical purposes a 
finite number would suffice to reach  some definite degree of precision). In the latter
case, perhaps some unknown new theory at a scale beyond the Planck scale
fixes  these parameters. It is  hard to dismiss these alternatives rigorously, but it is
clear that they are less attractive than the obvious one: that all  of these  problems
are caused by naive quantization of gravity, and that they should be solved by
a correct theory of quantum gravity. Since we already achieved the goal of
unlimited precision for three of the four forces  we know, it does not seem {\it a priori}
unreasonable to expect the fourth to be manageable as well.\fi
  
To achieve the goal of unlimited precision, we need to control  these corrections.
However, controlling these corrections
is not merely a problem limited to gravitational interactions. All other particles and
interactions
inevitable couple to gravity, and contribute quantum corrections of the same
order of magnitude as purely gravitational contributions. \ifPreprint This insight predates the
new string era, which started in 1984.
Very illustrative is the following quote of Howard Georgi, which
appeared on Jacques Distler's  blog ``musings" \cite{Distler},  and dates from 1982 or 1983: 

{\it ``There's no decoupling limit in which it is sensible to consider quantum
 gravity effects, while neglecting other interactions. Unless you know particle physics
 all the way up to the Planck scale, you can never hope to say anything predictive about quantum gravity".}
\fi

One  might hope that  this can be turned around, and that when we understand
quantum gravity, we will know particle physics all the way up to the Planck scale.
Perhaps a theory of quantum gravity will then uniquely determine the Standard Model?

\ifPreprint \else
The perturbative expansions of string theory suggest that it does indeed solve the problem of
quantum corrections by keeping its particle spectrum tightly under control. 
A simple exercise in string theory spectroscopy reveals that its spectrum consists
of an infinite series of particles with masses corresponding to quantized eigenmodes of 
oscillations of the fundamental string. Of all these modes, only the lowest ones are observable
to us as particles.
Such a spectrum is essentially unchangeable. All quantum
corrections are under control for precisely that spectrum, but
if one
changes the mass of any of the particles by the slightest amount, this property is lost immediately.\fi

\ifPreprint
Let us  now  see what string theory has to say about these matters. First of all,
what we  learn from the perturbative expansions described in the previous  
subsection suggests that the quantum corrections are indeed under control.
Infinite (and hence indeterminable) integrals in quantum gravity perturbation
theory turn into finite integrals yielding  definite numbers.
Furthermore the reason why this happens is precisely the one stated
above:  in string theory we know all particle physics all the way up to the Planck scale.
The infinities  of quantum field  theory  and naive quantum gravity are
identified  as  being due to an  infinite over-counting of a  finite  contribution. 
In the best understood case, theories of closed strings,
this is related to a property known as ``modular invariance". 
At lowest order,  the relevant  integral (an expression for the simplest quantity that
displays the problem)
takes the form (a few irrelevant details have been ignored here)
\beq
\int \frac {d^2\tau} {( \Im \tau)^2 } ( \Im \tau)^{(2-D)/2}\   \hbox{{\rm Tr}} \  e^{2 i\pi \tau H_0} \ ,
\eeq
where $D$ is the number of space-time dimensions and $H_0$ the perturbative
Hamiltonian. The trace is over all  states in the spectrum of $H_0$.
The integral in quantum field theory would be over the entire complex 
upper half plane, and is clearly divergent near $\tau=0$. But in string theory
some contributions to this integral can be shown to  be identical copies
of each other, and
we they would be over-counted
 if we were to integrate over the entire upper
half plane. These  identical copies are related by the following transformation
\beq
\tau \rightarrow \frac{a\tau+b}{c\tau+d},  \  \  \  a,b,c,d \in {\bf Z},\  \ \  ad-bc=1.
\eeq
It is an amusing exercise to work out how this divides the complex
upper half plane into an infinite number of copies. For this to be relevant the
integrand must be invariant under this  transformation, which implies strong
constraints  on the spectrum of $H_0$. These constraints are known as ``modular invariance".
To avoid the over-counting we can then
limit ourselves to one region, and in particular we may choose one
 that excludes $\tau=0$, thereby explicitly avoiding the field theory divergence.
The spectrum  of string theory consists of an
infinite ``tower" of excited states, corresponding to  quantized energy levels of the
various modes of the string. Any change in the spectrum
of such a tower destroys the  crucial property of modular invariance.\fi

\subsection{Non-Uniqueness in String Theory}\label{NUST}

It is understandable that this rigidity of the spectrum fueled the hope that string theory might
lead us to a unique gauge theory, and perhaps a completely unambiguous derivation
of the Standard Model from first principles.  This hope is very well described by the following
paragraph from the book ``The Problems of Physics" by A.J. Legget, which dates from 1987 
\cite{Leggett}\rlap.\footnote{This book also contains a remarkably prescient description of what
might be called an ``anthropic landscape", even with references to  an important r\^ole for 
higher-dimensional theories, a 
notion that also appeared in equally prescient 
work by Andrei Sakharov from 1984 \cite{Sakharov} about a possible anthropic solution to the cosmological
constant problem. However, precisely because of the cited text about string theory, this remained an overlooked link
in the idea for more than a decade.}
The author is not a string theorist (he received the Nobel Prize in 2003 for his work on
superfluidity) but echoes very accurately the atmosphere in part of the string community around that time:

\noindent
{\it The hope is that the constraints imposed on such theories solely by the need for
mathematical consistency are so strong that they essentially determine a single possible theory
uniquely, and that by working out the consequences of the theory in detail one might eventually
be able to show that there must be particles with precisely the masses, interactions, and so on, 
of the known elementary particles: in other words, that the world we live in is the only possible one.}

\noindent

If this had been true, this would have led us to straight to the  anthropic
dilemma explained in section (\ref{TGC}).  So how does string theory avoid this? 

The answer to that question emerged during two periods of revolutionary change
in our understanding, one occurring around 1986, and the the other during the first
years of this century. I will refer to these periods as the first and second string vacuum
revolution. Although string theorists love revolutions, these
two are usually not on their list.

It is important to distinguish two concepts of uniqueness:  uniqueness of the theory itself, 
or uniqueness of its ``ground states" or ``vacua". I will use these notions in a loose sense here, because
one of the issues under dispute is even how they are defined (which is especially problematic
in a universe with a positive cosmological constant, as ours seems to have). By ``vacuum" I will
simply mean anything that is suitable to describe our universe, and anything that merely differs from
it by being located in a different point in the Gauge Theory Plane. 
\ifPreprint I am not trying to argue that such
vacua exist, but merely that if they do exist there are likely to exist in huge quantities. The picture that seems
to emerge is that of a perhaps unique theory, but with a huge number of vacua. 
Although this picture  has started emerging more than twenty years ago, most people refused to accept it  as
the final outcome, and instead were (and in surprisingly many cases  still are) 
hoping that one of the many candidate vacua would  be
singled out by some still to be discovered mathematical principle.
%Some advocates of
%``uniqueness" may say that this is all they were talking about anyway, but having been a participant
%of the first string vacuum revolution myself, I know this is not true.  
%Somewhere along the way some people
%slowly changed their minds, but they forgot to say so, and missed the importance of this change.\fi

\subsubsection*{The first string vacuum revolution}

String theory was originally discovered in 1969 in an attempt to understand certain
features of the strong interactions. Around 1975 it was realized that it always contains
graviton-like particles and hence was an interesting candidate for a theory of gravity.
It has been under serious consideration as a theory of {\it all} interactions since 1984, when
a new kind of string theory  (called  the ``heterotic string")  was discovered. This became
the arena for the first string vacuum revolution. 

Heterotic strings are fundamentally ten-dimensional theories, and six of the ten dimensions
must be ``compactified" in order to make them macroscopically unobservable.
A rather trivial way to achieve that is to roll them up almost literally on a six-dimensional
torus. This leads to a set of four-dimensional string theories with continuous parameters that
specify the shape and size  of the torus. 
\ifPreprint
Note that already at this point the uniqueness of the
outcome is in doubt. Clearly this construction would work for any number of space-time
dimensions less than ten, and furthermore it leaves us with a set of undetermined continuous parameters. It is hard
to dismiss torus compactifications on theoretical grounds, but at least they are
ruled out for phenomenological reasons.\fi

\ifPreprint
\else
 Gauge theories obtained from torus compactifications have several fatal flaws, and one of
 them is that left-handed and right-handed particles couple to all gauge bosons in the
 same way, and hence cannot reproduce the parity violation in observed in the weak interactions.
 In 1984 and subsequent years this problem was overcome in a variety of ways. The best known
 one is compactification on six-dimensional spaces called ``Calabi-Yau manifolds". Within a few years
 it became clear that there was a huge number of possibilities, and many people wrote conclusions 
 similar to the following last line from a paper \cite{Strominger:1986uh} by Andy Strominger in 1986: 
 ``All this points to the overwhelming need to find a
dynamical principle for determining the ground state, which now appears more
imperative than ever". I was involved, with Wolfgang Lerche and
Dieter L\"ust,
 in another paper \cite{Lerche}
from that year, in which we made an attempt to quantify the meaning of ``huge". We 
quoted the number $10^{1500}$ as a precisely defined, but not saturated upper bound on the
kind of theories we were constructing.
We also stated ``Even if all that string theory could achieve would be a completely
finite theory of all interactions including gravity [...] it would be a considerable success".
 \fi

\ifPreprint
These theories {\it do} contain matter and gauge interactions in four
dimensions, in addition to Einstein gravity, but they suffer from a fatal flaw that makes them
unsuitable for describing the Standard Model interactions that we observe: they are unable
 to produce different couplings for left- and right handed particles, a feature known
 as ``chirality", and which is experimentally observed in the Weak interactions. This may seem
 like a minor point in view of many other unsolved problems, but it is in fact a crucial
feature of the Standard Model. Indeed, progress in string theory
 was severely hampered until 1984, precisely because of the problem of obtaining
 chiral interactions.
This problem could be solved
by using instead of a torus a special kind of six-dimensional manifold called a ``Calabi-Yau manifold".
In 1984 this subject had barely been studied by mathematicians, and hence the preprint version of the
first paper \cite{Candelas:1985en}
using them could state with confidence that ``very few are known". This remark was removed
in the published version of the paper, because by then it was clear that there were many possibilities.

Others tried to construct Heterotic Strings directly in four dimensions, by modifying the
ten-dimensional construction (this is now believed to be equivalent to space-time
compactification). One of the first papers following this strategy was by Narain in 1985 \cite{Narain:1985jj}.
He found in fact a continuous infinity of solutions, a result that was initially received with scepticism
by some people, but was soon realized to be absolutely correct. Most people comforted
 themselves with the fact that Narain's theories were unsuitable for describing the
 observed Standard Model interactions for a rather basic reason:  they could be identified as
generalized torus compactifications, suffering the same chirality problem mentioned above.

In subsequent years this explosion of possibilities continued. In particular in 1986 Andy Strominger 
\cite{Strominger:1986uh}
studied a generalization called
``Calabi-Yau
manifolds with torsion", and he found so many possibilities that he
remarked in desperation that ``all predictive power seems to have been lost". The last phrase of his paper 
reflects the attitude at that time: ``All this points to the overwhelming need to find a
dynamical principle for determining the ground state, which now appears more
imperative than ever".  The need for this elusive dynamical principle was mentioned
in many papers in subsequent years.

Other papers in 1986 generalized the direct construction of String Theories
in four dimensions further, to overcome the lack of chiral interactions of Narain's
theories. These approaches also suggested huge numbers of solutions, which could
not be dismissed so easily anymore. The authors of these papers commented on their
results in interestingly different ways, but they essentially all took a phenomenological
attitude: there are many possibilities, but it is sufficient if just one of them matches
the experimental data. The authors of \cite{Kawai:1986va}
conclude with ``We believe [...] a complete classification is a tractable problem and
that the relevance of string theory to nature can be tested". In \cite{Antoniadis:1986rn} one finds the statement
``The number of consistent four-dimensional string theories is so huge that classifying them
all would be both impractical and not very illuminating". At CERN I was involved, with Wolfgang Lerche and
Dieter L\"ust,
 in another paper \cite{Lerche}
from that year, in which we made an attempt to quantify the meaning of ``huge". We 
quoted the number $10^{1500}$ as a precisely defined, but not saturated upper bound on the
kind of theories we were constructing.
We also stated ``Even if all that string theory could achieve would be a completely
finite theory of all interactions including gravity [...] it would be a considerable success".

By that time everyone understood that at the level of approximation 
of the time, there were going to be huge numbers of solutions, which could
not be dismissed straightforwardly. The first vacuum revolution continued for many 
more years after 1986, and was not limited to Heterotic strings. This just made
the number of possibilities even larger.

However, there were (and still are) plenty of unsolved problems, and at least two of those  are
relevant to the problem of defining the ground state. The first is that most results
relied on a symmetry between fermions and bosons that is not observed in nature: supersymmetry.
Supersymmetry leads to a cancellation of quantum corrections between fermions and bosons,
and makes certain computations possible (well-defined and finite) that are otherwise undoable.
It is a great computational tool, but it is still not clear whether it is a symmetry humans need in order
to deal with quantum field theory, or whether nature itself makes use of it, at least approximately.
A popular idea, testable at the LHC, is that it will emerge as an approximate symmetry, valid for energy
scales larger than about a TeV. Whether that is true or not, in any case the supersymmetric
solutions  are at best approximately relevant for our universe.

The second problem is the so-called moduli problem. \fi Just as torus compactifications, Calabi-Yau
compactifications
 turn out to have continuous parameters called ``moduli". The same is true for all the other
 four-dimensional string constructions, although the geometric interpretation of their moduli 
 may be more complicated.

Essentially, what the first string vacuum revolution brought us was a huge 
(but still incomplete) list of
topologically distinct supersymmetric moduli spaces: four-dimensional string theories,
that are
not connected to each other by continuous changes of their moduli. 
The topological distinctions arise from various discrete choices one can
make when construction a four-dimensional string theory: the choice of the Calabi-Yau
manifold, and a variety of fields that wrap non-contractible cycles on such a manifold.
These choices give rise to distinct gauge theories in four dimensions: distinct
gauge groups, coupling to distinct sets of particles. These particles usually come out organized
in ``families" occurring with a topologically determined multiplicity.
Each of these topologically distinct
choices comes with a certain number of moduli. 
 \ifPreprint 
 
 The number of such topologically
distinct moduli spaces is huge indeed.  We do not know how many, but just to get an idea,
let us consider
Calabi-Yau manifolds. The topology of these spaces is characterized
in part by two integers called Hodge numbers. 
An extensive list of Calabi-Yau spaces is available \cite{Kreuzer}, containing 30.000 distinct pairs
of Hodge numbers. Since spaces with identical Hodge numbers may still be
topologically different, the actual number is much larger. If we consider more general
topological possibilities, it becomes much larger still. 

%By now we can say that it is indeed possible to get the correct gauge group,
% $SU(3)\times SU(2) \times U(1)$, coupling to particles with 
% the quantum numbers of quarks and leptons, occurring with the required multiplicity of three.
% There are in fact many examples, usually with some additional matter and/or interactions
% that does not couple to the Standard Model and hence might easily have escaped
% detection so far, and might even account for the non-baryonic matter in our universe. 
% But it is also clear that the Standard Model is by no means the only kind of matter that we can get.

On top of these topological choices one gets the continuous parameters, the moduli. \fi
A typical four-dimensional string vacuum may have hundreds of them. 
The physical parameters of the resulting four-dimensional
 theory depend on these moduli. This includes in particular the strength of the gauge
 couplings and \ifPreprint the Yukawa coupling  parameters, and hence ultimately \fi the masses of all
 quarks and leptons, in those cases where quarks and leptons can be identified.

\subsubsection*{Dynamical Parameters}

\ifPreprint In the supersymmetric approximation all values the moduli can take
are equally good. We might be able to get the Standard Model, but essentially
have no theoretical control over any of its parameters. They are functions of the moduli.\fi

This may seem to disagree with the expectation expressed in
section (\ref{QGPP}) that Quantum Gravity would
be  restrictive enough to fix  the entire spectrum. But there is no disagreement. The way
out is that the spectrum consists of infinite ``towers" of particles, which themselves are
rigid. All that can be changed is a {\it finite} number of parameters that determine the
{\it infinite} number of coupling constants of the particles within the towers. In other words, the 
set of variables is of measure zero in the space of  all possible couplings of all particles
in the string spectrum. It appears that the infinite number of excitations of the string spectrum
is an essential feature to satisfy on the one hand the rigidity of Quantum Gravity, and allow
on the other hand enough variability to end up with an interesting gauge theory. At the moment,
however,
 this is merely an observation about string theory, and not a general theorem about Quantum Gravity.

If the moduli were merely parameters, we would now be facing the
problem mentioned in section (\ref{DiscCont}): how can we move around
in parameter space, so that we can end up in different points in the parameter space 
when new universes get started?

Fortunately string theory provides a natural solution to this
problem: all parameters are dynamical. In string theory all parameters are
functions of scalar fields, that obey field equations. Classically, a scalar field is just
a functions of space and time without a preferred direction, and whose dynamics is
governed by a field equation.
There is such a scalar field for
each modulus.
Since scalar fields do not have preferred space-time directions, they can take constant values in
space-time patches without violating Lorentz invariance locally. The classical values
of these fields determine the values of the gauge theory parameters in each patch.
These classical values correspond to vacuum expectation values of the scalar fields
in a quantum theory.

\ifPreprint
This feature has often been heralded as one of the greatest triumphs of string theory:  ``string theory
has no free parameters" was a popular slogan. However, what used to be parameters had
now become vacuum expectation values of scalar fields. Nowadays, one sometimes
 hears the objection
that the anthropic principle is turning physics into an environmental science: some
parameters are claimed to be a feature of our environment, and not derived from first
principles. But how different is ``environmental variable" from ``vacuum expectation value"?
Turning  the Standard Model parameters into functions of scalar fields opens the door for
the possibility that they may be different in different parts of the universe/multiverse, and
hence makes it immediately a lot more plausible that their values depend on the observer
(or the presence of observers).
\fi

\subsubsection*{The second string vacuum revolution}

If the parameters  of the Standard Model are dynamical some obvious questions are:
how do they get fixed, and
why would they remain constant over large ranges of space and time?
\ifPreprint
Indeed, in the supersymmetric limit one can change their values without
any cost in energy. In fact, this energy is always exactly zero (in the absence of gravity).
But this turns out to be an artifact of supersymmetry. This
causes an exact cancellation between bosonic and fermionic contributions
to the vacuum energy which leads to a rather counter-intuitive result: physically
distinct systems can have identical vacuum energies. As soon as supersymmetry
is broken this degeneracy is lifted, and the result is as one would expect: any
change in the values of one of the parameters of the gauge theory leads
to a change of the vacuum energy.\else
This requires a discussion of vacuum energy density as a function of the vacuum expectation values of 
the scalar fields.\fi

   If we take into account the values of these energy densities, we get something
   like a potential for the values of the scalar fields, the moduli. 
   This is a function
   of tens or hundreds of such moduli. If we plot the energy density as
   a function of the values of the scalar fields, one may hope to get a curve with some
   minima. To obtain a more or less stable universe, one would have
   to end up in one of the minima. Such a universe may not be exactly stable: it may
   decay by quantum tunneling. But to fit the experimental data of our universe
   it just has to live at least $13.7 \times 10^9$ years. 
   
   However, it is by no means guaranteed that a minimum exists. What may happen 
   (and indeed does often happen in simple models) is there
   are directions in moduli space where the potential just gradually decreases, reaching 
its true minimum at infinity.  In this situation the moduli run away to physically
unacceptable values. This is known as the ``moduli stabilization problem".

   We encounter an even more serious problem if we couple the system to gravity.  Then not only
   the existence of minimum is relevant, but also the value of the potential at the
   minimum. This is the famous cosmological constant, discussed earlier. 
   
   In a given string theory vacuum the cosmological constant is a finite and calculable number. 
But since the natural scale of string theory is the Planck scale, the result of such a calculation
is some dimensionless constant $x$  times $M_{\rm Planck}^4$. Somehow this dimensionless
number $x$ has to get a value of about $10^{-120}$. 
\ifPreprint   

It is impossible to say how many non-supersymmetric vacua string theory could have
been expected to have, based on the information available before the end of last century,
but I think most people would have put this number between 1 and  $10^{20}$ or so, had they been
forced to mention a number. If that had been correct, it seems extremely improbable that a value
$x=10^{-120}$ could come out. 
\else
Clearly this would not happen by coincidence, unless the total number of solutions would turn out to
be much larger than $10^{120}$. But even the few people who, like me, believed in the existence of a large
ensemble as the final outcome would not have dared to dream  of numbers as large as that.\fi

\ifPreprint Indeed, most people took the attitude that 
presumably $x$ was exactly zero
even in non-supersymmetric string theories, due to some mechanism that was still to be 
discovered.  If such a mechanism had been found, 
the observation of a non-vanishing $x$ would have 
falsified string theory. But even with a distribution of $10^{20}$ non-zero values $x$, string theory
would have been falsified at an enormous confidence level, simply because the
distribution is not dense enough.\fi

But a discovery by Bousso and Polchinski \cite{Bousso:2000xa} in the year 2000 made string theory survive
this potential falsification. They identified a new topological feature that has the potential to
solve this problem at the expensive of
an explosive growth in the number of string vacua.

\subsubsection*{The Bousso-Polchinski mechanism}

%The 1986 string vacuum revolution gave us good indications for the existence of a huge 
%number of string vacua. Although most papers focused on supersymmetric string theory, which
%have a vanishing cosmological constant, already in the late eighties some computations were
%done of non-zero cosmological constants in non-supersymmetric string vacua (to lowest
%order in perturbation theory). As expected, the result is a certain distribution of positive and
%negative values for $x$. The number of theories sampled in these computations is of
%order thousands to millions. Obviously with such a small number of vacua the chance
%of hitting the number $10^{-120}$ is essentially zero.

We have seen that scalar fields can take constant values without breaking Lorentz
or translation invariance. Electric and magnetic fields cannot do that, because they
must point in a certain direction. But there is one other kind of field that can also
play this r\^ole. Electromagnetic fields are given by field strength tensors $F_{\mu\nu}$, which
are anti-symmetric in $\mu$ and $\nu$.
Such tensors have natural generalizations to tensors with an arbitrary number of anti-symmetrized
indices. The one of interest here is the one with the maximum number of indices
in four space-time dimensions, $F_{\mu\nu\rho\sigma}$. 
The reason we do
not usually hear much about such fields is that they are non-dynamical in four dimensions; they do not have
any propagating modes, and hence there are no ``photons" associated with them.  In higher dimensions such fields
are conceptually on equal footing with electromagnetic fields, and they do occur abundantly
in string theory. They appear as field strengths of three index anti-symmetric tensor fields $A_{\mu\nu\rho}$, obvious
generalizations of  the electromagnetic vector potential $A_{\mu}$, and the 
possible existence of fields of this kind is limited only by the number of space-time dimensions.

Such four-index fields {\it can} get constant values without breaking Lorentz invariance, namely
$F_{\mu\nu\rho\sigma} = c \epsilon_{\mu\nu\rho\sigma}$, where 
$\epsilon_{\mu\nu\rho\sigma}$ is the Lorentz-invariant completely anti-symmetric four-index tensor;  it is unique up to normalization, which is fixed in the standard way as $\epsilon_{\mu\nu\rho\sigma} \epsilon^{\mu\nu\rho\sigma} = -24$.
The presence of such a classical field strength in our universe is unobservable unless
we couple the theory to gravity. If we do, it turns out to gives a contribution similar to the
cosmological constant $\Lambda$, in such a way  that the latter is replaced by $\Lambda_{\rm phys} = \Lambda 
 - \frac{1 }{48} F_{\mu\nu\rho\sigma} F^{\mu\nu\rho\sigma} = \Lambda + \frac12 c^2$.

One might think that this solves the problem, since we can now choose $c$ in order to
tune  $\Lambda_{\rm phys}$ to any desirable value (provided  $\Lambda$ is negative, which it always
is in a supersymmetric theory coupled to gravity).
However, it turns out that 
in string theory $c$ is not an arbitrary real number: it is quantized. So the formula for 
the cosmological constant now looks something like this
 \beq
 \Lambda_{\rm phys} = \Lambda + \frac{1}{2} n^2 f^2 \ ,
 \eeq
where $f$ is some number derived from the string theory under consideration. If instead
of $F_{\mu\nu\rho\sigma}$ we were to consider an electromagnetic field, the quantization
of $c$ is akin the charge quantization, and 
$f$ would be
something like  the strength of the electromagnetic coupling $e$: some number of order 1.
This looks like bad news. For generic negative values of $\Lambda$ we would only
be able to tune $\Lambda_{\rm phys}$ to an extremely small value if 
$f$ is ridiculously small.

However, it turns out that string theory typically contains more than one field 
 $F_{\mu\nu\rho\sigma}$
Usually it contains tens or hundreds. Taking $N$ such fields into account, the result now
becomes 
 \beq
 \label{BPlambda}
 \Lambda_{\rm phys} = \Lambda + \frac{1}{2} \sum_{i=1}^N n_i^2 f_i^2 . \ 
 \eeq
 One would expect the values for the real numbers $f_i$ to be different. 
Again an analogy with electromagnetic fields is useful. Although for the Standard Model we need
just one such vector field, string theory may contain more than one. Generically, these will
all have different fine-structure constants, or in other words different values for the
electromagnetic couplings $e_i$.

If indeed the values of  $f_i$ are distinct, and in fact incommensurate, then Eqn.
(\ref{BPlambda}) defines a dense set of values. Bousso and Polchinski called
it a ``discretuum"\ifPreprint, a very descriptive though linguistically awkward name. \else. \fi It is an easy
exercise to show that with $N$ equal to a few hundred, values for $f_i$ of the order of
electromagnetic couplings and small integers $n_i$,  one can indeed obtain the required small value of $\Lambda_{\rm phys}$, given some negative $\Lambda$.
It is clear that in addition to the correct value for $\Lambda_{\rm phys}$ one will find a huge number
of wrong values. This is the price one pays if the cosmological constant is to be neutralized
in this way.

I should emphasize one important point. All the ingredients used in the foregoing
discussion are already present in string theory; nothing was added by hand.
In particular the fields $F_{\mu\nu\rho\sigma}$ are
present, and the quantization of the field strengths follows using standard arguments. 
The fact that there are so many of these fields is closely related to what used
to be viewed as an embarrassment: the large number of moduli. Furthermore
the mechanism could not even have worked by having a single, extremely small $f$. This would
allow the same fine-tuning of $\Lambda$, but one would end up in an empty universe \cite{Brown:1987dd}. 
\ifPreprint
%It is an unfortunate
%accident of history that this mechanism was discovered only after the first observational
%evidence for a non-zero $\Lambda$. Since all the ingredients were available much earlier,
%string theory -- together with the Barrows-Tipler-Weinberg argument --
%might have been used to predict the observation of a non-zero value.
%But predictions are better made
%before an observation.

%\subsubsection*{Approximations}

Although the Bousso-Polchinski mechanism can be derived in string theory,
this does not mean that no assumptions or approximations were made.  In reality, 
everything is quite a bit more complicated than explained above. 
In addition to the 
field strengths $F_{\mu\nu\rho\sigma}$ there are
analogous fields that live in the six (or seven) compactified dimensions. They play a r\^ole in the quantization
of the field strengths, in a way analogous to the old argument due to Dirac that relates the
existence of magnetic monopoles to charge quantization. Those fields are analogous to
electromagnetic flux lines that are wrapped
around cycles in the internal space, like rubber bands around a doughnut.
The arguments of \cite{Bousso:2000xa}  are done in an idealized situation, and
ignore the back-reaction of the non-zero values of these ``flux fields"
 on the compactification manifold. They treated the manifold as rigid, so that around any of its
cycles an unlimited number of ``fluxes" could be wound, without affecting the shape ({\it i.e}
the values of the moduli).  But in the supersymmetric limit, the manifold is
certainly not rigid. Quite the contrary, its shape can be changed without cost in energy by
changing the moduli. The  mechanism can therefore  not be made to work without solving
at the same time the problem of stabilizing those moduli.
Furthermore, one cannot simply add  arbitrary fluxes to a given string theory compactification.
If we start with an exact solution to the equations of motion, we cannot
simply add fields  and 
and expect that we still have a solution after doing that.
One has to reconsider the full set of equations of motion for every set of choices $n_i$. It turns out that those
flux lines contribute to the stabilization of the internal space, in other words to the moduli stabilization problem
explained earlier.
%In reality the winding of fluxes is more like wrapping elastic bands
%around a balloon: the shape changes as one adds more elastic bands, until a maximum
%is reached and the balloon pops. In more precise terms, the addition of fluxes generates
%a potential for the moduli fields. 
In practice, Bousso-Polchinski fluxes only stabilize part of the moduli.  In addition we have
to get rid of the symmetry that made the entire discussion technically possible, but that is
not observed in nature: supersymmetry. Other mechanisms
are needed (and available \cite{KKLT}) to achieve these goals.

The description of this full stabilization mechanism combined with
supersymmetry breaking  takes us to (or even beyond) the edge of the current
knowledge of string theory, and there is a lot of debate about the validity of the  approximation used.
The skeptics often express their doubts with the phrase ``no example of this is known
in a controlled approximation". This statement would probably hold up in court, at present, but I think
it is used mainly as a last straw by people who are still hoping for uniqueness. 

\subsubsection*{The cosmological constant  problem finally solved?}

The cosmological constant problem has been viewed during several decades 
as one of the most difficult
and profound problems in fundamental physics. Can we now declare it ``solved"?
Well, not yet. As I have emphasized earlier, string theory
is not yet properly a ``theory". And even if that problem is solved, we still do not know if it is a theory
of our universe.  It is not enough to know that a Bousso-Polchinski discretuum of sufficient
density can exist, but it must exist in combination with a realization of the Standard Model.
The difficulty with string theory is that these two issues cannot be decoupled. 
However, within the context of string theory the cosmological constant problem has lost its status
as the most profound problem we have to worry about. 
It is truly remarkable that string
theory guided us to this potential solution of the problem, despite the fact that this is
not the kind of solution most people hoped for.

\fi

%\ifPreprint
%\else

\subsection{The String Theory Landscape}

The combination of the results of the two string vacuum revolutions was called
the ``String Theory Landscape" by Susskind \cite{ALS} \cite{Susskind:2006eg}.  
It yields some distribution of points in the huge space formed by the
Gauge Theory Plane combined with a cosmological constant axis. 
One of those points should correspond
to our universe. To find out if that is true, we need a map of the landscape.

Naively, one might think  that these two revolutions could be combined in the following way.
The 1986 revolution produced a huge number of topologically distinct
supersymmetric moduli spaces, each with tens or hundreds of moduli. Each defines some piece
of the Gauge Theory Plane: a choice of particles and gauge symmetries, and a set
of continuous parameters. Let us assume that we can work out the vacuum energy as
a function of the moduli, after having solved the moduli stabilization problem  and the
supersymmetry breaking problem.
This then defines a kind of potential on the Gauge Theory Plane.
Let us furthermore assume that this potential has some discrete local minima.
These local minima fix the values of the moduli, and thereby the values
of all couplings and masses. 
Now we add the Bousso-Polchinski fields. The local minimum also fixes the numbers
$\Lambda$ and $f_i$ in formula (\ref{BPlambda}).
We can add a cosmological constant
axis orthogonally to the Gauge Theory Plane, and all the Bousso Polchinski
mechanism would do is to generate a infinite number of points along that axis, all
projecting down to the same point in the Gauge Theory Plane. If this were correct, all we have
to do is match the couplings and masses of one local minimum to the experimental data,
and then use the Bousso-Polchinski integers $n_i$ to make the physical cosmological constant
as small as $10^{-123}$ times its natural, Planckian size. Ideally we might also hope to determine
the integers $n_i$ that achieve this, but in order to do this we would have to compute $\Lambda$ and
all the $f_i$ with 123 digit precision, which is clearly impossible in practice.

But this is far too naive, as it completely ignores the back-reaction of the fluxes, as explained above.
Indeed, the potential postulated above may not even {\it have} any local minima in the absence
of fluxes. Furthermore,
the local minima themselves will start moving in the Gauge Theory Plane when we change the integers $n_i$,
and the range of allowed values for each $n_i$ becomes finite. So adding fluxes is not like wrapping rubber bands around
a rigid doughnut, but more like wrapping them around a balloon with many handles, which change shape
when we increase the wrapping number, until they finally snap. 

Consequently the total number of choices for the set of integers $n_i$ is finite. 
Their range and their total number depends on the string compactification under consideration.
For one particular one, the total number of possibilities was estimated to be around $10^{500}$ \cite{Ashok:2003gk}, an often
quoted  number that started leading a life of its own and is often quoted as an estimate for the total\ number of string vacua\rlap.\footnote{Most likely the actual number is not even finite, but
there are arguments \cite{Acharya:2006zw} suggesting that it is finite after imposing bounds on the parameter space.
In particular, the total number that would agree with all current experimental data would then be finite. In the
following, I will continue to use the number $10^{500}$ as a guess for the number of vacua, but this should
not be taken too literally.}
To get the total number of string vacua one would have
to sum such numbers over all possible compactifications, {\it i.e.} the results of the first string vacuum revolution.
It is impossible to estimate at this moment how many of those there are. It is know that the number of topologically
distinct Calabi-Yau spaces is at least 30.000, but this is a very conservative lower limit, and furthermore
a straightforward compactification on a Calabi-Yau manifold is only a very small fraction of the possibilities
discovered since 1984. 

\ifPreprint
So there will not be a clean separation between the results of the two string vacuum revolutions. The 
picture where the first gives a scattering of points in the Gauge Theory Plane and the second
a distribution of points on lines orthogonal to that plane is too naive. But it it might happen that 
all the points originating from the same moduli space, but with different values of $n_i$
cluster around a certain central value when we projected them on the Gauge Theory Plane.
In \cite{ArkaniHamed:2005yv} it was argued that this would happen for statistical reasons: there are many
more ways to get a value close to the central value than to get a point far away, for which
all contributions would have to add up coherently. Under the assumptions of this paper, only
dimensionful quantities (namely the cosmological constant  and the scale of the weak
interactions) would have a huge spread, whereas all others would form dense clusters of points, 
something like the interference points in the spot of a laser pointer.

If this is indeed true in the string theory landscape, we might still be able
to discover features of the underlying topological moduli space in the parameter
values, although only with limited precision. We might be able to find out
which of the   laser spots we live in.
In order to get unlimited precision we
would have to determine precisely which of the separate, distinct 
points within the laser spot corresponds to our world, and this 
is probably an impossible task.\fi

Is the Standard Model contained in this huge and extremely complicated ensemble?
This will be discussed in a little more detail in the next chapter.
At present we know enough to say that the very rough topological features of the Standard
Model ({\it i.e.} the gauge groups and the quarks and lepton representations, including the
number of families) can indeed be reproduced. Despite the size of the landscape, this is non-trivial:
infinitely many gauge theories are not realized.
The exploration of more detailed features, masses and couplings,
is still in its infancy. 
Furthermore a very small piece of the Landscape is accessible with currently
available methods: we are only scratching the surface.
%Even with such a huge Landscape, there is no
%guarantee that Standard Model is contained in it, and we might easily have failed already. It is not
%hard to think of possible results of the upcoming LHC experiment that would completely destroy 
%everything I said in this article, and in particular the idea that the Standard Model extrapolates to
%the Planck scale. 

Neither the gauge group, nor the structure
of a particle family, nor the number of families follow uniquely from string theory. The same will be true
for all the remaining details, in particular the couplings. There will not be a simple formula for $\alpha$ expressing
it in terms of $e$ or $\pi$, as Feynman may have hoped. Instead we should expect a dense set of
possible values, correlated in complicated ways with dense set of values of the other parameters.
In some sense the second string vacuum revolution smears out the results of the first, making it harder
to discover the Planckian topological origin of the Standard Model. 
But in the friendly landscape of \cite{ArkaniHamed:2005yv} enough of the underlying
topological structure may be preserved to allow us to solve that
puzzle nevertheless.

\fi

\ifPreprint

\section{Our place in the Landscape}

\subsection{Possible Landscapes}

Without committing ourselves to one particular fundamental theory, let us compare
some scenarios for the way the Standard Model might emerge from such a theory.
No matter what  the fundamental theory is, it will have a ``landscape": the points in the
Gauge Theory Plane it  allows. One may  distinguish several possibilities.

Scenario A: The Standard Model comes out uniquely. We can compute the gauge
group, the type and number of families, all masses and couplings, and they all agree
with experiment. There is a unique prediction for the value of the cosmological constant,
or some effect that mimics it. Let us furthermore assume that there are good indications that this value  
is as small as it should be; a  precise calculation might be too much to ask for.

Scenario B: The fundamental theory has a large, but still manageable number of vacua,
let us say about $10^{30}$ in some
finite region $R$ of the Gauge Theory Plane near the Standard Model.
Within the Gauge Theory Plane we can draw a region $E$ that constitutes the current
experimental  information, with  a size determined by the  experimental error bars. 
It is represented by the small circle in the figure in section \ref{TGC}.
If we assume that the $10^{30}$ points are smeared out more or less equally  in a large
region around the Standard Model (thus defining a measure), then $10^{30}$ is too  
small a number to expect one of them to land in the experimental
region $E$ just by chance. One estimate, due to Michael Douglas,  implies
that about $10^{80}$ vacua  would be needed (the number 80 is the number of digits
of information in the masses and couplings of the Standard Model particles, expressed
in natural units, plus a guess  about the likelihood for the discrete data, {\it i.e.} the gauge
group and  representations). Therefore, if we {\it  do} find a solution within $E$
(and with only $10^{30}$ to check we might just be able to find it), 
there would be
50 digits  worth of predictions  left over. This is less that the 80 digits we would have
in scenario A, but still more than enough to convince ourselves that we  had found the
correct theory. This  would at the same time prove, beyond any doubt, the mathematical
existence of the other $10^{30}$ vacua.

Furthermore, such a result would prove the version of the anthropic principle
advocated here. The relative size  of the anthropic
window  (assuming the same measure)  might be anywhere from $10^{-5}$ to $10^{-25}$ (depending
mainly on the  physical mechanism that determines the weak scale). If it is, for example,
 $10^{-10}$, we
may expect to find $10^{20}$ out of the $10^{30}$
vacua to lie within the anthropic region, but still the vast majority lies outside. 
Nobody would
dispute the statement that  some of the parameters of the Standard Model are anthropically
determined. Furthermore in scenario B (as well as in A) there would an unlimited amount of 
falsifiable information available, since after determining the vacuum corresponding to our
environment, everything would be fixed: we would be able to predict the properties of
all particles and all interaction all the way up to the Planck scale, with unlimited
precision.

After  achieving this tremendous success, some people would start discussing the question
why we find ourselves  in one of the $10^{20}$ vacua that allow life, rather than any other. Perhaps we would
be able to argue that among the vacua that allow life, ours is among the
most probable ones, but this would not
be a major issue anymore. It would be a bit like worrying if our planet is the most common one in
the set of all planets that allow life. That is not a major issue anymore because no-one doubts the
existence of a landscape of possible planets: we can see more than a single sample with our own eyes.

In my opinion, scenario B is  much more desirable than scenario A. Both would give us
a huge amount of confidence in the correctness of the underlying theory, but scenario B
has the advantage that it also explains why we ended up in such an apparently fine tuned
universe, something that scenario A converts into an eternal mystery.
 
However, there is no theoretical upper limit on the number of
vacua the fundamental theory may have. In this situation, hoping for  not more than $10^{30}$ vacua
is wishful thinking, just as hoping for  just one. Furthermore, scenario B is unlikely to  work if we include
the cosmological constant into the discussion. After having determined which of the
$10^{30}$ vacua we live in, this becomes a computable number, and it has
to have a value of about  $10^{-120}$, although it is naturally of order 1. This computation
requires an unachievable amount of  precision, but no-one would expect it to work
just by chance, and even if it did we would  once again be left with an eternal mystery.

The only way out would be to hope for a fundamental theory that by some mechanism 
dramatically reduces
the natural range of the cosmological constant. That might be acceptable: as I argued in
section (\ref{circ}),
it does not produce
the same kind of troublesome conflict between unrelated `gedanken' computations as a derivation
of the Standard Model would.
However, no such theory is in sight.

In string theory, it  seems that we end up with scenario C: The actual number of vacua is
too large to allow enumeration, let us say $10^{500}$ for definiteness. If they are spread 
around evenly over the relevant region, this would nicely explain all anthropic coincidences,
including the cosmological constant, but  there is no way to determine  exactly which one
is realized in our universe.  Indeed, in case of a structureless distribution, if we include the 80 digits of the Standard Model data
and the 120 of the cosmological constant, then there are still $10^{300}$ vacua left that
fit all the current data. Unfortunately those 120 digits are not even of any practical value. In order
to use them we would have to compute each of the two terms  on the right hand side of 
Eqn. (\ref{BPlambda}) with a precision of 120 digits.

In this scenario there are  not only more vacua than 
current experimental
data, but in addition they may be spread over the parameter space in such a way  that,
statistically speaking, 
there is no way the Standard Model could be absent. In fact, it could be present so
abundantly that no predictions can be made.
 In this case,
it seems that we are back where we started:  we end up with the Standard Model with all
its parameters taking practically any possible value.
But even if this were the final result,
it would still be tremendous progress.
First of all we would have succeeded
in coupling the Standard Model to a consistent theory of quantum gravity. 
Secondly, all parameters of the Standard Model are now
dynamical, allowing them to change in the very early  universe, or during the birth
of new ones, providing a mechanism to move around in the landscape. 
And last, but certainly
not  least, we would have gained a profound understanding of why we live in such a 
special  universe, apparently finely tuned to allow our existence.  In this case we would
not gain any insight in the Standard Model itself, nor would we be able to use it to
learn something about the theory of quantum gravity. 
But this was neither guaranteed nor required.

But the amount of
experimental data is monotonically increasing, so for any finite number of (relevant) vacua,
scenario C will eventually turn into scenario B.  Even if the landscape densely covers the
parameter space of the Standard Model, this may not be the case for extensions of
the Standard Model we may discover experimentally. Undoubtedly the  $10^{300}$  vacua 
would make entire different predictions about  such extensions. One popular extension
of the Standard Model, low energy supersymmetry, adds 105 parameters. Measuring
each of them with three digit precision would bring us to the highly desirable scenario B.
Obviously this is far beyond our present horizon.

The extreme limit of this might be called scenario D. This is a featureless landscape obtained by covering the
entire gauge theory plane densely (or even continuously) with points according to
some smooth distribution. The string theory landscape is probably not like this at all.
The little bit we know about the string theory landscape does not suggest that, and what
we know about the Standard Model does not suggest that it lives in such a landscape.

\subsection{String Spectra}

Before discussing how the Standard Model fits (or might not fit) in the string theory landscape, let us consider
what we know about the kind of spectra string theory produces. 

A string spectrum
consists  of a few massless particles plus an infinite ``tower" of particles with  Planckian masses. The latter cannot
be observed, because  they are out of range of any imaginable accelerator.
The Standard Model particles are of course not exactly massless. In Planckian units, there masses are of order $10^{-19}$, so
if they come out massless in a string spectrum, that is the correct answer to an excellent approximation, but not what
we ultimately want. 

Such spectra can be computed rather easily in a huge number of cases. It is usually  an algebraic exercise that can be
done by hand in a few cases, but is normally done by means of a computer. The result is described in terms
of some Lie-group (called the ``gauge group") and a set of representations of that group. The Lie group gives us a set of vector bosons,
like the photon or the gluon (the vector bosons that bind quarks together in a proton or neutron), and
the representations tell us how the massless particles couple to these vector bosons. They are essentially
the charges of the particles.

In nearly all published results the spectra
are obtained in supersymmetric string theory.  This means that every particle is part of  a boson/fermion pair
with identical masses.  In particular, all the Standard Model particles have massless supersymmetric partners, because
it turns out that the  already know particles
cannot be supersymmetrically paired.  The reason for computing  supersymmetric spectra is
that non-supersymmetric spectra are much more difficult to control in exact string theory. They tend to have instabilities which
are automatically absent or easily avoided in the presence of supersymmetry. In the absence of supersymmetry, these instabilities (``tachyons" and
``tadpoles") can all be avoided at the lowest order in perturbation theory in exact string theory, 
but at the price if a huge loss in statistics (the conditions for the absence of instabilities are very restrictive)
and even then one still has to worry about stability at higher orders in perturbation  theory. 
For this reason, the most common
approach is to compute supersymmetric spectra in exact string theory, and make further assumptions about low energy
physics from there on. Ultimately, all these extra assumptions will have to be derived from string theory.

These exact supersymmetric string spectra are not really points in the string landscape yet.  They are points in flat planes
in the landscape, the moduli spaces. After supersymmetry is broken and the moduli are stabilized, a true landscape of mountains
and valleys  is expected to rise up on these planes, and  we hope to live in one of those valleys. Ideally, we
would prefer to compute the bottom of one of these valleys exactly in string theory, but just as anywhere
else in physics, the only available way is to get as close as possible with exact methods, and then approach
the point of interest using a variety of approximations.

In this process  of landscape
formation, the  masses of all particles are expected to shift.
In order for the particles in a supersymmetric spectrum to get a realistic mass two things have to happen:
supersymmetry has to be broken, and the weak interaction symmetry (the $SU(2)$ of the Standard Model) has to be broken. 
In both cases this  is believed to happen because the true vacuum of the theory, the valley of the landscape
where we end up, violates these symmetries by a tiny  amount:
they are symmetries of the Hamiltonian but not of the ground state.
In the case of the weak interactions this goes by the name ``Higgs mechanism", and leads to the prediction that a neutral
scalar particle should exist, the famous Higgs boson. 

When supersymmetry is broken, the fermionic and bosonic partners of the Standard Model particles acquire a mass. There is
indeed an excellent reason why the as yet unobserved superpartners acquire a mass at this stage, and not  the Standard Model
particles themselves. The reason is that the Standard Model particles can only get  a mass once the weak interaction symmetries 
are broken as well. Because we have not observed any of the  supersymmetric particles yet, the scale 
associated with supersymmetry breaking (the mass splitting among the supersymmetric particles) is  apparently
larger than the scale of weak interaction symmetry breaking (related  the masses of the  $W$ and $Z$ bosons and the heaviest quarks).
Usually one takes something of order $1$ TeV for the former, whereas the latter is of order 100  GeV. There are no convincing fundamental
physics  arguments for the values of these scales, and in a landscape a large range of values is likely to exist, with an unknown
distribution.

Once one of these supersymmetric string spectra comes out of the computer, the first criterion we
can impose is that the set of vector bosons must include the ones of the Standard Model, in other words
that the gauge group contains $SU(3)\times SU(2) \times U(1)$. It may contain more than that; there may
exist vector bosons we have not seen yet, either because in our valley they have acquired a mass out
of reach of current accelerators, or because they do not couple to any known matter. The next thing to check
is that there are other particles (supersymmetric multiplets of fermions and bosons) that couple to the
$SU(3) \times SU(2) \times U(1)$ vector bosons precisely  as three families of quarks and leptons, {\it i.e.}
three times Eqn. (\ref{SMrep}). Also in this  case there may exist additional particles, called ``exotics", which
may be a blessing or a curse, depending on their coupling to the Standard Model. Just as superfluous vector
bosons, exotics may acquire a mass when we move out of the supersymmetry plane into our own valley. 

Supersymmetric spectra containing (at least) the Standard Model with three families of quarks and leptons 
exist in abundance in many regions of the string landscape. There are examples with just the
Standard Model gauge group and nothing else, and there are examples without exotics, but as far as
I know there is not yet an exact string theory example which has the exact supersymmetric Standard Model spectrum without anything else.  This seems just a matter of statistics: if we work out enough examples, eventually we
will find this. But it may well be that this is not what we should be looking for anyway: there appears to exist
dark matter that does not belong in the Standard Model. Many people hope that this dark matter
is made up of some supersymmetric partners of the Standard Model particles, but it could well be
something entirely different. 

The next step is to compute the coupling constants between the various particles. In particular
the Yukawa couplings between fermions and the Higgs boson are of great interest. Knowing them, we
can work out the mass ratios of quarks and leptons, and compare with observations. However, this is
considerably harder than working out groups and representations. It is not just harder because
the computation of coupling constants is technically more complicated; the more important problem
is that these quantities (unlike groups and representations) depend on the moduli, in other words
they vary over the supersymmetric plane. This implies that they will also take distinct values in the
various valleys of the landscape the rises up on top of that plane. This is good, because it is
precisely this variation that will be needed to end up in one of the small anthropic regions. 
Coupling constants have been computed
in a few cases, but little is known about the distribution of their values over the landscape.

\subsection{String Phenomenology}

The field that deals with the way the Standard Model is realized in string theory is called
``String Phenomenology". It
is an essential  activity: all matter, including the Standard Model has to fit into string theory, or else
this theory has nothing to do with our universe.  
One may think that the explosion of string
vacua should have had a major impact on this field. I have been active in this are around 1987
(which led me to the conclusions presented here)  and again in the last few years, and to me the
similarities are more striking than the differences. There has certainly been progress: we can obtain
string solutions that are more similar to the Standard Model  than twenty years ago, and we have
more
methods to construct them. There has been  major progress in moduli stabilization and supersymmetry
breaking. There is more interest in ``landscape statistics".
But very little seems to have changed in the way many people view the problem we are facing. 
Although many
of my string phenomology colleagues
claim that it was clear to them a long time ago that there are many solutions, I cannot
help noticing  that they still talk about their most  recent ``model"\footnote{The way particle physicists use the world ``model" may be confusing to outsiders.  In this
context it simply means some string theory spectrum, which is not really modelling anything else.
The use of this word dates back to the days the Standard Model
was constructed as an approximate, idealized description (model) of the forces of nature. Attempts to build
something with similar properties were called ``model building". In string theory one does not really
build anything; one simply finds what is there.}  
 as if it would actually have a chance to
be {\it the} Standard Model.  
And even nowadays one still hears the occasional expression of hope for the unknown
and elusive dynamical principle  that will select the vacuum. The most common way of dealing with the
large vacuum degeneracy is to say ``I do not care about the other $10^{500}$ vacua, I only care about the
one that  describes our universe". That may sound reasonable, and fact it may sound like
the very definition of phenomenology, but it is actually an escape from reality.

First of  all, if indeed there are $10^{500}$ vacua, it is highly unlikely that anyone will find 
``{\it the} Standard Model" in string theory. One should expect to find a huge number that satisfy all
current  experimental constraints.  In addition, although we now have many techniques at our
disposal to construct string theories in four dimensions, it is quite clear that we are just scratching the surface.
Statistically speaking, our chances of finding even one of the expected huge number of Standard Model
realizations is essentially zero. Furthermore, even if we do  find one, we can only make predictions  about novel
phenomena if we know {\it all}  the other solutions and their predictions for the same phenomena. This is
a crucial change in comparison to the state of the art about ten years ago: with $10^{20}$ solutions
(the largest number anyone may have expected), if one
is found that agrees with all current data, 
the probability that there is
a second one  is extremely small.  With $10^{500}$, the same probability
is astronomical. 
So we should forget about the idea of finding {\it the}  Standard Model and then making predictions based
on it.

The second reason why the aforementioned ``phenomenological" point of view is  unreasonable is that the phenomenological
wish list has always included the so called ``why" problems: why $SU(3) \times SU(2) \times U(1)$, why three
families, etc.  If the Standard Model is part of a huge ensemble, the only way to answer such questions is
to understand  the distribution of that ensemble. We {\it have} to care about more than just our own universe.
An inevitable consequence of a huge ensemble is that anthropic arguments  will play a part in answering some
of the ``why" questions. But in the field of string phenomenology this is still nearly equally unmentionable as twenty years ago. 
%There are yearly
%conferences on string phenomenology, and recently  I was in the advisory committee of one of them. I
%suggested to devote a session to anthropic arguments. As I expected, I did not even get an answer. During the
%last such conference (in Philadelphia, May 2008) the word ``anthropic" was mentioned precisely once, by the
%Russian cosmologist Mukhanov, who never fails to express his disgust for the idea. 
There is even more 
attention to anthropic arguments in traditional phenomenology ({\it i.e.} not based on string theory) than there is in the
field that cries out for it, string phenomenology.
On the other hand, I often
hear talks containing statements like ``property X is observed in our universe, but occurs rarely in string theory. Here
we (proudly) present an example with property X". Examples of property X are the absence of exotics (see below) or the presence of
exactly three families. In this situation I ask myself the question: ``if property X is so rare in string theory, why do we observe it
in our universe?", but few other people seem to worry about that. 

Of course most of the $10^{500}$ vacua are utterly irrelevant to us.
It may be possible to identify regions in the landscape where Standard-Model-like
spectra occur more abundantly, and other regions that are essentially barren. It may well be that most
of the huge number of vacua are in such barren regions. However, we cannot expect that to
reduce the number that really matters to something extremely small. If that were true, the cosmological
constant problem is still not solved, nor are the anthropic fine-tunings of the Standard Model explained.
 If the Standard Model were an essentially isolated
point, the anthropic reason of existence of the landscape would lose its meaning. We must expect to
find ourselves in a dense region of the landscape.
There is no reason why the densest region of the landscape should coincide with the anthropic region
-- in fact, if it would that  would itself be a mystery -- but {\it within} the anthropic region most universes with observers
obviously occur in the densest regions.

So what {\it can} realistically be achieved in string phenomenology? A fairly modest goal is to
use string theory as a guide to new ideas in phenomenology
beyond the Standard Model, which keep their validity outside that context;
there are already plenty of examples of that, but this is not my main interest here. 
A goal that seems within
reach is to arrive at a statistical proof that what we are looking for, a string vacuum that agrees with
all current data,
must exist, even if we cannot find
a single explicit example.  It may be possible to identify features of the Standard Model with those of
certain classes of string vacua, and extract generic predictions for that class. But the most important point
I am trying to make here is that we should not focus too strongly on the observed Standard Model, but
explore the region around it.
Even if we have convinced ourselves that the Standard Model
is present in the landscape, there are serious challenges left, and I will discuss some of them
in the next section.

\subsection{Challenges and worries}

The Standard Model may look complicated, but 
 it does not look like someone was throwing darts at random into 
in the Gauge Theory Plane until an observer window  was hit. 
In section (\ref{Oon})  those
aspects of Standard Model structure having to do with our own existence were sketched.
Here I will focus on the rest. The {\it anthropic} features of the Standard model need not be
typical in the full ensemble of possibilities; some parameters may be outliers, taking some
rather extreme values in our universe. However, the {\it non-anthropic} structure has to be
understood in terms of  a landscape of a fundamental theory.

The denser the string theory landscape, the less information we will get from the necessary
condition that it must contain the Standard Model. In this limit the more pressing question becomes:
it is really plausible that the Standard Model structure we see comes from the string theory landscape?
Ultimately, this would involve computing relative probabilities of various features, which may be
beyond our present capabilities.  
However, if some clearly non-anthropic feature is dominant in the
accessible part of the landscape, and is not seen in the Standard Model, this is a reason for concern.
An accumulation of such concerns will end up removing our confidence in the hypothesis that the
string theory landscape is the right one. Here ``dominant" means that the number of vacua with a given
property vastly outnumbers the number of vacua without that property. In ensembles as large as the
string theory landscape, such relative multiplicities can easily be huge. Of course the multiplicity
of vacua is not the final answer. One has to combine this with the probability that we end up in a particular
vacuum when a new universe is created. Such probabilities are notoriously hard to define, not to mention compute.
One could still try to argue that
such an initial probability might compensate for the  dominance of a wrong feature, but it seems implausible
that cosmic probabilities have large  variations over the gauge theory parameter space, and even less plausible
that such a variation would precisely compensate an undesirable surplus or absence of vacua. 

We are assuming
here that the Standard Model is directly embedded in this fundamental theory, without
layers of unknown physics in between. Otherwise it is simply to early to ask the question.
A bit of intermediate structure may be acceptable, provided it is a sufficiently transparent layer.
However, it is certainly necessary that the quarks and leptons  are fundamental degrees of
freedom, and not some sort of composites. All of the following features are serious
challenges for any landscape, and most of them will come out wrong in a featureless landscape.

\subsubsection*{The $\theta$-parameter}

Perhaps the most obvious challenge is the so-called
 $\theta$-parameter  of QCD, one if the 28 parameters of the Standard Model.
 This is an angular parameter that is experimentally
consistent with zero, for no known reason.
If non-zero, it would generate an electric dipole moment for the neutron, which
is not observed\rlap.\footnote{The parameter $\theta$ also violates CP-invariance  (which is 
essentially the same as invariance under time-reversal), but since this symmetry
 is violated in the
weak interactions, it is not a symmetry of nature, and hence not a valid argument for
the vanishing of $\theta$.}
The current limit on $\theta$  is about $10^{-9}$. Unlike the cosmological constant, there is
no anthropic constraint  on this number. Even if a  non-zero value were  measured in one
of the ongoing efforts to measure the neutron electric dipole moment, it would still be
unacceptable that we live in a universe with $\theta \approx 10^{-9}$ purely  by non-anthropic
coincidence.  It is  noteworthy that there does exist a mechanism in which $\theta=0$ automatically.
This mechanism requires the existence of a new particle, the ``axion" which may have
anthropic implications \cite{Wilczek:2004cr}. So then the challenge is to demonstrate that this mechanism is
realized naturally  in a candidate landscape, or that we are forced into it by a chain of indirect anthropic
requirements.

 \subsubsection*{Gauge groups and representations}

The Standard Model gauge group,
$SU(3)\times SU(2)\times U(1)$ with its family structure looks like one of the simplest
gauge theories with a sufficiently interesting spectrum 
(although one can argue about the need for the $SU(2)$ factor \cite{Harnik:2006vj}). 
Since in the string theory landscape large gauge groups are statistically suppressed, 
and smaller ones may not allow life, the Standard Model looks like a reasonable choice: perhaps
among all the anthropic ones it is the most abundant one. 

One may distinguish string theories
 in which all strings are closed, and string theories in which both
 open strings (strings with two end-points)  and closed strings occur (this distinction is valid if we
 are in the neighbourhood of one of the perturbative regions).
  In both cases, gravity is described
 by closed strings. In the former case, all other interactions are described by closed strings as well, but
 in the latter case all non-gravitational interactions are described by open strings. There seems to be nothing
 fundamentally wrong with either approach. Indeed, the existence of both is required by a complicated network of 
 ``dualities", amazing relations between apparently totally
 different string theories. It is easy to get the basic standard
 model structure 
 to come out right in either case.
  Historically this was first achieved in 1985 \cite{Candelas:1985en}
 in closed string theories, in particular
 the heterotic string. One can say without exaggeration that the Standard Model comes out rather
 naturally in these theories. By simply imposing the condition that space-time should have four
 flat dimensions one gets a theory with a certain number of families, although not directly
 with the group $SU(3)\times SU(2)\times U(1)$ but rather with one of the unified gauge groups (see below)
 that were first proposed in the mid-seventies. Around the turn of the century \cite{Ibanez:2001nd} it was
 shown that the same feat could be achieved with open strings as well.  
The fact that this took so much longer is due only to the fact that open strings are technically a bit
more difficult to deal with. Both Standard Model realizations, closed as well as open, are among
the simplest constructions one can make in either case. It is straightforward
to write down examples of gauge theories and matter that would be 
essentially impossible to obtain from string theory, and that at the very least would  require some
extremely contrived and unnatural tricks. It is perhaps even slightly disturbing that one can
get the Standard Model so easily in {\it two} totally different ways. To some people, this may
smell like a consequence of the aforementioned duality. If one could make that precise, it
would be additional circumstantial evidence that the structure of the Standard Model
suggests a string theory origin.

The structure of a single family is
also one of the simplest non-trivial choices. It is strongly restricted by consistency conditions 
called ``anomaly cancellation", leading to cubic sum rules for charges. 

\subsubsection*{The number of families}

But why do we observe {\it three} families of quarks and leptons? For our existence, one family might
seem all that is needed: we are made of up-quarks, down-quarks and electrons.
The top quark may play an essential r\^ole in triggering weak
symmetry breaking, which might justify the existence of two families. Why there are three families is
somewhat of a mystery. Already for a long time there exists a qualitative argument: with three families one
can construct couplings that violate CP symmetry in the quark sector \cite{Kobayashi:1973fv}. CP violation is 
one of the famous three necessary conditions formulated by the Russian physicist Andrei Sakharov
for generating a surplus of baryons over anti-baryons in the universe. Such a surplus is, at least locally,
 clearly necessary 
for our
existence. But using CP-violation in the quark sector does not appear to work quantitatively, 
and even if it did, it would work
just as well with four or more families. One can also generate a surplus of baryons using 
CP violation in the lepton sector, by first generating a surplus of leptons (``leptogenesis") \cite{Fukugita:1986hr} \cite{Buchmuller:2003gz}. However, this uses the so-called Majorana masses of neutrinos,
a feature that has no analog in the quark sector. This in its turn introduces new possibilities for
CP violation which would allow  leptogenesis with just two families.

The number of families affects the physics of our universe in other ways. One of them is neutrino
processes, for  example Big Bang nucleosynthesis (see below). All families enter in the running
of coupling constants through the coefficients $b_0$ in Eqn. (\ref{RunC}).  
If we  simply add a  family or remove one while keeping everything else fixed at the Planck scale (while making some choices for the masses
of the families to be added or removed), the strong and electromagnetic coupling constant will move far enough from their
observed values to endanger our anthropic environment. To some extent this can still be compensated by changing the initial values
at the Planck scale. But if there are more than 16 quark flavors ({\it i.e.} eight families) in a non-supersymmetric version of the
Standard Model even that  does not work anymore: in that case the QCD coupling constant decreases towards lower energies, and
hence the strong force becomes a weak force.  If one considers  other couplings and higher order quantum corrections even
stronger limits may emerge.
In \cite{Pirogov:1998ws} it is shown that
with four families the Standard Model couplings cannot be extrapolated consistently to
the Planck scale anymore. If we turn that around, it would imply
that a four family analog of the Standard Model (keeping everything else as much as
possible fixed) cannot be obtained from a fundamental theory at the Planck scale. To  turn any of these statements into true anthropic
bounds requires a careful and systematic study of possible initial conditions at the Planck scale, to see exactly under which modified conditions
we can still reach our current anthropic island in parameter space.

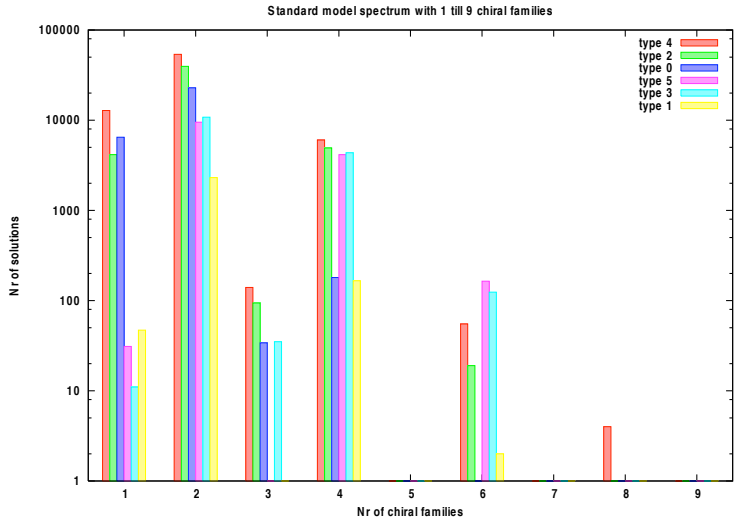
\begin{center}
\includegraphics[width=10cm]{logfam123456789.pdf}
\end{center}

In string theory, the fact that there is a repetition of families is easily reproduced. In the simplest heterotic
models, one obtains a number of families equal to the Euler number of some six-dimensional
compactification manifold. This allows many possibilities, up to hundreds of families. 
The distribution of the number of families
has been studied in certain cases for open strings.
({\it e.g.} \cite{Dijkstra:2004cc}, \cite{Gmeiner:2005vz}). Some care is needed in
drawing conclusions from these results: they are obtained for samples of supersymmetric models, and not for  true
landscape distributions. Furthermore they only cover a very limited region of the landscape. 
It turns out that
vacua with larger number of families are exponentially
suppressed. This is clearly visible in the  figure above, which is from the first of these papers (the colors represent different ways
of realizing the Standard Model; the vertical scale represent the number of models found with a 
given number of families within a certain sample; the absence of five families is simply a result of having
a small enough sample size to keep the number of two-family models reasonably small).
Note that odd numbers are suppressed with respect to even ones, and
as a result four families are one or two orders of magnitude more common than three.

We may conclude that it is  a minor puzzle why  there are  three families in our universe:   
cases with two families are more  common by a factor of a hundred to a thousand, while there is no convincing
anthropic argument against  having just two families. 

\subsubsection*{Quark and lepton masses}
 
 Although they are poorly understood, quark and charged lepton masses certainly do not look
 structureless, and are not fully determined by anthropic considerations either. 
 The distribution
 of their
 values must tell us something about the nature of the underlying theory. 
 Donoghue \cite{Donoghue:2007zz} has suggested that they are
 more or less  randomly distributed on a logarithmic scale, a feature that is
 indeed produced in a class of string models. If the masses were more or less random
 on a linear scale, one would expect most of them to be distributed around  some
 natural scale for quark or lepton masses, which is clearly not the case. On the other hand, the
 masses may be hierarchical, expressible in powers of some small parameter. These hierarchies may
 be anthropically tuned, as they help in pushing down the up and down quark masses and the electron masses
 to small values.
 It is quite common in string models
 that some of the quark or lepton masses vanish to first order in some perturbative expansion parameter, 
 and appear in higher order
 or by non-perturbative effects.  
 Too little is known about masses and their distributions in the string theory
 landscape to see quark and lepton masses as a serious worry at the moment.  
 
\subsubsection*{Neutrinos}
 
Every Standard Model family contains two quarks, with charges differing by
one unit (the charges are $\frac23$ and $-\frac13$), and two leptons, also with charges differing by one unit. If we order
the families by mass, the quark masses within a family differ by less than two orders 
of magnitude. But for the leptons this is radically different. Leptons can have charge $-1$ or $0$.
The charged leptons (electron, muon and
tau) have masses ranging from $.511 {\rm~MeV}$ to about $1888 {\rm~MeV}$. Neutrinos are so light that for a long time
they were thought to be massless. The experimental upper bound on the mass of the
electron neutrino is about $2~{\rm eV}$, at least five orders of magnitude below the lightest charged lepton.
Since about a decade we know that the differences of the squares of the neutrinos masses are
non-zero. This is a consequence of the observation that a 
quantum-mechanical linear combination of distinct mass eigenstates changes its composition periodically
with time (``neutrino oscillations"), a process that vanishes if the masses are the same.
These differences are tiny: the two that have been measured
are of order $10^{-4}$ and $10^{-2} {\rm~ eV}^2$ respectively.
If we make the assumption
that the actual masses are of the same order as the differences, we get masses of order $.01 {\rm eV}$
to $.1 {\rm eV}$, about 6 to 10 orders of magnitude smaller than typical charged lepton masses (depending
on which charged lepton one considers ``typical").

This extreme lightness of neutrinos might appear to be a deep mystery, 
but a potential way out is suggested by the fact that
neutrinos, unlike all other quarks and leptons, have zero charge. This means that they have the
same charge as their anti-particle. This in its turn implies that there might exist a possibility
of quantum-mechanical mixing between neutrinos and anti-neutrinos. For a single family, this
mixing results in a two-by-two mass matrix with two parameters, of the form
\beq
\pmatrix{0 & m \cr m & M \cr }
\eeq
Here $m$  (called the Dirac mass parameter) is generated in the same way as all other quark and lepton masses ({\it i.e.} by means
of the Higgs mechanism). Therefore it may be expected to be of the same order of magnitude
as those masses (which still leaves a huge range from $.511 {\rm ~MeV}$ for the electron to $174$ GeV for
the top quark).  The parameter $M$ (called the Majorana mass parameter) is special for neutrinos and does not occur for
 charged leptons and quarks.
For $M=0$ a neutrino behaves like any other lepton, and has mass $m$. For  $M \gg m$ there are
two neutrino mass eigenstates per family, one with a large mass $M$, and one with a tiny mass $m^2/M$
(this is know as the ``see-saw mechanism").

There is a priori no reason to expect any particular value for $M$. In other words, its natural value,
without any theoretical prejudice, would be a number of order 1 times the Planck mass, about 
$10^{19}$ GeV. Even for the largest acceptable values for $m$, about $200$ GeV,  
this would give neutrino masses that are
too small to yield the observed differences. Although some people are advocating using the unification
scale (see below) of about $10^{16}$ GeV for $M$, this also seems uncomfortably large. This
would just barely give reasonable neutrino masses if we assume $m$ to be of order the top quark mass, whereas
it seems to be more plausible that $m$ is like a typical charged lepton mass, {\it i.e.} at least two
orders of magnitude below the top quark mass. This implies that $M$ must have a value at some 
intermediate scale between the weak scale and the unification scale.

Although the neutrino interacts so weakly that it cannot have any relevance for our everyday life,
it does affect our existence in many ways. It plays an essential r\^ole in Big Bang Nucleosynthesis
(the creation of the light elements). Indeed, this led to a rather tight upper limit on the total
number of neutrino species (which is meanwhile superseded by the result of the LEP experiments, 
$2.994 \pm .012$ neutrino species). Neutrinos are important in the functioning of the sun, in supernova 
explosions (which liberate the heavy elements made in stars), in the decay of the neutron (which would be stable
if the electron neutrino had a typical leptonic mass), and they contribute to the mass density of the universe,
just to name a few examples. The latter fact leads to an anthropic bound on neutrino masses: if the
sum of their masses exceeds about $40$ eV they would ``over-close" the universe. This means that
the mass density of the universe is so large that it collapses. The anthropic window might be even smaller
if neutrinos are involved in generating the matter-antimatter asymmetry in our universe through a process called
``leptogenesis". This works only if  all neutrinos are lighter than $0.1$ eV \cite{Fukugita:1986hr}\cite{Buchmuller:2003gz}. 
The anthropic nature of this bound requires more
discussion:
they are derived under the assumption that only the neutrino mass is varied, while everything
else is kept fixed and one would have to systematically investigate all alternative mechanisms before
claiming that without light neutrinos we cannot exist.

Neutrinos provide thus an interesting battle ground between fundamental and anthropic explanations.
The string theory landscape will certainly provide a wide range of possible values for both $m$ and $M$.
There is a tendency to say that their values will be ``generically non-zero". This would be true in the
Gauge Theory Plane: if we pick a random point we will find non-zero values. But is that also true in
the string theory landscape? One might hope that string theory is not generic in all respects. 
If  string theory only produces
generic points in the Gauge Theory Plane, there is little hope left to extract a prediction from it that 
could not have been made just as well without string theory.

In explicit examples one often finds that $m$ or $M$ vanish. Perhaps
this is because we only have access to rather simple examples, which might be atypical. There
is no anthropic {\it lower} limit on neutrino masses; if there were it would have been possible a long
time ago to predict that they would be massive.  This means that the anthropic requirements are
fulfilled if all neutrino masses vanish. If vanishing neutrino mass turns out to be a dominant feature
in the string theory landscape, this would be sign that something is seriously wrong. Unfortunately
the attitude of most string phenomenologist is to simply reject models without a mechanism to
generate neutrino masses; they do not agree with experiment, so they are of no interest. Likewise,
in explicit models it is not easy to get non-zero values for $M$. If $M$ vanishes, but $m$ does not,  
neutrinos have typical leptonic masses. If $M=0$ turns out to be dominant in the string theory landscape, this
would be a smaller problem, as long as no direct evidence for Majorana masses exists.
One could then argue that the small masses we observe are Dirac masses,
which are just exceptionally small for anthropic reasons.  

In 
\cite{Tegmark:2003ug} an anthropic probability argument was presented in favor of neutrino masses
of the observed order of magnitude.  This argument is based  on the effect of neutrinos on galaxy formation.
However, a prerequisite for applying such an argument  to the string theory landscape is a better understanding
of mass distributions.

Another interesting issue is the mass hierarchy
we observe for the charged quarks and leptons. In the Standard Model this hierarchy
helps in pushing the anthropically important electron, up-quark and down-quarks masses to small
values. For neutrinos, this is not needed. It would therefore be very interesting to know
if there is a hierarchy in the values of the neutrino Dirac masses $m$. And from a theoretical
point of view it is obviously extremely important to find out if mass hierarchies for standard
model particles of different charges are correlated.

There are many other issues that could be discussed here, but I just want to re-emphasize
the main point: we should not just aim at finding ``the Standard Model" in string theory, but
explore all its variations, and ask  if we live in the most plausible one.

 \subsubsection*{Apparent unification}
 
 There are two remarkable facts about the Standard Model whose precise r\^ole in
 the story is not yet clear. 
 
One family of Standard Model particles, Eqn. (\ref{SMrep}), fits precisely in
a single representation of a larger gauge group, namely $SO(10)$, provided that we extend it with just
one particle: a right-handed neutrino. This amounts to adding $({\bf 1}, {\bf 1}, 0)$ to Eqn. (\ref{SMrep}), 
a particle that does not couple to any of three Standard Model interactions, not even the weak one.
Nothing we know excludes the existence of such a particle, and the observation of neutrino
masses even makes its existence plausible.

Another remarkable fact is the apparent convergence of the three Standard Model couplings,
if we extrapolate these couplings to higher energies.  Such a extrapolation can only
be done if we know the exact particle spectrum, all the way up  to
the energy scale of interest. This includes particles with masses much larger
than are currently experimentally accessible. All these particles contribute to the coefficients
$b_0$ in Eqn. (\ref{RunC}). If one assumes that, in addition to the know particles, the only
ones that contribute are their supersymmetric partners (a new boson for every known fermion
and vice-versa) plus the Higgs and its partner, then it is found that the three lines 
go roughly through one point at an energy of about $10^{16}$ GeV, with a precision of about one 
percent\rlap.\footnote{This result depends on the masses of the supersymmetric partners. This is
because the coefficients $b_0$ are actually functions of energy. To first approximation they are step
functions that receive contributions from all particles with a mass smaller than a given energy. The convergence
works if we assume that the supersymmetric partners start contribution to $b_0$ for energies larger
than about a TeV.}

This will be subject to experimental verification at the LHC. If the LHC
experiments  do not find these supersymmetric partners as well  as the Higgs, we must conclude
that the seeming convergence of these three lines was indeed nothing but a coincidence, based on an
invalid extrapolation. On the other hand, if these particles are found
and if there are no surprises, then
the three lines may go through one point at even higher precision, making it
harder to dismiss as a mere coincidence. It is then natural to assume that the coupling constants do not just
go through one point, but that at energies larger than their meeting point there is a single theory with a single
coupling constant. This is called ``Grand Unification". This new theory could have a gauge group $SO(10)$
(although there are other possibilities), which
would nicely explain why one family of quarks and leptons looks like it fits in $SO(10)$.

The idea of Grand Unification goes back more than thirty years \cite{Georgi:1974sy} and has
had several ups and downs. It is partly responsible for the hope for uniqueness of the laws of
physics. I will comment on that in Appendix B.

 If the convergence of coupling constants survives the LHC, we have
 two new kinds of Standard Model structures to wonder about, namely
 not only Grand Unification but also supersymmetry. Most string theorists are very excited about that
 possibility, but is that excitement justified? If we did not know about string theory the answer would obviously
 be positive. We could say that we had discovered new symmetries that point to the existence of a beautiful new
 theory. But string theorists already know what that theory is supposed to be, and it seems to me that  they have
 to look at these discoveries (if indeed they are made) in a totally different way.
 
 String theory does not really need Grand Unification.  It already unifies all interactions with gravity, and
 there is no particular reason why the three gauge interactions should unify first, at an energy scale
 of about $10^{16}$ GeV, three orders of magnitude below the Planck scale. 
  
 Certainly there exist string constructions that produce supersymmetric Grand Unification. This occurs most
 naturally in the case of fully closed string models (in particular the Heterotic String). There are some difficulties
 with understanding the relative low scale of $10^{16}$ GeV, since a Planckian theory like string theory
 naturally favors a higher scale (slightly below the Planck scale, in fact).  But even in Heterotic string theories,
 and certainly in theories based on open  strings, it is at least as easy to get the  Standard Model without any
 intermediate step of
 Grand Unification. 
 
 But then why, if the string theory landscape is correct,  would we
observe  Grand Unification  in our universe? 
Would it not be easier to end up in an observer region if one could  tune
all three couplings separately and independently, rather then being constrained by a relation
between them? 

One possibility would be that Grand Unified solutions numerically dominate the string theory
landscape, if we impose the condition that we should get the Standard Model at low energies.
They were certainly the first kind of solution to be found around 1984, but the first
type of needle found in a haystack is not necessarily the most common one. A proper discussion would require
an actual counting of the valleys in the landscape. If it can be shown that heterotic models
dominate the subset of the landscape that contains the Standard Model, it would still seem
most plausible that the distribution peaks at a value  of the unification scale that is a bit closer
to the Planck scale than $10^{16}$ GeV. But perhaps here the  anthropic principle can help. Perhaps this is
what pushes the unification scale to slightly smaller values in the tail of the distribution, since otherwise
the strong interactions would end up being too strong.

If Grand Unified solutions do not dominate the Standard Model part of the string theory landscape, we may
have to look for anthropic reasons why we see evidence for  it in our universe (if indeed such evidence emerges).
Most anthropic arguments in the Gauge Theory Plane have to do with chemistry and nuclear physics. They
depend on the properties if quarks and leptons at very low energies, less than a GeV.
If we consider the three Standard Model interactions, 
the strength of two of them (the electromagnetic and strong interactions) at  those low energies is important, but the strength
of the weak interactions seems less crucial.
The fact that the three coupling constants are identical at a very
short distance scale (corresponding to $10^{16}$ GeV) has no direct relevance for our existence. 
But Grand Unification may affect the cosmology of the very early universe. Depending on how its realized, it
may imply a phase transition, when we move from the symmetric phase with a unified symmetry to the broken phase
corresponding to the Standard Model. This phase transition may have some relevance for the early evolution of our universe. 
In addition, Grand Unification adds a mechanism for violating the
conservation of baryon number. Indeed, in these
theories protons decay with a lifetime of about $10^{36}$ years. Stability of the proton has obvious anthropic implications,
but that is not at stake here. Life can easily exist even when the proton lives much shorter than $10^{36}$ years. However,
the baryon number violation itself may  play a crucial r\^ole in generating the surplus of baryons  over
anti-baryons  we observe (``baryogenesis"), and which does have anthropic implications. One of the
three aforementioned Sakharov conditions
for generating this surplus is -- not surprisingly --  violation of baryon number.
However, baryogenesis  related to grand unified theories has fallen out of favor as a plausible mechanism, for reasons to
complicated to  explain here. So even though Grand Unified Theories have potentially anthropic ingredients, I don't think anybody
has put these ingredients together in a convincing way so far.

The discovery of low-energy supersymmetry, which would go hand-in-hand with
the discovery of grand unification, poses even more questions. Low-energy supersymmetry means that the not yet
observed bosonic
and fermionic partners of the Standard Model particles have a small mass (in Planckian units), say about 1000 GeV.
Some people claim that string theory predicts supersymmetry\rlap.\footnote{But not everyone agrees with that,  see {\it e.g.} \cite{Dienes:2008tg}. However, a fair comparison of non-supersymmetric and supersymmetric string vacua requires better theoretical control over the former, and
a way of dealing with relative measure factors.}
In presently known string theories supersymmetry plays an important r\^ole in controlling the gravitational quantum corrections
at high energies (although in my opinion the purely string theoretic concept of modular invariance, mentioned in section (5.1),
is even more important). 
The fact that  we seem to need supersymmetry may simply be due to our limited abilities in dealing with {\it any} problem in quantum
field theory, but let us assume that it is true. 
There is a  convincing anthropic argument that supersymmetry cannot remain unbroken, {\it i.e.} that it cannot be
an exact symmetry of nature: electrons would be degenerate with their bosonic partners, and all electrons in excited states
of atoms could decay to their bosonic partners in the ground state; being bosons, they are not subject to the Pauli principle.
Then there would be no atoms of any interest. So supersymmetry cannot be an exact symmetry of nature. This is of course
also what we observe, but if there did not exist any anthropic argument against exact supersymmetry, its absence in our universe
could be a serious worry for any theory -- such as string theory --  that  seems to have exact supersymmetry built in. 
If supersymmetry  is broken, this means
that it can only be observed as a  symmetry at high energies: the degeneracy between bosons and fermions is lifted, their masses differ by
a quantity of order $M$, which may be ignored for energies much larger than $M$. If string theory has anything to do with
our universe (anthropically and observationally), it must remain consistent when supersymmetry is broken at some scale $M$. But
nothing in string theory, at least as far as anyone knows, predicts  the value of $M$.
So even  if string theory predict supersymmetry, it still does not follow that
it predicts {\it low-energy} supersymmetry. In other words, the mass differences between
the super-partners could be anything.

  \subsubsection*{The weak scale}

A related, and very  important point is the weak scale, and in particular its smallness with respect to the
Planck scale: 246 GeV versus $10^{19}$ GeV.   
There is no doubt that this small ratio has essential implications
for our existence \cite{Agrawal:1997gf}, given the rest of the Standard Model.

The smallness of the weak scale has long been regarded as a deep mystery that required a 
solution in a fundamental theory, like the smallness of the cosmological constant. Both quantities 
appear to be ``fine-tuned", which means that the natural size of all quantum corrections
is much larger than the quantity itself. For the weak scale quantum corrections contribute to
the square of the mass so that the observed value,  $(246 {\rm~GeV})^2$, is a factor of about $10^{-33}$
smaller than the ``naive" quantum corrections from quantum gravity, of order $(M_{\rm Planck})^2$.
For the cosmological constant the natural size of the quantum corrections is $(M_{\rm Planck})^4$, and
the observed value is 120 orders of magnitude smaller.
Decades of theoretical
and experimental efforts have been focused at understanding the fundamental physics that determines 
the smallness
of the weak scale, and
many people expect the upcoming LHC experiments to give us crucial information. There is no
convincing theoretical and {\it non}-anthropic argument that determines this scale, although it is  possible
that we will eventually stumble one a mechanism similar to the one the ``explains" the QCD scale: 
a logarithmic modification of the measure. On the other hand, during all those years people have wondered 
whether we should worry about the weak scale hierarchy problem, as long as we had no clue about 
a much more serious one: the cosmological constant problem. If we accept the string theory
landscape and in particular the Bousso Polchinski mechanism to neutralize the cosmological
constant,  this puts us in an interesting situation: there is no mechanism that ``explains" the smallness
of the cosmological constant, other
than the fact that we would not exist if it were a few orders of magnitude larger. But if we accept that solution
for the most horrendous hierarchy problem, then why would we need any special mechanisms for
the lesser one? Perhaps the answer lies in the fact that the two problems are not completely analogous:
the weak scale is associated with a symmetry breaking mechanism, and cannot be arbitrarily small, whereas
the cosmological constant is just a number without any implications except in a very far future. Furthermore,
the cosmological constant may just as well be zero, unlike the weak scale. The theoretical physics
community is very much divided on this issue, with some people still insisting on the need for
a
fundamental mechanism, whereas others would agree with the point of view expressed by Weinberg
 \cite{Weinberg:2005fh} 
``{\it If the electroweak symmetry breaking scale is anthropically fixed, then 
we can give up the decades long search for a natural solution of the hierarchy 
problem.}".

The most popular physical idea associated with the weak scale is supersymmetry.
In supersymmetric theories, quantum correction due to bosons exactly cancel quantum corrections
due to fermions. As already mentioned above, the weak scale is determined by a parameter $m^2$ of dimension
(mass)$^2$, which acquires quantum corrections of order $M_{\rm Planck}^2$. The physical weak
scale $m^2$ is an infinite sum of such corrections, but its observed value should only be about 
$(100~{\rm GeV})^2$. This is where supersymmetry comes in. It nicely cancels all these large
quantum corrections, so that it seems less bizarre that such a small value comes out in the end. 
Supersymmetry does not,
however,  {\it  explain} the actual value, and it  
has a mass parameter, named $\mu$, which in principle can take
any value. The real benefit lies in the fact that $\mu$ does not get Planckian corrections, unlike $m$.
Nevertheless, understanding why $\mu$ is much smaller than $M_{\rm Planck}$ has worried people
so much that this problem has been given a name, the ``$\mu$-problem". 

We have encountered supersymmetry before as a possible fundamental symmetry of string theory,
and also as a mechanism to provide extra particles that make the three 
Standard Model coupling constants converge. However,
the most common argument in favor of supersymmetry is the one of the previous
paragraph, the cancellation of corrections to $m^2$. In order to do its job it does not have to be
an exact symmetry. If the mass differences between the superpartners is roughly equal to $m$, or
a little larger, the quantum corrections to $m^2$ are themselves of order $m^2$ and there is no fine-tuning.
So the absence of fine-tuning requires low energy supersymmetry, broken at a scale of about $1 {\rm~TeV}$. 
Remarkably, this is precisely the mass the superpartners must have 
to make the coupling constants converge. Could this be just a coincidence? In the last few
years, as experimental falsification of low energy supersymmetry might be approaching, there
have been remarkable shifts of opinion.
The lack of confidence that
some people have in the fine-tuning arguments is perhaps best illustrated by an idea called
``split supersymmetry" \cite{ArkaniHamed:2004fb}, in which supersymmetry is mutilated in such a way that it only contributes
to the converge of the couplings, but does {\it not} solve the fine-tuning problem. 

One could say that supersymmetry is a non-solution to a non-problem: the large weak scale hierarchy is
already understood anthropically, and supersymmetry by itself does not even explain it.

If there is low energy supersymmetry is realized in nature, the upcoming LHC experiments should find
the superpartners. This is the favorite scenario
of many theorists. 
Some will say that they have correctly predicted supersymmetry, and they would be right. 
But if it is found, we will have to ask why it is present in our universe. 
Surely the fine-tuning argument is inadequate. Anyone with a basic knowledge of
quantum mechanics should have realized that this argument is not even correct: we can only measure
the sum of amplitudes, and it makes no sense to give significance to individual terms in the sum. 
If supersymmetry is discovered, it was predicted for the wrong reason, and
we will have to find a better one. A large landscape
may ultimately provide a rationale for the fine-tuning argument.
Perhaps 
supersymmetric vacua dominate
the ensemble of Standard-Model-like string vacua
with small values for the weak
scale, because it is ``easier" to get a small value in supersymmetric vacua?
It is not as simple as that. One has to deal at the same time with the fine-tuning
of the cosmological constant. Low energy supersymmetry also helps with that, so that one might
think that it is even more beneficial to have it. But
some simple statistical models suggest that
in fact supersymmetry breaking at a {\it high} scale may dominate the statistics, simply because
they are more numerous, allowing them to overcome their disadvantage. It is not yet
clear whether that is the correct conclusion for the full string theory landscape.
For a longer discussion 
of this issue and further references see \cite{Douglas:2006es}.  As these authors point out, 
if the correct conclusion is that a high supersymmetry breaking scale is statistically favored, 
``{\it the discovery of TeV scale supersymmetry would in some 
sense be evidence against string theory}". It is noteworthy that many string theorists see the
discovery of TeV scale supersymmetry as evidence {\it in favor} of string theory.

With the start of the LHC just months away (at least, I hope so), this is more or less the last
moment to make a prediction. Will low energy supersymmetry be found or not? I am convinced that
without the strange coincidence of the gauge coupling convergence, many people (including
myself) would bet against it. It just seems to have been hiding itself too well, and it creates the need for
new fine-tunings that are not even anthropic (and hence more serious than the one supersymmetry is
supposed to solve).

But even if evidence for low energy supersymmetry emerges at the LHC, in the
context of a landscape it will not be the explanation for the smallness of the weak scale. 
The explanation will in any case be anthropic. The landscape will undoubtedly allow a 
distribution of values for the weak scale, including values outside the anthropic window.

 \subsubsection*{Features that are not observed}
 
 Most string phenomenologists 
 seem to focus exclusively on one point: finding an (approximate)   string vacuum that
 resembles the Standard Model as closely as possible. Of course this is important: if the Standard
 Model is not present in the string landscape, the string landscape is wrong. 
However, as I emphasized earlier, there is another important
 question: why does some feature appear abundantly in some sample, whereas it is not 
 observed in out universe.

 \vskip .5truecm
 \leftline{\em \underline{B-L}}
 \vskip .5truecm
 
  A  good example is $B-L$. In many cases (namely if there are
 right-handed neutrinos), it turns out to be possible to add an extra gauge boson to the
 Standard Model that couples to the difference between Baryon and Lepton number (for
 technical reasons, ``chiral anomalies", it cannot be any other linear combination of $B$ and $L$).
 In the Standard Model one has the option to add such a gauge boson or omit it.
 In string theory, one cannot choose: one gets such an additional massless vector boson, or one
 does not. Most phenomenologists simply reject the cases with and extra massless $B-L$ vector boson
 without asking further questions. In a scan of part of the string theory landscape we found that
 in only $3\%$ of all the cases  this massless boson was absent \cite{Dijkstra:2004cc}.   In such a situation it seems natural
to ask: why do we live in such a relatively rare universe?

In this particular case there is a plausible answer. An additional vector boson coupling to $B-L$ would
change chemistry in a way that is hard to analyze. It would couple to neutrinos, which already by itself
would have drastic consequences for neutrino physics. 
If neutrinos had such a coupling, this would forbid Majorana masses, and hence make it
unlikely that they are light, which in its turn has serious consequences for neutrino processes.
All of this makes it already clear that an additional massless $B-L$ photon
would throw a spanner in the finely tuned works of our universe, but one could still ask whether this
would inhibit  {\it any} kind of life. Perhaps the most convincing argument is that $B-L$ photons would
produce an additional repulsive force in nuclei, between protons as well as neutrons. In the string theory
examples with a $B-L$ photon the strength of its coupling is comparable to that of the normal photon.
This would therefore very seriously affect the stability of heavier nuclei, including Carbon. It is unlikely
that any interesting chemistry would result.

Whereas in this case it is rather easy to  find plausible anthropic arguments against a 
Standard Model alternative, this is not always true. Of course one can always claim
that perhaps we just live in an unlikely universe, but if one has to appeal to such an argument
repeatedly it becomes harder and harder to believe that we really live in a point in the
string theory landscape. 

\vskip .5truecm

\leftline{\em \underline{Exotics}}
\vskip .5truecm

By exotics we mean in general particles that occur in a given string spectrum, but not in the Standard Model. It is  convenient
to exclude particles that do not couple to any of the Standard Model vector bosons from the ``exotics" category. Such particles (``Standard Model singlets")
can only be observed through their gravitational interactions, and it is quite likely that they exist in our universe in the
form of ``dark matter". Furthermore, most string spectra that have been computed so far are supersymmetric, and hence all particles
have a massless supersymmetric partner -- a boson for each fermion and a fermion for each boson. Since they always occur in
supersymmetric spectra,
it is not convenient to label them as ``exotics". 

If one imposes the requirement that the Standard Model with
its three families is contained in a spectrum, one finds essentially always, in addition
to the quarks and leptons, large numbers of exotics. We may distinguish two kinds of exotics,
chiral and non-chiral exotics. Roughly speaking, non-chiral exotics are particles that can
acquire a mass without violating any gauge symmetries, whereas chiral ones can only become
massive 
after a gauge symmetry has been broken. 
Since exotics, by definition, have not (yet) been seen, we must assume that when we move from
the supersymmetric plane to our valley of the landscape, these particles become sufficiently massive.
Clearly, chiral exotics are then a bigger worry than non-chiral ones.

The word ``chiral" comes from the Greek word for ``hand", and refers to the projection of the spin
of a half-integer spin fermion on its direction of motion: a particle is left-handed if its spin points
in the direction of motion, and right-handed if it points in the opposite direction. If a particle is massive
we can overtake it by boosting ourselves to a velocity that exceeds the velocity of the particle. Then
the relative velocity of the particle reverses, while the spin continues to point in the same direction:
it flips from left-handed to right-handed, or vice versa. Since we merely changed the Lorentz frame,
it follows that there cannot be an essential difference between the left- and right-handed particle:
in particular, they must have the same couplings to all vector bosons. No such requirement
exists for massless particles, because we cannot overtake them. Hence the left- and right-handed
components may have distinct interactions; indeed, it is not even necessary that both exist.

Since we compute exact string spectra for massless particles, we can invert the argument:
particles which exist both in left- and right-handed varieties can, in principle, acquire a mass.
These particles are called {\it non-chiral}, because they have no preferred handedness. On the
other hand, particles whose putative left- and right-handed components couple in different ways
to vector bosons cannot become massive, unless we remove the offending vector boson from
the low energy spectrum by giving it a mass. These particles are called chiral. The process of giving
a mass to a vector boson amounts to breaking the gauge symmetry corresponding to that vector boson.

It is usually assumed that non-chiral exotics appear in massless string spectra only by accident.
They could become massive if we just move around in the supersymmetric plane by changing
the moduli. However, when we compute string spectra we do not land on a generic point in the
moduli space, but in special points that are computationally accessible. This accessibility is related
to the existence of symmetries in the string description, and it is believed that these symmetries
are responsible for the extra, non-chiral massless states. So the idea is that the explicit computations
we are able to do are too special to correctly represent the generic situation. There is some evidence
for this point of view, but there is also counter-evidence in certain examples: particles that remain
massless even if one changes the moduli. The latter phenomena
are sometimes referred to as ``string miracles", because
these are particles that one would not expect to be exactly massless outside the context of string theory.
If non-chiral exotics are generically massive outside special points, then they are only a minor problem. 
The fact that we encounter them so frequently would merely be an artifact of the computational 
techniques at our disposal.
But
if this turns out to be false, we have to ask ourselves why the Standard Model spectrum as we
observe it is so much cleaner ({\it i.e.} completely free of exotics that do not fit in one
of he three families) as what we typically get out of
string theory.  Again the word ``generic" appears here, and the same remarks apply as I made
earlier regarding neutrino masses: if everything in string theory is generic, there is little left to
recognize its specific features.

Chiral exotics are a more serious problem. 
They are usually avoided  by simply rejecting spectra that
contain them, since it is not very likely that such particles exist.
Phenomenologically that is reasonable.
All quarks and leptons we have observed
so far are chiral with respect to the $SU(2)$ and $U(1)$ groups of the Standard Model. If there were
any other matter that is chiral with respect to the Standard Model gauge group, it seems reasonable
that we should have seen it already. If such matter can become massive, it needs the Higgs
mechanism to do so, and then the mass would be of the same order as the quark and lepton
masses. The top quark mass (174 GeV) is just about the maximum acceptable value for such matter.
If nature is holding
additional chiral matter up its sleeve, then the entire exercise of computing string spectra and
comparing their chiral matter with observations becomes essentially pointless. 

If the entire gauge group is just the Standard Model group $SU(3)\times SU(2)\times U(1)$ then there
are not that many possibilities for additional chiral matter. One may add a fourth family, but that
is essentially ruled out experimentally. 
But if the gauge group of nature contains additional components,
there are numerous possibilities. This is true for open string realizations of the Standard Model, but
 it may be true for the closed string realizations as well. If we insist on getting the observed three
families of quarks and leptons, but do not restrict the spectrum in other ways, the results are
dominated by spectra that contain extra chiral exotics. This extra matter couples to Standard Model
vector bosons, in particular the photon, and often has half-integer charge. In addition it is charged
under some additional gauge symmetry that is not part of the Standard Model.

Sometimes people admit spectra with chiral exotics out of desperation: without enough statistics
this is often the only way to get something reasonably close to the Standard Model. People who
do have a large enough sample can afford themselves the luxury to simply eliminate spectra
with chiral exotics from their database. There will still be plenty of others left. But, once again,
nobody asks the question why we do not see any evidence for these particles in nature if they 
occur so abundantly in string theory. 

There are a few possible answers. It could be that this problem occurs mainly in open string
realizations (which is what my own experience is based on). Perhaps open string realizations are
simply sub-dominant in the full string theory landscape. Perhaps the statistics in the supersymmetric
cases that have been studied
is not representative for the actual landscape.
In some specific cases one can
invent dynamical mechanisms that break the gauge symmetries that prohibit a 
 mass. 
Perhaps these particles then simply become massive after all, or occur only in massive bound states.
The trouble is that, unlike the 
case of non-chiral exotics, this does not work ``generically", but only in special cases.  Chiral exotics
that do not get a mass at all would certainly have anthropic implications: since they couple to
Standard Model forces, they would get involved in chemical and nuclear bound states. But it would
be hard to argue that this would make life impossible; it simply adds extra uncontrollable complications.
At this moment I do not have a plausible argument why we do not see such exotics in nature. The best
thing that could happen is that the LHC provides experimental evidence for such particles, having acquired
just enough mass to have escaped detection so far.

 \subsection{Misguiding principles?}

For the last thirty years, attempts to understand what might exist beyond the Standard Model
have been guided by four basic principles:
\begin{itemize}
\item{Consistency of the theory}
\item{Agreement with experiment}
\item{Naturalness}
\item{Beauty, elegance, simplicity}
\end{itemize}

The first two guiding principles are unquestionable, but the last
two may well turn out to be misguiding principles. 

The issue of ``naturalness" arises in cases where
dimensionless numbers are small. The principle of naturalness says that
one would not simply expect such a number to come out
of a fundamental theory as number of order one that just ``happens" to be small, but that
there should exist some underlying mechanism that explains its smallness. 
The most serious naturalness problem is the smallness of the cosmological constant
by 120 orders of magnitude. In the string theory landscape this is solved by
 having a distribution of values over a wide
range,  including the anthropic window. This brutally violates the idea of naturalness, and one
should be prepared for more examples  whenever naturalness meets the anthropic principle.
The next frontier is the hierarchy between the weak scale of 100 GeV  and the Planck scale
of $10^{19}$ GeV, as discussed above.

The fourth guiding principle is also of dubious value.
In the string theory landscape, there is not much reason the expect the point
in the Gauge Theory Plane where we live to be especially beautiful or simple if there
are $10^{500}$ other competitors in the beauty contest. Perhaps string theory itself can be
called beautiful, or perhaps a truly elegant description will be found
in the future, but that has nothing to do with beauty of its ground states. Nevertheless, within
the string community it is standard practice to 
distinguish string vacua on the basic of esthetic criteria. Perhaps arguments {\it can}
be found to justify that, but it would have to involve some knowledge about the
distribution of vacua in the landscape. Perhaps certain ``esthetic" features are simply
more common. 

It is also imaginable that in typical string vacua most matter acquires large
masses, and that the Standard Model, with many light particles is very atypical. In that
case, one would expect that the only those particles are light that are required
for our existence. Having any additional light particles would simply lower the
statistics without any benefits. This would indeed imply a certain degree of simplicity.
Of course the Standard model seems to contain lots of inessential particles, but 
one can only remove an entire family at once.

\subsection{Natural selection?}

If there is a huge landscape, did we end up in our place just be chance after a humongous number of
tries, or is there a better way?
An interesting idea put forward by Lee Smolin \cite{Smolin:1994vb} makes an analogy with the theory of evolution. 
An essential feature of the multiverse is some process  of eternal creation of new universes. The
currently most
popular process is eternal inflation, in which bubbles of new universes   are formed by quantum tunneling.
Smolin considered the possibility that new universes are created in the interior of
black holes. This idea has been criticized, but 
let us not focus on a concrete mechanism, but on the idea of natural selection itself.
Regardless of the precise mechanism,
one could contemplate the possibility that a child universe inherits some property from its parent. If that property
furthermore enhances the number of child universes, then it is to be expected that in the total sample universes with that
property will dominate. This would be the analog of natural selection in evolution. In evolution, there is also
an analog of the landscape of possible universes.  It is the entire set
of all possible biochemical molecules and combinations thereof, including all possible forms of life. It is clear
that  in that biological landscape natural selection is absolutely essential in order to understand how
we managed to reach the complexity of life on our planet in a few billion years. 

All though I have some sympathy for this idea, Smolin proposed it as a remedy against the
anthropic principle. Perhaps that is correct for some forms of that principle, but certainly not for the point
of view I have been advocating
in this article: the likely existence
of a huge landscape.  Clearly natural selection can only work if such a landscape exists in the
first place, just as biological evolution can only work if there is a biological landscape. 
None of the arguments I gave in chapter (\ref{TGC}) are invalidated
if there is a natural selection process that guides us to our valley exponentially faster than  a random process.

Another objection is that creation of universes  is  necessarily a process on a cosmic scale, and
it seems preposterous that such a process should care about the minor chemical accident we call life. 
In the concrete case of Smolin's proposal, there is one important ingredient for our kind of life that plays
an important r\^ole, namely the formation of stars and the collapse of some of them into black holes. But these
are mainly nuclear processes, and therefore I would expect universes  to exist where black holes
are formed, but the chemistry for life is just not available. In that case we would still have to appeal  to the anthropic
principle to explain why we live in this particular universe, within the set of those with abundant stars or black holes.
In other words, cosmic natural selection, if it can be realized at all, might produce a  large peak in the distribution somewhere
in a  large neighbourhood around the anthropic region, but it would be miracle if it were to coincide with it. It just
might guide us a little faster to the right region, but  it can never pinpoint the right spot. 

An essential ingredient of such a process is that small changes in parameter values take place at the birth of a new
universe. Exactly as in the case of evolution, large changes don't work; they would randomize the process.
However, in the string theory landscape, an extremely small change in  a Standard Model parameter implies
a huge change in the cosmological constant. This may be compensated by a discrete change of the Bousso-Polchinski
fluxes, but that would put us in an entirely different part  of the landscape. In other words, points in the
Gauge Theory Plane that are close  to each other are generally very far from each other in the full string theory
landscape. Therefore natural selection by small ``mutations" of universes seems impossible in the string theory
landscape.

There is another important, and somewhat disturbing, difference with biological evolution, namely the
absence of an obvious time interval required to reach our universe. The adjective of  ``eternal inflation" says it all.
What is worrisome about this is that  ``eternal" implies an infinity, and infinities can have nasty
consequences. In this case they convert any option, no matter how improbable, to  a real possibility. This
even tends to undermine the argument that biological natural selection is needed to explain how the
complexity of live could emerge in such a short time. With enough universes  to try, in one of them it may
have worked all in one step! This raises the question why had to pass through the complicated process of evolution
if it could have happened so much more easily? The answer to questions like these requires a proper
understanding of the thorny issue of probabilities, and may well be beyond the edge of our current
knowledge.

\fi

\section{Conclusions}

\ifPreprint
%\subsection{The moral}
One of my main points is nicely summarized by the following version of Hans Christian Andersen's
fairy tale:

{\it 
Many years ago, there lived some physicists who cared much about the uniqueness of their theories. 
One day they heard from about a beautiful theory,  which was absolutely unique. 
They were told that  only people stupid
enough to  be open-minded about the anthropic principle would not be able to see that.

Of course, everyone wildly praised the magnificent unique theory, 
afraid to admit that it had ``anthropic principle" written all over it, 
until Susskind shouted:

``String Theory has an anthropic landscape"

\noindent Then some people claimed to have observed this already.}\fi

\subsection{Will we ever know?}

As I have emphasized repeatedly, the use of anthropic reasoning here is not the 
simple statement that some property of our universe could be different, but involves
a fundamental theory that specifies precisely {\it how} it could be different. Although those
differences are presumably realized in some other part of a multiverse, it is not likely
that we can check that. Perhaps it might be possible to do that {\it in principle}; it has
been suggested that information about ancestors of our own universe might be
encoded in the microwave background radiation, in the same way as black hole
radiation is now believed to contain the information that went in it; it has also been
suggested that signals of other parts of the multiverse might be visible in the future
(which of course is measured in billions of years)
when our horizon has expanded sufficiently; there might even be a dramatic confirmation
when our universe tunnels to another minimum in the landscape. But in all these cases
{\it in principle} really means {\it in principle}. 

This is not an acceptable state of affairs, if indeed it is the final word. But it is a completely
acceptable outcome if we can establish the correctness of string theory 
(or some other fundamental theory) beyond any 
reasonable doubt. 
During the last two decades there was some reason to hope that we might be able to
do that my means of some prediction of a Standard Model feature. That hope is fading now.
I am not saying that this will never happen, but I have seen too much wishful thinking
to make an optimistic statement about this.
Essentially,
we came to that conclusion already in 1986 \cite{Lerche}. We are dealing with a theory of gravity. Getting information
about it through the back door of particle physics is a luxury that we once had good reasons to hope for, but that may not exist. 
Rejecting a theory of gravity that makes no particle physics prediction may be like
rejecting the theory of continental drift because it does not predict the shape of Mount Everest. 

One cannot count on any
direct experimental check of a theory of quantum gravity, since any observable consequences it might have
are extremely small, unless we are extremely lucky.
Indeed, when 't Hooft en Veltman tackled Quantum Gravity in 1974 \cite{'t Hooft:1974bx} they
wrote about their motivation: ``{\it One may ask why one would be interested in quantum gravity.
The foremost reason is that gravitation undeniable exists. [ ....]
Mainly, we consider the present work as a kind of finger
exercise without really any further underlying motive}".
The problem of quantizing gravity is a problem
of consistency, which can be checked by theoretical reasoning. Perhaps that is indeed the only
way. As I have indicated in chapter (\ref{StringTheory}), in string theory there is still a long way to go before
success can be claimed on that point. The best imaginable outcome would be a proof
that quantum gravity {\it requires} string theory; on the other hand, it is
possible that we end up with a huge ensemble of unrelated theories of quantum gravity, each with
its own landscape, and without any way of distinguishing between them.

Although the Standard Model may not provide us with a {\it verification} of string theory, it
might easily have falsified string theory in many ways. It might have been impossible
to get chiral fermions, the Standard Model gauge group, quarks and leptons, the right
number of families or non-trivial couplings. All of these could have easily  gone wrong,
but did not. For comparison, if instead of the Standard Model we would have tried
to get the Periodic System directly out of string theory (treating nuclei as fundamental
particles), we would have failed convincingly (see appendix A). A failure to stabilize the moduli, or
to obtain a small, but non-zero  cosmological constant would also eventually have led
to abandoning the whole idea. It is not hard to imagine future experimental results
that would cast serious doubts on the entire landscape idea. If the upcoming LHC results point
towards new strong interactions around the TeV scale, we have to revise the entire
extrapolation of the Standard Model to the Planck scale; the same is true if the  Higgs boson
 weighs more than 180 GeV. If there is no fundamental Higgs scalar at all, we might feel
 less confident about the r\^ole of fundamental scalars 
 as moduli, and we certainly have to rethink the Standard Model itself, before even
 attempting to embed it in string theory. New particles with unusual charges
 or representations could make such an embedding essentially impossible.  
 
 There is yet another, extremely optimistic possibility for the LHC, namely that
 it might  find direct evidence for string theory. This could 
 happen if our naive estimates for the Planck scale are wrong. It could in fact be directly
 accessible for LHC experiments if there are extra dimensions that are large
  \cite{ArkaniHamed:1998rs}
  \cite{Antoniadis:1998ig}
  or warped
   \cite{Randall:1999ee}.
In the first of these ideas the weakness of gravity (or in other words the smallness
of the weak scale)
is explained because at short
distances of order $.1$ mm or smaller (but much larger than the Planck length) space-time has
more than four dimensions, allowing the field lines of gravity to get rapidly diluted.
If we are really lucky, we might observe  excited states of string theory and even produce
black holes. 
However, although the idea of large extra dimensions found its origin in string theory, it 
is not clear if it is a really a dominant feature of the string theory landscape, or at least that
part of the landscape that can accommodate a small weak scale. Studies of different classes
of string theories have led to 
opposite conclusions on this issue \cite{Douglas:2004zg} \cite{Cicoli:2008va}.

 There is an impressive experimental effort going on in the area of astro-physics, which
 could modify our current ideas about gravity and the cosmos. 
 If any deviations from General Relativity are observed, this would have profound
 repercussions on our current understanding of string theory. A very
 important constraint is the possible variation of constants of nature. 
If it is found that some of them are different in some part of the visible universe, this would
imply that this part of the universe should have a very different vacuum energy, and
hence should have collapsed or rapidly expanded \cite{Banks:2001qc}. If we observe
such a variation experimentally, then either its contribution to the vacuum energy
is compensated by another, unobservable variation, or our entire understanding of
vacuum energy and its relation with the cosmological constant is wrong. In the former case,
the landscape has a valley along which physics can change without a change in vacuum
energy, which would imply the existence of a (nearly) massless scalar field with significant
couplings to observable quantities. In the latter case, a basic
principle behind of the string landscape idea would have been falsified. \ifPreprint
This may already be the case: in a recent paper
\cite{Reinhold:2006zn}\ the observation of a small time variation of the proton/electron mass ratio (of order $10^{-5}$ over 12 billion years) is claimed. If correct, this would be very hard to reconcile with the idea that the
cosmological constant is fine-tuned with 120 digit precision, and therefore extremely sensitive to even the smallest imaginable variations.

Most of these results would wreak havoc in other areas than just string theory, and one cannot
rule out the possibility that some reformulation or renewed understanding of string theory
might revive it. But in any case it underscores the fact that string theory is not immune to falsification,
and that it is in fact a small miracle that it has not been falsified already. Indeed, it is truly remarkable
that something like the Bousso-Polchinski mechanism exists to avoid a falsification by the
cosmological constant.
In string theory, this is not something one can
engineer: it is either already there, or it is not.  I find it difficult to believe that a wrong theory 
would produce such spectacularly correct results. 

The foregoing  is just
a small selection of observations that would change our point of view completely and that
might occur in the near future. On longer timescales, 
it is clearly ridiculous to pretend that what we currently know will be the state of the art forever.
When Darwin formulated his theory of evolution he was unaware of Mendel's results on
inheritance, and could not even have imagined DNA. It may well be that these are the kind
of time-scales we have to think about, but I am convinced that eventually mankind will gather enough information
to arrive at a definitive conclusion.

\subsection{Final remarks}

\fi

Finally, after 15 years  the debate has started that should have started around the mid-eighties of last century,
but was stifled by irrational opposition against the notion that our observation of the  Standard Model could be
biased by our own existence. \ifPreprint To me at least one thing seems absolutely obvious: the idea that the Standard
Model is (even approximately) unique will eventually find its place in history next to Kepler's attempt
to compute the orbits in the solar system: understandable at its time, but terribly anthropocentric when
properly thought about.

\fi Far from ``giving up", within the context of string theory and its landscape
our new mission is exactly the opposite. We have to understand the landscape, find our place in it, and determine if it is really plausible
that we find ourselves in that particular place and not elsewhere. Despite the huge size of the landscape, it
is not guaranteed that the Standard Model is in it. 
Because of the huge size, there is a risk
that within the ensemble of universes that allow life, ours is  incomprehensibly rare. 

%A proof -- most likely by statistical methods -- that the Standard Model is realized somewhere in the string theory landscape may be within reach.  But such a proof would be based on what we know today.
%Future experiments, and in particular the LHC,
%will extend our knowledge of the Standard Model, and it is not hard to think of experimental results that
%would ruin the possibility of extrapolating directly to the Planck scale. New particles or interactions
%may be discovered that make an embedding in string theory impossible or extremely awkward. Tens
%of ongoing and upcoming astrophysics experiments may change our view of the universe so dramatically that
%most of what I wrote here would be wrong. 
%In the next two decades, confrontations with experiment and observation could easily fail
%so badly that we will have to dismiss the string theory landscape and look for another one. 

Every theory
of quantum gravity has a landscape: the matter it can couple to. There are other ideas for quantizing gravity,
but the reason I called string theory the ``only candidate" earlier in the article is that it is the only
approach that is sufficiently advanced to provide us some insight into its landscape. It would be interesting
to know what alternatives other approaches have to offer. I think that a unique solution cannot be
expected in any approach, and that a continuous
infinity distributed over the Gauge Theory Plane would not be the right answer.

%\ifPreprint
%If everyone adopted the anthropic point of view, this might
%be bad for our field, as some critics argue. But this is equally true for any other point of view.
%Progress in science thrives on debate about conflicting ideas. It is precisely the almost
%exclusive dominance of anti-anthropic, pro-uniqueness thinking that has cut off important avenues of
%progress in the last two decades.
%It is not hard to see how much damage this can cause. It may lead to a lack of interest or even a rejection
%of a theory that gives perfectly reasonable answers: string theory. It would not be the first time. Around
%1975 some people realized that string theory might be a theory of gravity. This is not what people hoped
%or expected, and as a result interest dropped dramatically. Around 1984 there was a huge surge in popularity
%of string theory precisely for the same reason: that it was a candidate theory of quantum gravity.
%History may repeat itself. Bad expectations lead to disappointment and wrong conclusions. Of course in
%due course a future generation will liberate themselves from the prejudices of the previous one, and
%the subject will have another revival.
%\fi

I have presented string theory here as concept that is still very much under construction, but which makes a
very clear suggestion about what a ``final theory" might look like. 
To me, what is emerging looks very appealing. It fulfills and even exceeds the hopes I expressed
in 1998. It is has been amazing to see this theory leading us in the right direction, sometimes even against the initial
expectations  of most
of the people working on it. We should continue to follow its lead, and do everything in our
power to strengthen its theoretical underpinnings. The emergence of a huge
``landscape" makes this more worthwhile then ever before. 

String was expected to answer some of the big
questions in science. Indeed it does. What bigger questions can a scientist dream of than those concerning
our own existence? We got an answer to one such question,
but unfortunately not everyone is ready for it.
 One can easily justify the past two and a half
decades of research in string theory by the theoretical and mathematical spin-offs alone, but I think
we are on our way to something much bigger than that.

But will this road really lead us to the ``final theory"? Was this the last time we were mesmerized by
misguided ideas about our special place in the multiverse? Was this the last time the emperor
turned out to have no clothes? Could string theory just turn out to 
be the first item in a huge ensemble of theories with similar properties? 
I have no answers to these questions. History does tend to repeat itself, but unfortunately
not in a predictable way.

\vskip 1.truecm

\centerline{\bf Acknowledgements}
\vskip .2truecm

This article is based on theoretical physics colloquia in DESY, Hamburg, and at the university of Utrecht.
I would like to thank Jan Louis and Stefan van Doren for the invitation, and Peter Zerwas for inviting me
to write this up for ``Reports on Progress in Physics". This work has been partially 
supported by funding of the Spanish Ministerio de Educaci\'on y Ciencia, Research Project
BFM2002-03610. Thanks to Joe Conlon, David Maybury and Gordy Kane for their constructive criticisms of the
first version of this manuscript.

\ifPreprint

\appendix

\section{The Standard Model versus the Periodic System}

I have presented the Standard Model as a high precision theory with clearly defined
parameters, a property not shared by the old theories of nuclear and hadronic physics. 
Based on this fact, and on the apparent anthropic coincidences in the Gauge Theory
Plane, I have argued that it is plausible that a Landscape of possibilities should 
survive in any more fundamental theory. Such an argument would clearly be meaningless when
applied to the space of nuclear or hadronic theories, which we cannot even define. 
There is, however, one predecessor of the Standard Model that might qualify as an
alternative, namely the Periodic System of Elements. 

One often hears the statement 
that physics would never have progressed if people had surrendered to the anthropic principle too soon. 
So let us do a historic thought experiment. Let us imagine that string theory 
and gauge theory had reached its
current state many decades earlier, as a mathematical theory. 
Is it conceivable that
someone would have tried to apply it to understand the Periodic System, the Standard Model
of their time, and arrive at incorrect
anthropic conclusions?

By  ``Periodic System" I mean here concretely a theory with nuclei treated as point-like elementary
particles, with an electron, and with only electromagnetic interactions. This is a gauge theory,
so it is in fact a point in the Gauge Theory Plane.  

In order to do this though experiment we have to assume that the physicists of that time were
foolish enough to overlook some obvious problems, such as the fact that the nuclei are not stable,
that we need to understand their abundances, and that this purely electromagnetic theory does not
explain how the sun works. This alone would be enough to answer my rhetorical question. 
Perhaps there are problems today that I am foolishly overlooking as well, something that so
obviously requires entirely new physics that it is premature to even consider a fundamental theory.
I am taking that risk.

So what can be varied in the Periodic System, treated as a point in the Gauge Theory plane? Obviously
this includes the electron mass, the strength of the electromagnetic coupling and the overall
mass scale of all the nuclei. These are less strongly anthropically constrained than in the
case of the Standard Model, because we decided not to consider abundances, and because
there is no competition between the strong and electromagnetic interactions. But anyway, any
anthropic considerations for these parameters are contained in those of the Standard Model,
and hence would not be premature, if the latter are correct.

This leaves us with the charges and the nuclear mass ratios. The charges are sometimes mentioned
as a candidate for anthropic consideration: if the electron charge is not equal to the proton charge
with enormous precision, atoms would have a net charge and we would not exist. This is an excellent
example of misuse of the anthropic principle for at least three reasons.

First of all, it is not natural to assume that these charges are variables. 
It is true that the nuclear charges can be treated as true variables
in the description of the Periodic System. However, anyone would immediately have noticed two facts:
that the charges are integers within experimental precision, and that atomic physics does not change
smoothly if the charges are allowed to vary over the real numbers: one cannot interpolate
smoothly between a hydrogen and a helium atom, for example. Hence the most natural assumption would be
that they belong to the set of integers, not to the set of real numbers. Indeed, it is unimaginable
that someone would propose a dense landscape of Periodic Systems with arbitrary real charges for the
nuclei, just to explain why all their observed charges are indistinguishable from integers.

The second fallacy is the ``anthropocentric trap". Whether {\it we} would exist if the charges
were non-integer is irrelevant. One would have to show that nothing of comparable intellect can
exist. It is beyond our knowledge of atomic and molecular physics to conclude that nothing
of equal complexity could exist if we change some of the charges.    

The third argument can only be made with current insights, and is relevant because the
integer charge anthropic argument is still heard from time to time. In the Standard Model
we already know that the proton charge is the exact opposite of the electron charge with
absolute precision, because the consistency of the theory demands that. I am referring to
``chiral anomalies", which are cubic sums of charges that must vanish exactly. Although there are
several ways of satisfying them, changing for example the electron charge while keeping
everything else fixed is not an option. 
These anomaly constraints are only relevant if left- and right-handed fermions have different
couplings, as is the case for the weak interactions. Hence this idea would not have occurred
to a physicist studying only the periodic system. However, I was assuming that String Theory and
gauge theory had already been developed theoretically, and anyone with knowledge of these theories 
would have understood that a fundamental theory explaining charge quantization was entirely feasible.

The other parameters to be considered are the nuclear masses. With enough experimental
precision these would have been distinguishable from integers, because of binding energies 
and isotope averaging. But even if they would have been treated as free parameters, they have
little anthropic relevance. Again, what matters is not whether we would exist, but if anything
of comparable complexity would exist. The complexity of atomic and molecular physics is determined
almost entirely by the charges. Indeed, even if we multiply the mass of the hydrogen by two, we 
get heavy water, a substance with physical and chemical properties very similar to normal water. 
So it seems plausible that chemistry of comparable complexity will exist over a wide range of
the Periodic System plane, if we vary the masses. So no incorrect anthropic conclusions are
possible here. 

The imaginary string theorists in the Periodic System epoch would encounter serious problems
if they tried to do what present-day string theorists are attempting: obtaining their
low-energy theory from string theory. The first problem is that it is in fact not true that the
Periodic System extrapolates to the Planck scale. The reason is that the Landau pole 
of electrodynamics in such a theory
occurs
at energies well below the Planck scale. The coefficient $b_0$ in eqn (\ref{RunC}) receives contributions
from all elementary particles, and these contributions are proportional to the square
of the charge. The assumption we were making about the Periodic System is
that all nuclei  ({\it i.e.}  all those observed) are elementary. But then they will all make separate contributions to the quantum
corrections that make the gauge coupling increase. The contribution of
Hydrogen en Helium alone is already equal to that of all 
three quarks of charges $\frac{2}{3}$ and three of charge $\frac{1}{3}$; by the time we have included
all nuclei up to Carbon the Landau pole has already moved below the Planck scale.

Even if one ignores that, it would be practically impossible to obtain a set of particles from string theory
with charges ranging from 1 to, for example, 20, in sharp contrast to getting the gross  
features of the  Standard Model, which come out easily.

\section{Grand Unification, little uniqueness}\label{GUT}

Grand Unification is the unification of the three known non-gravitational
forces into one a gauge theory based on a single gauge group. 
It is sometimes mentioned as a sign of uniqueness of the Standard Model, or
at least as the origin of a misguided expectation of uniqueness. 

Grand Unification is based
on two miraculous facts. The Standard Model group and representations
fit nicely in a larger group, and the three Standard Model couplings, when
extrapolated to higher energies, seem to meet at a common point at an energy
just below the Planck scale. The last miracle requires another speculation:
supersymmetry. The couplings only merge if we assume that every Standard Model
boson/fermion has a hitherto unobserved fermionic/bosonic counterpart\rlap.\footnote{Plus an additional
Higgs scalar and the fermionic partners of the two Higgs scalars.}, contributing
to the coefficients $b_0$ in (\ref{RunC}). 
These ``super-partners" must have a mass of at least a few hundred times the proton mass to have 
escaped detection up to now. Furthermore, the normalization of the  $U(1)$ coupling constant
must be modified by a (theoretically motivated) factor $\sqrt{3/5}$ in order for this to work.

The merging of coupling constants may just be a coincidence. The chance that it is, based
on current experimental errors, is about one percent. In other words, if we assume that the value
of the weak coupling varies over the multiverse and has no anthropic relevance, 
the inhabitants of about one in a hundred universes would be puzzled by this coincidence.
Fortunately we will not be puzzled forever. If the necessary superpartners exist they must be found
at the LHC. If the apparent merging of the couplings survives the LHC results and becomes even
more accurate, it will be hard to argue that it is a coincidence.

The second grand unification miracle is that the group $SU(3)\times SU(2) \times U(1)$ fits exactly
in the group $SU(5)$, in such a way that one family (see Eqn.(\ref{SMrep})) fits precisely in the representation
$(\bar {\bf 5})+({\bf 10})$ of $SU(5)$. One can go one step further and embed $SU(5)$ into $SO(10)$. Then one
family (including a right-handed neutrino) fits into a single representation $({\bf 16})$ of $SO(10)$.

Even readers who are not familiar with the group-theoretical details will understand the 
$({\bf 16})$ looks a lot simpler than Eqn. (\ref{SMrep}). In fact, there {\it is} something mathematically unique about the $({\bf 16})$ 
of $SO(10)$: it is the smallest complex irreducible cubic-anomaly-free representation of any Lie group 
(please ignore the precise meaning of this mathematical jargon).
Doesn't this point towards uniqueness of the Standard Model? Not really. First of all, families are repeated
three times, so that we really have a representation $3 \times ({\bf 16})$. This ruins both the ``irreducible" and
the smallest in the statement. Secondly, there are other particles in these models: Higgs particles needed
to break symmetries. They must in strange representations like $({\bf 45})$, $({\bf 54})$ or $({\bf 126})$. There is nothing
mathematically unique about these. In string theory, the symmetries may sometimes be broken in a
different way, but certainly not more uniquely.

But there is a more serious problem with the concept of uniqueness here. The groups $SU(5)$ and $SO(10)$ also
have other subgroups beside $SU(3)\times SU(2) \times U(1)$. In other words, after climbing out of our
own valley and reaching the hilltop of $SU(5)$, we discover another road leading down into a different valley
(which may or may not be inhabitable). When I was confronted with this around 1977, it started my doubts about
the uniqueness of the Standard Model. I studied the parameters that determine the valleys, and tried to
find values such that ours was the lowest one. This was possible, but I understood that I was just fooling myself:
this just shifted the problem into the choice of parameter values.  The thought also occurred to me that not
only the choice of gauge group, but also the Standard Model parameters might be fixed
at the start of the universe as a sort of boundary condition, and could not be computed from first principles.
I am not sure if I attached any anthropic implications to that at the time, but I remember being deeply
troubled by this.

In other words, people who associate Grand Unification with the idea of uniqueness are simply confusing
the two concepts. 

The story continues. Even if there  is something conceptually nice about the
fact  one family of the Standard Model fits into
the $(16)$ of $SO(10)$, it still is a choice out of many possibilities.  
Is there a natural way of realizing this in a fundamental theory of physics?
Indeed, there is a fundamental theory that naturally realizes this: string theory, more precisely
the heterotic string. This theory has a gauge group $E_8 \times E_8$ in ten dimensions. The group $E_8$ 
contains $SO(16)$ as a maximal subgroup (the other $E_8$ is not used), and $SO(16)$ has a  natural
decomposition into $SO(6) \times SO(10)$. There is also a natural reason to  split it in precisely this
way: The $SO(6)$ is the rotation group in the six extra dimensions\rlap.\footnote{This
can be made precise most easily in the supersymmetric case: $E_8$
decomposes to $SU(3) \times E_6$, and $SU(3)$ is the holonomy group of a Calabi-Yau manifold. The
group  $E(6)$ contains $SO(10)$.} In this way we do indeed get $SO(10)$, but not necessarily with three families.
The number of families  depends on the topology of the compactification manifold, and there is a huge
number of possibilities.

However, even though the appearance of $SO(10)$ and the family structure is rather natural in Heterotic
string theory,  it is by no means unique, and this  gets worse if one considers further breaking of $SO(10)$
to the Standard Model. In landscape terminology: if we continue climbing from the $SU(5)$ hill to $SO(10)$
and finally reach the summit, $E_8  \times E_8$ Heterotic string theory, we find that there is a humongous
number of other trails leading to other valleys than the one we came from. And higher up, partly in the clouds,
we see the outline of a still higher and rather mysterious summit, called M-theory, from which
even more paths descend down into even more different valleys.

\section{Dimensional transmutation and the measure}

The Standard Model has two mass scales:
The QCD scale and the weak scale.
The existence of two small scales implies two problems: why is the QCD scale about twenty
orders of magnitude smaller than the Planck scale, and why is the weak scale seventeen
order of magnitude smaller than the Planck scale?  Both of these small ratios have
anthropic implications.

The second of these problems is often referred to as the gauge hierarchy problem, and there
is a huge number of papers addressing it. 
The first problem, on the other hand, is often regarded as ``solved",
and this is used as a counter argument against anthropic reasoning. 

The argument goes as follows \cite{Gross}. Dirac already knew that the ratio of the proton mass and
the Planck mass could be regarded as anthropic, but to his credit -- the opponents of
anthropic reasoning say -- he did not use that argument to explain it. Instead he proposed
that this ratio could be time dependent, and become smaller and smaller as time goes by.
The large lifetime of the universe would then supply the large number needed 
to explain the small ratio. While this was a nice idea at the time, it was of course wrong, but
according to opponents of anthropic arguments even a wrong idea is better than invoking
the anthropic principle. In my opinion, it would have been equally reasonable for Dirac
to have said: ``maybe in a more fundamental theory we will understand that the proton mass
has many possible values, and we see a proton with a small mass because otherwise we
would not exist to observe it". The words ``in a more fundamental theory" are crucial.
As I have already emphasized many times, one cannot simply assume that some parameter
can take many values. There must exist a theory that tells us what the allowed variations are. This is
of course what the string theory landscape provides us with. 

Meanwhile we understand the proton mass in terms of the QCD scale, and asymptotic
freedom of QCD seems to give a convincing, non-anthropic answer:
\beq
\label{DimTrans}
{M_{\rm proton} \over M_{\rm Planck} }= e^{-{c\over \alpha_s(M_{\rm Planck})}}
\eeq
This relation is obtained by inverting Eqn. (\ref{RunC}). The proton mass is
of the same order as the QCD scale, which is the mass scale where  (\ref{RunC}) has
a pole. We can use (\ref{RunC}) to express the position of the pole in terms of
the value of the coupling constant at any scale, and we choose the Planck scale.
This relation is sometimes referred to as ``dimensional transmutation", since it
seems to generate a dimensionful parameter (the QCD scale) out of a dimensionless
number, $\alpha_s$.
Here $\alpha_s$ is the strong coupling constant, the analog of the fine structure constant
of QED. It is a function of energy, which is evaluated here at the Planck energy (which is
how the mass dimension enters into the equation). The coefficient
$c$ is a  positive numerical constant of order 1. The ratio on the left hand
side is small if the value of  $\alpha_s(M_{\rm Planck})$ is small, which of course it is.
But that, by itself, cannot yet be called an explanation of the smallness of the ratio unless
we can compute $\alpha_s(M_{\rm Planck})$. What the equation actually does is turn a major
miracle into a minor miracle: instead of having to understand why in our universe we measure
a number as small as $10^{-19}$, we now only have to understand why it is of order $10^{-1}$. 

Of course this does take the urgency out of the problem, but unless someone actually
computes $\alpha_s(M_{\rm Planck})$ from first principles, it is not the end of the story. If we end up with a 
fundamental theory that has a landscape, what one would expect to find is a distribution of
possible values of  $\alpha_s(M_{\rm Planck})$. This distribution will undoubtedly
 have values outside
the anthropic window, in which case we will once again have to resort to anthropic arguments
to understand why we find ourselves {\it within} that window.
Eqn. (\ref{DimTrans}) {\it does} tell us something very important, namely that before we
worry about the smallness of a parameter, we should worry first about the measure on
the parameter space. But in a landscape picture it only {\it changes} the measure, it does
not provide us with the definitive measure on the space of couplings. This we will only know
once we know the actual distribution of the values of $\alpha_s(M_{\rm Planck})$ in the
landscape.

There might exist a way to improve this argument. Perhaps the smallness of $\alpha_s(M_{\rm Planck})$
could be related to the smallness of the fine structure constant $\alpha$, which is also
a function of energy.
This is indeed true
in Grand Unified Theories (see appendix B), which at this moment is still a theoretical speculation
awaiting experimental evidence.
In those theories both functions are actually equal at some
GUT mass scale $M_{\rm GUT}$, below which the strong coupling increases and
the QED coupling decreases with decreasing energy. This would  indeed look like a counter indication
to anthropic reasoning, because it relates two quantities with anthropic implications\rlap.\footnote{Gauge coupling unification relates {\it three} couplings to two input parameters,
$M_{\rm GUT}$ and the value of the unified coupling at that scale. The third one, the
strength of the weak coupling, has far less severe anthropic implications, and hence is not likely to
enhance the amount of anthropic coincidence by much.} 
 However,
note that the two quantities are  determined in terms of two other parameters, the GUT
mass scale and the value of the unified coupling at that scale. Mathematically speaking this
does not imply a relation at all. In reality, it does imply a limitation of the allowed values,
if one imposes the condition that $M_{\rm GUT}$ should be smaller than 
$M_{\rm Planck}$. This would be a reasonable assumption
if $M_{\rm Planck}$ is the largest possible mass scale in physics. 

If the GUT picture survives future experiments, it is undeniably true that there is something
remarkable about the fact that $M_{\rm GUT}$ is close to $M_{\rm Planck}$. Had it been equal
to $M_{\rm Planck}$, one could claim a direct relation between the QCD scale 
and fine-structure constant. But it is {\it not} equal, and hence $M_{\rm GUT}$ is
 just a free parameter with  a value that is remarkably close to the Planck scale.

\fi

\end{document}